\documentclass[12pt]{article}
\pdfoutput=1

\usepackage[a4paper,text={16.8cm,22.4cm}]{geometry}
\usepackage{amsmath,amsfonts,braket,slashed,amssymb,tikz,bm,psfrag,graphicx,color,dsfont}
\usepackage{multicol}
\usepackage{ctable}
\usepackage[small,labelfont=bf]{caption}
\usepackage{hyperref}

\RequirePackage[sort&compress,square,comma,numbers]{natbib}
\allowdisplaybreaks
\newcommand{\hhref}[1]{\href{http://arxiv.org/abs/#1}{arXiv:#1}}

\begin{document}

\begin{titlepage}

\begin{flushright}
\normalsize
MITP/17-047\\ 
August 1, 2017\\
Revised: November 22, 2017
\end{flushright}

\vspace{0.5cm}
\begin{center}
\Large\bf\boldmath
Collider Probes of Axion-Like Particles
\end{center}

\vspace{0.5cm}
\begin{center}
Martin Bauer$^a$, Matthias Neubert$^{b,c}$ and Andrea Thamm$^b$\\
\vspace{0.7cm} 
{\sl ${}^a$Institut f\"ur Theoretische Physik, Universit\"at Heidelberg\\
Philosophenweg 16, 69120 Heidelberg, Germany\\[3mm]
${}^b$PRISMA Cluster of Excellence \& Mainz Institute for Theoretical Physics\\
Johannes Gutenberg University, 55099 Mainz, Germany\\[3mm]
${}^c$Department of Physics \& LEPP, Cornell University, Ithaca, NY 14853, U.S.A.}\\
\end{center}

\vspace{0.8cm}
\begin{abstract}
Axion-like particles (ALPs), which are gauge-singlets under the Standard Model (SM), appear in many well-motivated extensions of the SM. Describing the interactions of ALPs with SM fields by means of an effective Lagrangian, we discuss ALP decays into SM particles at one-loop order, including for the first time a calculation of the $a\to\pi\pi\pi$ decay rates for ALP masses below a few GeV. We argue that, if the ALP couples to at least some SM particles with couplings of order $(0.01-1)\,\mbox{TeV}^{-1}$, its mass must be above 1\,MeV. Taking into account the possibility of a macroscopic ALP decay length, we show that large regions of so far unconstrained parameter space can be explored by searches for the exotic, on-shell Higgs and $Z$ decays $h\to Za$, $h\to aa$ and $Z\to\gamma a$ in Run-2 of the LHC with an integrated luminosity of 300\,fb$^{-1}$. This includes the parameter space in which ALPs can explain the anomalous magnetic moment of the muon. Considering subsequent ALP decays into photons and charged leptons, we show that the LHC provides unprecedented sensitivity to the ALP--photon and ALP--lepton couplings in the mass region above a few MeV, even if the relevant ALP couplings are loop suppressed and the $a\to\gamma\gamma$ and $a\to\ell^+\ell^-$ branching ratios are significantly less than~1. We also discuss constraints on the ALP parameter space from electroweak precision tests. 
\end{abstract}

\end{titlepage}

\tableofcontents
\newpage

\section{Introduction}

New pseudoscalar particles with masses below the electroweak scale appear frequently in well-motivated extensions of the Standard Model (SM). Examples are axions \cite{Peccei:1977hh,Peccei:1977ur,Weinberg:1977ma,Wilczek:1977pj,Kim:1979if,Shifman:1979if,Zhitnitsky:1980tq,Dine:1981rt} addressing the strong CP problem or pseudoscalar mediators of a new interaction between dark or hidden sectors and the SM \cite{Dolan:2014ska}. Further, various anomalies can be explained by the presence of new spin-zero states with pseudoscalar couplings. Examples are the longstanding deviation of the anomalous magnetic moment of the muon from its SM value \cite{Chang:2000ii,Marciano:2016yhf}, or the excess in excited Beryllium decays ${^8\text{Be}}^*\to {^8\text{Be}}+e^+ e^-$ recently observed by the Atomki collaboration \cite{Krasznahorkay:2015iga,Feng:2016ysn,Ellwanger:2016wfe}. Dark-matter portals with a pseudoscalar mediator lighter than the Higgs boson can address the gamma-ray excess observed in the center of the galaxy by the Fermi-LAT collaboration, while avoiding constraints from direct detection and collider searches \cite{Boehm:2014hva,Berlin:2014tja}.

Axion-like particles (ALPs) have triggered interest way beyond their potential relevance in the context of the strong CP problem \cite{Jaeckel:2010ni,Arias:2012az}. Pseudo Nambu--Goldstone bosons arise generically in models with spontaneous breaking of a global symmetry. Due to an (approximate) shift symmetry they can naturally be light with respect to the electroweak or even the QCD scale. Low-energy observables, cosmological constraints and ALP searches with helioscopes probe a significant region of the parameter space in terms of the mass of the ALP and its couplings to photons and electrons. Future helioscope experiments like the International Axion Observatory (IAXO) \cite{Irastorza:2013dav}, and beam-dump experiments such as the facility to Search for Hidden Particles (SHIP) \cite{Alekhin:2015byh,Dobrich:2015jyk}, will further improve these constraints for ALP masses below a GeV. Collider experiments have searched directly and indirectly for ALPs \cite{Kleban:2005rj}. Besides ALP production in association with photons, jets and electroweak gauge bosons \cite{Mimasu:2014nea,Jaeckel:2015jla,Knapen:2016moh,Brivio:2017ije}, searches for decays of the $Z$ boson into a pseudoscalar $a$ and a photon at LEP and the LHC provide limits for ALPs with up to electroweak scale masses \cite{Kim:1989xj,Djouadi:1990ms,Rupak:1995kg,Jaeckel:2015jla}. Constraints from flavor-violating couplings have recently been summarized in \cite{Izaguirre:2016dfi}. Utilizing Higgs decays to search for light pseudoscalars has been proposed in \cite{Dobrescu:2000jt,Dobrescu:2000yn,Chang:2006bw,Draper:2012xt,Curtin:2013fra}. Several experimental searches looking for the decay $h\to aa$ have been performed, constraining various final states \cite{Chatrchyan:2012cg,CMS:2015iga,CMS:2016cel,Aad:2015bua,Khachatryan:2015nba,CMS:2016tgd,Khachatryan:2017mnf}. Surprisingly, no experimental analyses of the decay $h\to Za$ exist, even though analogous searches for heavy resonances decaying into a $Z$ boson and a pseudoscalar $a$ \cite{Khachatryan:2016are} as well as a search for a light $Z'$ boson in $h\to ZZ'$ decays \cite{Aad:2015sva} have been performed. The reason is, perhaps, the suppression of the $h\to Za$ decay in the decoupling limit in two-Higgs-doublet models in general and supersymmetric models in particular \cite{Branco:2011iw}. In models featuring a gauge-singlet ALP, there is no dimension-5 operator mediating $h\to Za$ decay at tree level, and hence this mode has not received much theoretical attention either (see, however, a recent brief discussion in \cite{Brivio:2017ije}). 

In this paper we present a comprehensive analysis of the on-shell Higgs decay modes $h\to Za$ and $h\to aa$ as well as the on-shell $Z$-boson decay $Z\to\gamma a$ starting from a general effective Lagrangian for a gauge-singlet ALP interacting with SM fields. We show that these decays can be used to probe the ALP couplings to SM particles in regions of parameters space inaccessible to any other searches. A first exposition of the main ideas of our approach has been presented in \cite{Bauer:2017nlg}. In the present paper we extend this discussion in several important ways, both as far as technical details are concerned and also regarding the number of relevant observables. The extraordinary reach of on-shell $h\to Za$, $h\to aa$ and $Z\to\gamma a$ searches in constraining the ALP couplings to photons, charged leptons and heavy quarks allows us to improve existing bounds derived from searches for $e^+ e^-\to\gamma a$ at LEP and $pp\to\gamma a$ at LHC \cite{Mimasu:2014nea,Jaeckel:2015jla,Knapen:2016moh} by up to six orders of magnitude. This improvement results from the fact that we consider decays of on-shell Higgs or $Z$ bosons in a parameter region where the ALP decays in SM particles before it leaves the detector. The best sensitivity is obtained for ALP masses above a few tens of MeV, which are almost unconstrained by low-energy observables. In particular, the parameter space in which an ALP can provide the explanation of the anomalous magnetic moment of the muon can be probed by these searches, assuming at least one of the relevant ALP--Higgs and ALP--$Z$--$\gamma$ couplings is larger than about $(100\,\mbox{TeV})^{-1}$. We emphasize that the decay $h\to Za$, which naively is mediated by a dimension-7 operator, can also originate from a non-polynomial operator of dimension-5, which receives a loop contribution from the top-quark and moreover could receive contributions from new heavy particles, as long as they receive (most of) their mass from the electroweak scale \cite{Bauer:2016ydr}. This makes the corresponding searches particularly interesting, because an observation of $h\to Za$ decay could reveal highly non-trivial information about the structure of the UV completion of the SM. 

The phenomenology of the decay modes $h\to Za$, $h\to aa$ and $Z\to\gamma a$ varies drastically for different ALP masses. Heavier ALPs can lead to clean di-photon, di-lepton, $b\bar b$ or di-jet final states, which will be easy to reconstruct. Lighter ALPs in the sub-GeV range can decay into strongly boosted photon pairs, which appear as ``photon jets'' in the detector \cite{Toro:2012sv}, effectively enhancing the $h\to Z\gamma$, $h\to\gamma\gamma$ and $Z\to\gamma\gamma$ rates (the absence of an interference term makes a suppression of these rates impossible). The smaller the ALP mass and couplings are, the more likely it is that the ALP decay is not prompt, but takes place at a displaced vertex. We therefore take the possibility of a macroscopic decay length of the ALP carefully into  account and discuss in which regions of parameter space this effect becomes important. For the case where the ALP decays at a displaced vertex inside the detector, the resulting signature is almost background free and hence can be reconstructed with high efficiency. Very light or very weakly coupled ALPs can predominantly decay outside the detector and could either be observed by a future surface detector specifically designed to search for long-lived particles (MATHUSLA) \cite{Chou:2016lxi,Curtin:2017izq} or through missing-energy signatures, which can be probed using mono-$X$ searches, with $X=Z,W,\gamma,h$ or a jet $j$. The case of long-lived ALPs has recently been discussed in \cite{Brivio:2017ije} for the special case where the ALP--photon coupling is set to zero. It was found that with 300\,fb$^{-1}$ of integrated luminosity at the LHC the relevant ALP couplings to $W$ and $Z$ bosons can be constrained up to roughly $(0.1-0.3)\,\mbox{TeV}^{-1}$. In our analysis we give special consideration to the region of parameter space in which the anomalous magnetic moment of the muon, which receives contributions from the ALP--muon and ALP--photon couplings, can be explained \cite{Chang:2000ii,Marciano:2016yhf}. 

This article is structured as follows: In Section~\ref{sec:Lag} we introduce the most general effective Lagrangian describing the ALP couplings to SM fields at dimension-5 order and discuss selected higher-dimensional operators relevant for Higgs physics. A detailed discussion of the possible two-particle decays of ALPs is presented in Section~\ref{sec:DR}, where we consistently include the tree-level contributions and one-loop corrections to the decay amplitudes. For ALP masses below a few GeV, we calculate the $a\to\pi\pi\pi$ decay rates and the effective ALP--photon couplings using a chiral Lagrangian. We also survey present constraints on the ALP--photon and ALP--electron couplings and point out that, under the assumption that the ALP couples at least to some SM particles with couplings larger than about $(100\,\mbox{TeV})^{-1}$, its mass must be above 1\,MeV. In Section~\ref{sec:amu} the preferred region of parameter space in which an ALP can explain the anomalous magnetic moment of the muon is derived. Section~\ref{sec:Higgs} is devoted to a detailed discussion of the exotic Higgs decays $h\to Za$ and $h\to aa$. We discuss which regions of parameter space can be probed with 300\,fb$^{-1}$ of integrated luminosity in Run-2 of the LHC, and which regions can already be excluded using existing searches. In Section~\ref{sec:Zpole} we extend this discussion to the exotic decay $Z\to\gamma a$, and we study $Z$-pole constraints from electroweak precision tests. We conclude in Section~\ref{sec:Conclusions}. Technical details of our calculations are relegated to four appendices.

\section{Effective Lagrangian for ALPs}
\label{sec:Lag}

We assume the existence of a new spin-0 resonance $a$, which is a gauge-singlet under the SM gauge group. Its mass $m_a$ is assumed to be smaller than the electroweak scale. A natural way to get such a light particle is by imposing a shift symmetry, $a\to a+c$, where $c$ is a constant. We will furthermore assume that the UV theory is CP invariant, and that CP is broken only by the SM Yukawa interactions. The particle $a$ is supposed to be odd under CP. Then the most general effective Lagrangian including operators of dimension up to~5 (written in the unbroken phase of the electroweak symmetry) reads \cite{Georgi:1986df} 
\begin{equation}\label{Leff}
\begin{aligned}
   {\cal L}_{\rm eff}^{D\le 5}
   &= \frac12 \left( \partial_\mu a\right)\!\left( \partial^\mu a\right)
    - \frac{m_{a,0}^2}{2}\,a^2 
    + \frac{\partial^\mu a}{\Lambda} \sum_F\,\bar\psi_F\,\bm{C}_F\,\gamma_\mu\,\psi_F \\[-1mm]
   &\quad\mbox{}+ g_s^2\,C_{GG}\,\frac{a}{\Lambda}\,G_{\mu\nu}^A\,\tilde G^{\mu\nu,A}
    + g^2\,C_{WW}\,\frac{a}{\Lambda}\,W_{\mu\nu}^A\,\tilde W^{\mu\nu,A}
    + g^{\prime\,2}\,C_{BB}\,\frac{a}{\Lambda}\,B_{\mu\nu}\,\tilde B^{\mu\nu} \,,
\end{aligned}
\end{equation}
where we have allowed for an explicit shift-symmetry breaking mass term $m_{a,0}$ (see below). $G_{\mu\nu}^A$, $W_{\mu\nu}^A$ and $B_{\mu\nu}$ are the field strength tensors of $SU(3)_c$, $SU(2)_L$ and $U(1)_Y$, and $g_s$, $g$ and $g'$ denote the corresponding coupling constants. The dual field strength tensors are defined as $\tilde B^{\mu\nu}=\frac12\epsilon^{\mu\nu\alpha\beta} B_{\alpha\beta}$ etc.\ (with $\epsilon^{0123}=1$). The advantage of factoring out the gauge couplings in the terms in the second line is that in this way the corresponding Wilson coefficients are scale invariant at one-loop order (see e.g.\ \cite{Bauer:2016lbe} for a recent discussion of the evolution equations beyond leading order). The sum in the first line extends over the chiral fermion multiplets $F$ of the SM. The quantities $\bm{C}_F$ are hermitian matrices in generation space. For the couplings of $a$ to the $U(1)_Y$ and $SU(2)_L$ gauge fields, the additional terms arising from a constant shift $a\to a+c$ of the ALP field can be removed by field redefinitions. The coupling to QCD gauge fields is not invariant under a continuous shift transformation because of instanton effects, which however preserve a discrete version of the shift symmetry. Above we have indicated the suppression of the dimension-5 operators with a new-physics scale $\Lambda$, which is the characteristic scale of global symmetry breaking, assumed to be above the weak scale. In the literature on axion phenomenology one often eliminates $\Lambda$ in favor of the ``axion decay constant'' $f_a$, defined such that $\Lambda/|C_{GG}|=32\pi^2 f_a$. Note that at dimension-5 order there are no ALP couplings to the Higgs doublet $\phi$. The only candidate for such an interaction is
\begin{equation}
   O_{Zh} = \frac{(\partial^\mu a)}{\Lambda} \left( \phi^\dagger\,iD_\mu\,\phi + \mbox{h.c.} \right)
   \to - \frac{g}{2c_w}\,\frac{(\partial^\mu a)}{\Lambda}\,Z_\mu\,(v+h)^2 \,,
\end{equation}
where $c_w\equiv\cos\theta_w$ denotes the cosine of the weak mixing angle, and the last expression holds in unitary gauge. Despite appearance, this operator does not give rise to a tree-level $h\to Za$ matrix element; the resulting tree-level graphs precisely cancel each other \cite{Bauer:2016ydr}. Indeed, a term $C_{Zh}\,O_{Zh}$ in the Lagrangian is redundant, because it can be reduced to the fermionic operators in (\ref{Leff}) using the equations of motion for the Higgs doublet and the SM fermions \cite{Bauer:2016ydr}. The field redefinitions 
\begin{equation}
   \phi\to e^{i\xi a}\,\phi \,, \qquad
   u_R\to e^{i\xi a}\,u_R \,, \qquad
   d_R\to e^{-i\xi a}\,d_R \,, \qquad
   e_R\to e^{-i\xi a}\,e_R \,, 
\end{equation}
with $\xi=C_{Zh}/\Lambda$, eliminate $O_{Zh}$ and shift the flavor matrices $\bm{C}_F$ of the $SU(2)_L$ singlet fermions by\footnote{In addition, the coefficient $C_{ah}$ of the Higgs-portal operator in (\ref{LeffD>5}) is shifted by $C_{ah}\to C_{ah}-(C_{Zh})^2$.}
\begin{equation}\label{CFshifts}
   \bm{C}_u \to \bm{C}_u - C_{Zh}\,\bm{1} \,, \qquad
   \bm{C}_d \to \bm{C}_d + C_{Zh}\,\bm{1} \,, \qquad
   \bm{C}_e \to \bm{C}_e + C_{Zh}\,\bm{1} \,, 
\end{equation}
while the matrices $\bm{C}_Q$ and $\bm{C}_L$ of the $SU(2)_L$ doublets remain unchanged. There are no additional contributions to the operators in (\ref{Leff}) involving the gauge fields, because the combination of axial-vector currents induced by the shifts in (\ref{CFshifts}) is anomaly free.

In this work we will be agnostic about the values of the Wilson coefficients. We will show that ALP searches at high-energy colliders are sensitive to couplings $C_i/\Lambda$ ranging from $(1\,\mbox{TeV})^{-1}$ to $(100\,\mbox{TeV})^{-1}$. In weakly-coupled UV completions one expects that the operators describing ALP couplings to SM bosons have loop-suppressed couplings (see e.g.\ \cite{Chala:2017sjk} for a recent discussion). This is in line with estimates based on naive dimensional analysis, which we briefly discuss in Appendix~\ref{app:NDA}. Departures from these estimates can arise in models involving e.g.\ large multiplicities of new particles in loops. It is common practice in the ALP literature to absorb potential loop factors that may arise into the Wilson coefficients $C_i$. As we will discuss in Section~\ref{sec:amu}, the puzzle of the anomalous magnetic moment of the muon can be resolved within our framework if $C_{\gamma\gamma}/\Lambda={\cal O}(1/{\rm TeV})$. Probing this region at colliders is thus a particularly well motivated target \cite{Bauer:2017nlg}. We emphasize, though, that by using the search strategies developed here it will be possible to probe even loop-suppressed couplings as long as the new-physics scale $\Lambda$ is in the TeV range.

The ALP can receive a mass by means of either an explicit soft breaking of the shift symmetry or through non-perturbative dynamics, like in the case of the QCD axion \cite{Weinberg:1977ma,Wilczek:1977pj}. In the absence of an explicit breaking, QCD dynamics generates a mass term given by \cite{Bardeen:1978nq,Shifman:1979if,DiVecchia:1980yfw} 
\begin{equation}
   m_{a,\,{\rm dyn}}\approx 5.7\,\mbox{$\mu$eV} \left[ \frac{10^{12}\,\mbox{GeV}}{f_a} \right]
   \approx 1.8\,\mbox{MeV}\,|C_{GG}| \left[ \frac{1\,\mbox{TeV}}{\Lambda} \right] .
\end{equation}
When an explicit symmetry-breaking mass term $m_{a,0}$ is included in the effective Lagrangian (\ref{Leff}), the resulting mass squared $m_a^2=m_{a,0}^2+m_{a,\,{\rm dyn}}^2$ becomes a free parameter. We will assume that $m_a\ll v$. At dimension-6 order and higher, several additional operators can arise. The ALP couplings to the Higgs field are those most relevant to our analysis. They are 
\begin{equation}\label{LeffD>5}
   {\cal L}_{\rm eff}^{D\ge 6}
   = \frac{C_{ah}}{\Lambda^2} \left( \partial_\mu a\right)\!\left( \partial^\mu a\right) \phi^\dagger\phi
    + \frac{C_{ah}'}{\Lambda^2}\,m_{a,0}^2\,a^2\phi^\dagger\phi
    + \frac{C_{Zh}^{(7)}}{\Lambda^3} \left( \partial^\mu a\right) 
    \left( \phi^\dagger\,iD_\mu\,\phi + \mbox{h.c.} \right) \phi^\dagger\phi + \dots \,.
\end{equation}
The first two terms are the leading Higgs portal interactions, which give rise to the decay $h\to aa$. Note that the second term, which explicitly violates the shift symmetry, is allowed only if the effective Lagrangian contains an explicit mass term for the ALP. Its effect is suppressed, relative to the first term, by a factor $m_{a,0}^2/m_h^2$. The third term is the leading operator mediating the decay $h\to Za$ at tree level \cite{Bauer:2016ydr}. These decay modes will be of particular interest to our discussion in Section~\ref{sec:Higgs}. 

After electroweak symmetry breaking (EWSB), the effective Lagrangian (\ref{Leff}) contains couplings of the pseudoscalar $a$ to $\gamma\gamma$, $\gamma Z$ and $ZZ$. The relevant terms read
\begin{equation}\label{gammaZ}
   {\cal L}_{\rm eff}^{D\le 5}
   \ni e^2\,C_{\gamma\gamma}\,\frac{a}{\Lambda}\,F_{\mu\nu}\,\tilde F^{\mu\nu}
    + \frac{2e^2}{s_w c_w}\,C_{\gamma Z}\,\frac{a}{\Lambda}\,F_{\mu\nu}\,\tilde Z^{\mu\nu}
    + \frac{e^2}{s_w^2 c_w^2}\,C_{ZZ}\,\frac{a}{\Lambda}\,Z_{\mu\nu}\,\tilde Z^{\mu\nu} \,,
\end{equation}
where $s_w=\sin\theta_w$ and $c_w=\cos\theta_w$, and we have defined
\begin{equation}\label{Cgagadef}
   C_{\gamma\gamma} = C_{WW} + C_{BB} \,, \qquad
   C_{\gamma Z} = c_w^2\,C_{WW} - s_w^2\,C_{BB} \qquad
   C_{ZZ} = c_w^4\,C_{WW} + s_w^4\,C_{BB}\,.
\end{equation}
The fermion mass terms resulting after EWSB are brought in diagonal form by means of field redefinitions, such that $\bm{U}_u^\dagger\,\bm{Y}_u\,\bm{W_u}=\mbox{diag}(y_u,y_c,y_t)$ etc. Under these field redefinitions the matrices $\bm{C}_F$ transform into new matrices
\begin{equation}\label{KFdef}
\begin{aligned}
   \bm{K}_U &= \bm{U}_u^\dagger\,\bm{C}_Q\,\bm{U}_u \,, \qquad
    \bm{K}_D = \bm{U}_d^\dagger\,\bm{C}_Q\,\bm{U}_d \,, \qquad
    \bm{K}_E = \bm{U}_e^\dagger\,\bm{C}_L\,\bm{U}_e \,, \\
   &\hspace{2cm} \bm{K}_f = \bm{W}_f^\dagger\,\bm{C}_f\,\bm{W}_f \,; \quad f=u,d,e \,.
\end{aligned}
\end{equation}
In any realistic model these couplings must have a hierarchical structure in order to be consistent with the strong constraints from flavor physics. We will discuss the structure of the flavor-changing ALP couplings in a companion paper \cite{inprep}. For now we focus on the flavor-diagonal couplings. Using the fact that the flavor-diagonal vector currents are conserved, we can rewrite the relevant terms in the Lagrangian in the form
\begin{equation}\label{diagferm}
   {\cal L}_{\rm eff}^{D\le 5} 
   \ni \sum_f \frac{c_{ff}}{2}\,\frac{\partial^\mu a}{\Lambda}\,\bar f\gamma_\mu\gamma_5 f \,,
\end{equation}    
where the sum runs over all fermion mass eigenstates, and we have defined (with $i=1,2,3$)
\begin{equation}\label{cffdef}
   c_{u_i u_i} = (K_u)_{ii} - (K_U)_{ii} \,, \qquad
   c_{d_i d_i} = (K_d)_{ii} - (K_D)_{ii} \,, \qquad
   c_{e_i e_i} = (K_e)_{ii} - (K_E)_{ii} \,.
\end{equation}

ALP couplings to neutrinos do not arise at this order, because the neutrino masses vanish in the SM, and hence the neutrino axial-vector currents are conserved. The leading shift-invariant coupling of an ALP to neutrino fields arises at dimension-8 order from an operator consisting of $\Box a$ times the Weinberg operator. Even in the most optimistic case, where no small coupling constant is associated with this operator, the resulting $a\to\nu\bar\nu$ decay rate would be suppressed, relative to the $a\to\gamma\gamma$ rate, by a factor of order $m_a^2\,v^4/\Lambda^6$. Alternatively, if Dirac neutrino mass terms are added to the SM, the corresponding couplings in (\ref{diagferm}) yield a $a\to\nu\bar\nu$ decay rate proportional to $m_\nu^2$. In either way, for $\Lambda$ in the TeV range or higher, this decay rate is so strongly suppressed that if an ALP can only decay into neutrinos (e.g.\ since it is lighter than $2m_e$ and its coupling to photons is exactly zero for some reason) it would be a long-lived particle for all practical purposes.

\section{ALP decay rates into SM particles}
\label{sec:DR} 

The effective Lagrangian (\ref{Leff}) governs the leading interactions (in powers of $v/\Lambda$) giving rise to ALP decays into pairs of SM gauge bosons and fermions, while the additional interactions in (\ref{LeffD>5}) are needed to parametrize the exotic decays of Higgs bosons into final states involving an ALP. In computing the various decay rates, we include the tree-level and one-loop contributions from the relevant operators. We find that fermion-loop corrections can be numerically important, and they can even be dominant in new-physics models where the coefficients $C_{VV}$ in (\ref{Leff}) (with $V=G,W,B$) are loop suppressed.

\subsection{ALP decay into photons}
\label{sec:agaga}

\begin{figure}
\begin{center}
\includegraphics[width=\textwidth]{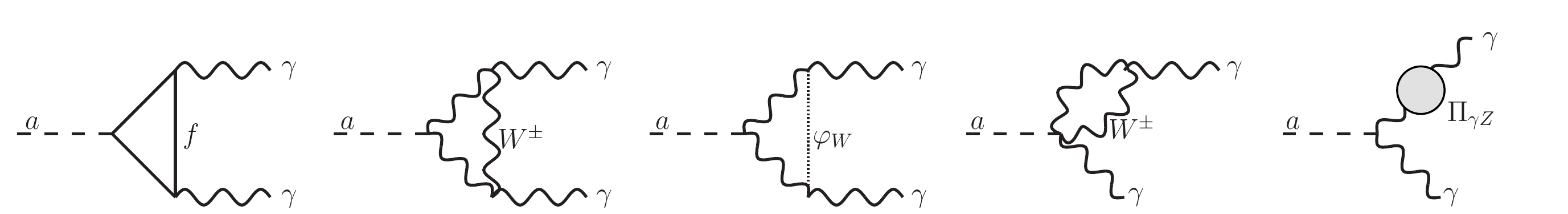}
\end{center}
\vspace{-2mm}
\caption{\label{fig:a2gamma} Representative one-loop Feynman diagrams contributing to the decay $a\to\gamma\gamma$. The internal boson lines represent charged $W$ bosons and the associated charged Goldstone fields. The last diagram contains the (gauge-dependent) self-energy $\Pi_{\gamma Z}(0)$. One also needs to include the on-shell wave-function renormalization factors for the external photon fields.}
\end{figure}

In many scenarios, the di-photon decay is the dominant decay mode of a light ALP. Because of its special importance, we have calculated the corresponding decay rate from the effective Lagrangian (\ref{Leff}) including the complete set of one-loop corrections. The relevant Feynman diagrams are shown in Figure~\ref{fig:a2gamma}. We define an effective coefficient $C_{\gamma\gamma}^{\rm eff}$ such that
\begin{equation}\label{di-photonrate1}
   \Gamma(a\to\gamma\gamma) 
   \equiv \frac{4\pi\alpha^2 m_a^3}{\Lambda^2}\,\big| C_{\gamma\gamma}^{\rm eff} \big|^2 \,.
\end{equation} 
To an excellent approximation (apart from a mild mass dependence in the loop corrections) the $a\to\gamma\gamma$ decay rate scales with the third power of the ALP mass. For a very light ALP with $m_a<2m_e$ this is the only SM decay mode allowed, and with decreasing ALP mass the decay rate will eventually become so small that the ALP will leave the detector and appear as an invisible particle. 

The expression for $C_{\gamma\gamma}^{\rm eff}$ depends on the ALP mass. If $m_a\gg\Lambda_{\rm QCD}$, then all loop corrections, including those involving colored particles, can be evaluated in perturbation theory. We obtain
\begin{equation}\label{di-photonrate}
   C_{\gamma\gamma}^{\rm eff}(m_a\gg\Lambda_{\rm QCD})
   = C_{\gamma\gamma} + \sum_f \frac{N_c^f Q_f^2}{16\pi^2}\,c_{ff}\,B_1(\tau_f) 
    + \frac{2\alpha}{\pi}\,\frac{C_{WW}}{s_w^2}\,B_2(\tau_W) \,,
\end{equation} 
where $\tau_i\equiv 4m_i^2/m_a^2$ for any SM particle, and $N_c^f$ and $Q_f$ denote the color multiplicity and electric charge (in units of $e$) of the fermion $f$. The loop functions read
\begin{equation}
   \begin{array}{l}
    B_1(\tau) = 1 - \tau\,f^2(\tau) \,, \\
    B_2(\tau) = 1 - (\tau-1)\,f^2(\tau) \,, 
   \end{array}
    \qquad \mbox{with} \quad
   f(\tau) = \left\{ \begin{array}{ll} 
    \arcsin\frac{1}{\sqrt{\tau}} \,; &~ \tau\ge 1 \,, \\
    \frac{\pi}{2} + \frac{i}{2} \ln\frac{1+\sqrt{1-\tau}}{1-\sqrt{1-\tau}} \,; &~ \tau<1 \,.
   \end{array} \right. 
\end{equation} 
The fermion loop function has the property that $B_1(\tau_f)\approx 1$ for light fermions with masses $m_f\ll m_a$, while $B_1(\tau_f)\approx-\frac{m_a^2}{12m_f^2}$ for heavy fermions ($m_f\gg m_a$). Thus, each electrically charged fermion lighter than the ALP adds a contribution of order $c_{ff}/(16\pi^2)$ to the effective Wilson coefficient $C_{\gamma\gamma}^{\rm eff}$, while fermions heavier than the ALP decouple. The calculation of the electroweak loop corrections to the decay rate is far more involved than that of the fermion loops. We have evaluated the relevant diagrams shown in Figure~\ref{fig:a2gamma} in a general $R_\xi$ gauge. After some intricate cancellations, the main result of these corrections is to renormalize the fine-structure constant $\alpha$ in the expression for the rate, which is to be evaluated at $q^2=0$, as appropriate for on-shell photons. As mentioned earlier, the Wilson coefficient $C_{\gamma\gamma}$ is not renormalized at one-loop order. The remaining finite correction in (\ref{di-photonrate}) is strongly suppressed, since the loop function $B_2(\tau_W)\approx\frac{m_a^2}{6m_W^2}$ is proportional to the ALP mass squared. An interesting feature of our result for the effective ALP--photon coupling in (\ref{di-photonrate}) is that the loop-induced contributions from both fermions and $W$ bosons vanish in the limit $m_a\to 0$. This is an advantage of our choice of operator basis. 

It is interesting to compare our result for the fermionic contributions to the $a\to\gamma\gamma$ decay rate with the corresponding effects on the di-photon decay rate of a CP-odd Higgs boson. In this case the Higgs boson couples to the pseudoscalar fermion current, and one finds an expression analogous to (\ref{di-photonrate}), but with the loop function $[B_1(\tau_f)-1]$ instead of $B_1(\tau_f)$ \cite{Spira:1995rr}. The difference can be understood using the anomaly equation for the divergence of the axial-vector current, which allows us to rewrite the ALP--fermion coupling in (\ref{diagferm}) in the form
\begin{equation}
   \frac{c_{ff}}{2}\,\frac{\partial^\mu a}{\Lambda}\,\bar f\gamma_\mu\gamma_5 f 
   = - c_{ff}\,\frac{m_f}{\Lambda}\,a\,\bar f\,i\gamma_5 f 
    + c_{ff}\,\frac{N_c^f Q_f^2}{16\pi^2}\,\frac{a}{\Lambda}\,e^2 F_{\mu\nu}\,\tilde F^{\mu\nu} + \dots \,,
\end{equation}    
where the dots represent similar terms involving gluons and weak gauge fields. The first term on the right-hand side is now of the same form as the coupling of a CP-odd Higgs boson to fermions, while the second term has the effect of subtracting ``1'' from the function $B_1(\tau_f)$. 

At one-loop order, relation (\ref{di-photonrate}) involves all Wilson coefficients in the effective Lagrangian (\ref{Leff}) except for $C_{GG}$. Even if the original coefficient $C_{\gamma\gamma}$ vanished for some reason, these loop contributions would induce an effective coefficient $C_{\gamma\gamma}^{\rm eff}$ at one-loop order. The ALP--gluon coupling would first enter at two-loop order. Using results derived in the following section, its effect can be estimated as
\begin{equation}\label{deltac}
   \delta C_{\gamma\gamma}^{\rm eff}(m_a\gg\Lambda_{\rm QCD})
   \approx - \frac{3\alpha_s^2(m_a^2)}{\pi^2}\,C_{GG}\,\sum_q Q_q^2\,B_1(\tau_q)\,\ln\frac{\Lambda^2}{m_q^2} \,,
\end{equation} 
where for the light quarks $q=u,d,s$ one should use a typical hadronic scale such as $m_\pi$ instead of $m_q$ in the argument of the logarithm. Numerically, this two-loop contribution can be sizable due to the large logarithm. 

If the ALP mass is not in the perturbative regime, i.e.\ for $m_a\lesssim 1$\,GeV, the hadronic loop corrections to the effective ALP--photon coupling can be calculated using an effective chiral Lagrangian. This is discussed in detail in Appendix~\ref{app:ChPT}. Including interactions up to linear order in the ALP field, and working at leading order in the chiral expansion, one obtains \cite{Georgi:1986df}
\begin{equation}\label{LeffChpT}
\begin{aligned}
   {\cal L}_{\chi PT} 
   &= \frac12\,\partial^\mu a\,\partial_\mu a - \frac{m_a^2}{2}\,a^2
    + e^2 \left[ C_{\gamma\gamma} - \frac23\,(4\kappa_u+\kappa_d)\,C_{GG} \right] 
    \frac{a}{\Lambda}\,F_{\mu\nu}\,\tilde F^{\mu\nu} \\
   &\quad\mbox{}+ \frac{f_\pi^2}{8}\,\mbox{tr}\big[ D^\mu\Sigma\,D_\mu\Sigma^\dagger \big] 
    + \frac{f_\pi^2}{4}\,B_0\,\mbox{tr}\big[ \Sigma\,m_q^\dagger + m_q\Sigma^\dagger \big] 
    + \frac{if_\pi^2}{4}\,\frac{\partial^\mu a}{2\Lambda}\,
    \mbox{tr}\big[ \hat c_{qq} (\Sigma^\dagger D_\mu\Sigma - \Sigma\,D_\mu\Sigma^\dagger)\big] \,.
\end{aligned}
\end{equation} 
Here $f_\pi\approx 130$\,MeV is the pion decay constant, $\Sigma=\exp\big(\frac{i\sqrt2}{f_\pi}\,\tau^A\pi^A\big)$ contains the pion fields and $B_0=\frac{m_\pi^2}{m_u+m_d}$ is proportional to the chiral condensate. For simplicity we restrict ourselves to flavor $SU(2)$ with just one generation of light quarks. The hermitian matrices $m_q=\mbox{diag}\,(m_u,m_d)$ and $\hat c_{qq}=\mbox{diag}\,(c_{qq}+32\pi^2\,C_{GG}\,\kappa_q)$ are diagonal in the quark mass basis. The parameters 
\begin{equation}\label{kappaqvals}
   \kappa_u = \frac{m_d}{m_u+m_d} \,, \qquad \kappa_d = \frac{m_u}{m_u+m_d}
\end{equation} 
have been chosen such that there is no tree-level mass mixing of the ALP with the $\pi^0$ \cite{Georgi:1986df}. Note the unusual appearance of a ``tree-level'' contribution proportional to $C_{GG}$ to the coefficient of the ALP--photon coupling in (\ref{LeffChpT}). When higher-order corrections (including the effects of the strange quark) are taken into account, the coefficient of $C_{GG}$ inside the bracket is reduced by about 5\% and one obtains $[C_{\gamma\gamma}-(1.92\pm 0.04)\,C_{GG}]$ \cite{diCortona:2015ldu}. This large effect is a consequence of the axial-vector anomaly leading to enhanced $\pi^0,\eta,\eta'$ couplings to two photons combined with a mass-mixing of the ALP with these mesons \cite{Bardeen:1986yb}. 

QCD dynamics generates a mass for the ALP given (at lowest order) by \cite{Bardeen:1978nq,Shifman:1979if,DiVecchia:1980yfw} 
\begin{equation}
   m_{a,\,{\rm dyn}}^2 = \frac{f_\pi^2\,m_\pi^2}{2\Lambda^2} \left( 32\pi^2\,C_{GG} \right)^2 \frac{m_u m_d}{(m_u+m_d)^2} \,.
\end{equation} 
A possible explicit shift-symmetry breaking mass term $m_{a,0}^2$ would have to be added to this expression. The last term in (\ref{LeffChpT}) gives rise to a kinetic mixing between the ALP and the neutral pion. The physical states are obtained by bringing the kinetic terms into canonical form and rediagonalizing the mass matrix. This changes the mass eigenvalues for $\pi^0$ and $a$ by tiny corrections of order $f_\pi^2/\Lambda^2$ relative to the leading terms. At the same time, the state $\pi^0$ receives a small admixture of the physical ALP state, such that
\begin{equation}\label{admixture}
   \pi^0 = \pi_{\rm phys}^0 - \frac{\epsilon\,m_a^2}{m_\pi^2-m_a^2}\,a_{\rm phys} + {\cal O}(\epsilon^2) \,,
\end{equation} 
where
\begin{equation}
   \epsilon = \frac{f_\pi}{2\sqrt{2}\Lambda} \left[ (c_{uu}-c_{dd})
    + 32\pi^2\,C_{GG}\,\frac{m_d-m_u}{m_d+m_u} \right] .
\end{equation} 
Relation (\ref{admixture}) holds as long as $|m_\pi^2-m_a^2|\gg 2\epsilon\,m_\pi m_a$. In the opposite limit one would obtain $\pi^0=\frac{1}{\sqrt2}\,(\pi_{\rm phys}^0+a_{\rm phys})+{\cal O}(\epsilon)$, but such a large mixing requires a fine-tuning of the masses that is rather implausible. In the presence of the mixing in (\ref{admixture}), the SM $\pi^0\to\gamma\gamma$ amplitude mediated by the axial-vector anomaly induces an additional contribution to the $a\to\gamma\gamma$ amplitude. Combining all terms, we obtain (assuming $m_a\ne m_\pi$)
\begin{equation}\label{di-photonrate2}
\begin{aligned}
   C_{\gamma\gamma}^{\rm eff}(m_a\lesssim 1\,\mbox{GeV})
   &\approx C_{\gamma\gamma} - (1.92\pm 0.04)\,C_{GG} 
    - \frac{m_a^2}{m_\pi^2-m_a^2} \left[ C_{GG}\,\frac{m_d-m_u}{m_d+m_u} + \frac{c_{uu}-c_{dd}}{32\pi^2} \right] \\
   &\quad\mbox{}+ \sum_{q=c,b,t} \frac{N_c Q_q^2}{16\pi^2}\,c_{qq}\,B_1(\tau_q)
    + \sum_{\ell=e,\mu,\tau} \frac{c_{\ell\ell}}{16\pi^2}\,B_1(\tau_\ell) 
    + \frac{2\alpha}{\pi}\,\frac{C_{WW}}{s_w^2}\,B_2(\tau_W) \,.
\end{aligned}
\end{equation}
The contribution from the coefficient $c_{ss}$ not shown here would be suppressed, for light ALPs, by a factor of order $m_\pi^2/m_\eta^2$ relative to the contributions from $c_{uu}$ and $c_{dd}$.

\subsection{ALP decays into charged leptons}
\label{sec:cll}

\begin{figure}
\begin{center}
\includegraphics[width=0.42\textwidth]{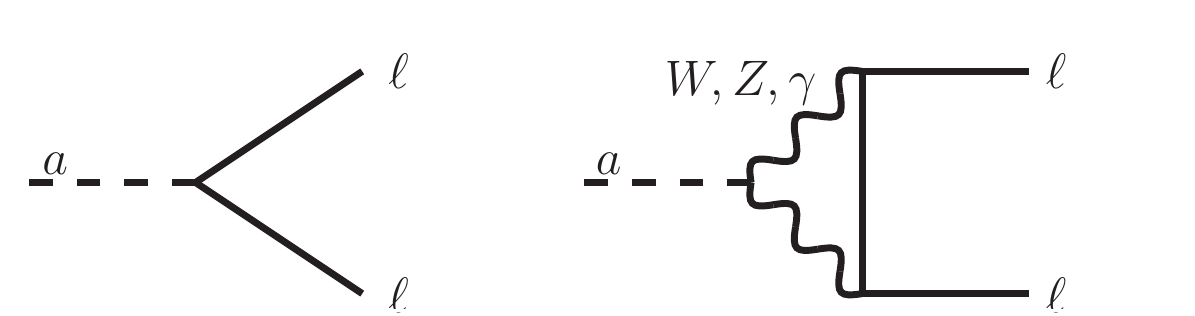}
\end{center}
\vspace{-2mm}
\caption{\label{fig:a2ee} Representative one-loop Feynman diagrams contributing to the decay $a\to\ell^+\ell^-$.}
\end{figure}

If the ALP mass is larger than $2m_e\approx 1.022$\,MeV, the leptonic decay $a\to e^+ e^-$ or decays into heavier leptons (if kinematically allowed) can be the dominant ALP decay modes in some regions of parameter space. We have calculated the corresponding decay rates from the effective Lagrangian including the complete set of one-loop mixing contributions from the bosonic operators in (\ref{Leff}) and (\ref{gammaZ}), see Figure~\ref{fig:a2ee}. In analogy with (\ref{di-photonrate1}), we write the result in the form (with $\ell=e,\mu,\tau$) 
\begin{equation}
   \Gamma(a\to\ell^+\ell^-) 
   = \frac{m_a m_\ell^2}{8\pi\Lambda^2} \left| c_{\ell\ell}^{\rm eff} \right|^2
    \sqrt{1-\frac{4m_\ell^2}{m_a^2}} \,,
\end{equation}
which is approximately linear in the ALP mass. At one-loop order, the effective Wilson coefficient $c_{\ell\ell}^{\rm eff}$ receives contributions from $c_{\ell\ell}$ as well as from the diboson coefficients $C_{WW}$ and $C_{BB}$. Using the linear combinations of Wilson coefficients defined in (\ref{Cgagadef}), we find 
\begin{equation}\label{clleff}
\begin{aligned}
   c_{\ell\ell}^{\rm eff} 
   &= c_{\ell\ell}(\mu) \left[ 1 + {\cal O}\big(\alpha\big) \right]
    - 12 Q_\ell^2\,\alpha^2 C_{\gamma\gamma} 
    \left[ \ln\frac{\mu^2}{m_\ell^2} + \delta_1 + g(\tau_\ell) \right] \\
   &\quad\mbox{}- \frac{3\alpha^2}{s_w^4}\,C_{WW}
    \bigg( \ln\frac{\mu^2}{m_W^2} + \delta_1 + \frac12 \bigg) 
    - \frac{12\alpha^2}{s_w^2 c_w^2}\,C_{\gamma Z}\,Q_\ell \left( T_3^\ell - 2 Q_\ell s_w^2 \right)\!
    \bigg( \ln\frac{\mu^2}{m_Z^2} + \delta_1 + \frac32 \bigg) \\
   &\quad\mbox{}- \frac{12\alpha^2}{s_w^4 c_w^4}\,C_{ZZ} 
    \bigg( Q_\ell^2 s_w^4 - T_3^\ell Q_\ell s_w^2 + \frac18 \bigg)
    \bigg( \ln\frac{\mu^2}{m_Z^2} + \delta_1 + \frac12 \bigg) \,.
\end{aligned}
\end{equation}
Here $Q_\ell=-1$ is the electric charge of the charged lepton, and $T_3^\ell=-\frac12$ is the weak isospin of its left-handed component. In the limit where $m_\ell^2$ is either much smaller or much larger than $m_a^2$, the loop function in the photon term is given by 
\begin{equation}\label{gfunasy}
   g(\tau_\ell) = \left\{
    \begin{array}{cl} 
    \displaystyle - \frac16 \left( \ln\frac{m_a^2}{m_\ell^2} - i\pi \right)^2 + \frac{2}{3} 
    + {\cal O}\bigg( \frac{m_\ell^2}{m_a^2}\bigg) \,; &~ m_\ell^2\ll m_a^2 \,, \\[4mm]
    \displaystyle \frac73 + {\cal O}\bigg( \frac{m_a^2}{m_\ell^2}\bigg) \,; &~ m_\ell^2\gg m_a^2 \,. \\
    \end{array} \right.
\end{equation}
The exact expression is given in Appendix~\ref{app:A}. In (\ref{clleff}) we have regularized the UV divergences of the various contributions using dimensional regularization in the $\overline{\rm MS}$ scheme. Only the sum of all contributions is scale independent, i.e.\ the scale dependence of $c_{\ell\ell}(\mu)$ compensates the scale dependence of the other terms. We do not show the one-loop corrections proportional to the tree-level coefficient $c_{\ell\ell}$ itself. They contain IR divergences, which cancel in the sum of the decay rates for $a\to\ell^+\ell^-$ and $a\to\ell^+\ell^-\gamma_{\rm soft}$ with a soft photon in the final state. The scheme-dependent constant $\delta_1$ in (\ref{clleff}) arises from the treatment of the Levi--Civita symbol in $d$ dimensions, as we also discuss in Appendix~\ref{app:A}. We obtain $\delta_1=-\frac{11}{3}$. In a scheme where instead the Levi--Civita symbol is treated as a 4-dimensional object, one would have $\delta_1=0$. 

Relation (\ref{clleff}) shows two important facts: first, at one-loop order ALP couplings to fermions are induced from operators in the effective Lagrangian coupling the ALP to gauge bosons; and second, it would be inconsistent to set $c_{\ell\ell}$ to zero in (\ref{Leff}), since this scale-dependent coefficient mixes with the coefficients of bosonic operators under renormalization. Hence it must contain $\mu$-dependent terms, which cancel the explicit scale dependence in the above result. Because of the presence of such terms, the only information that can conclusively be extracted from the calculation of the low-energy contributions performed above are the coefficients of the large logarithms obtained by identifying the factorization scale $\mu$ with the UV cutoff $\Lambda$. The result for these logarithmic contributions simplifies when one adds up the various terms in (\ref{clleff}), since they can be derived in the unbroken phase of the electroweak theory. We obtain
\begin{equation}
   c_{\ell\ell}^{\rm eff} 
   = c_{\ell\ell}(\Lambda) - 6\alpha^2 \bigg[ \frac{C_{WW}}{s_w^4}\,\mbox{tr}(\tau^A\tau^A)
    + \frac{C_{BB}}{c_w^4} \left( Y_{\ell_L}^2 + Y_{\ell_R}^2 \right)\! \bigg] \ln\frac{\Lambda^2}{m_W^2}
    - 12 Q_\ell^2\,\alpha^2 C_{\gamma\gamma} \ln\frac{m_W^2}{m_\ell^2} + \dots \,,
\end{equation}
where the first two terms arise from the loops of $SU(2)_L$ and $U(1)_Y$ gauge bosons, for which $\mbox{tr}(\tau^A\tau^A)=\frac34$ and $Y_{\ell_L}=-\frac12$, $Y_{\ell_R}=-1$. The last term contains the finite large logarithm related to the long-distance photon contribution, with $C_{\gamma\gamma}$ given in (\ref{Cgagadef}).

\subsection{ALP decays into hadrons}

At the partonic level, the pseudoscalar $a$ can also decay into colored particles. At tree-level the relevant modes are $a\to gg$ and $a\to q\bar q$. In the hadronic world these decays are allowed only if $m_a>m_\pi$. However, below 1\,GeV the number of possible hadronic decay channels is very limited, because the two-body decays $a\to\pi\pi$ and $a\to\pi^0\gamma$ are forbidden by CP invariance and angular momentum conservation, while the three-body modes $a\to\pi\pi\gamma$, $a\to\pi^0\gamma\gamma$ and $a\to\pi^0 e^+ e^-$ are strongly suppressed by phase space and powers of the fine-structure constant $\alpha$ \cite{Dobrescu:2000jt}. The dominant decay modes in this region are $a\to3\pi^0$ and $a\to\pi^+\pi^-\pi^0$. As long as the ALP is sufficiently light, so that the energy of the final-state mesons is much less than $4\pi f_\pi\approx 1.6$\,GeV, the calculation of the decay rates for exclusive modes such as $a\to\pi\pi\pi$ can be performed using the effective chiral Lagrangian (\ref{LeffChpT}). ALP couplings to three pions arise from each of the three terms shown in the second line of this equation, where in the first two terms one must substitute relation (\ref{admixture}) for the $\pi^0$ fields. Working consistently at leading order in the chiral expansion, we obtain 
\begin{equation}
   \Gamma(a\to\pi^a\pi^b\pi^0) 
   = \frac{\pi}{6}\,\frac{m_a m_\pi^4}{\Lambda^2 f_\pi^2}
    \left[ C_{GG}\,\frac{m_d-m_u}{m_d+m_u} + \frac{c_{uu}-c_{dd}}{32\pi^2} \right]^2 
    g_{ab}\bigg(\frac{m_\pi^2}{m_a^2}\bigg) \,,
\end{equation}
where (with $0\le r\le 1/9$)
\begin{equation}
\begin{aligned}
   g_{00}(r) &= \frac{2}{(1-r)^2} \int_{4r}^{(1-\sqrt{r})^2}\!\!dz\,\sqrt{1-\frac{4r}{z}}\,
    \lambda^{1/2}(1,z,r) \,, \\
   g_{+-}(r) &= \frac{12}{(1-r)^2} \int_{4r}^{(1-\sqrt{r})^2}\!\!dz\,\sqrt{1-\frac{4r}{z}}\,(z-r)^2\,
    \lambda^{1/2}(1,z,r) \,.
\end{aligned}
\end{equation}
Both functions are normalized such that $g_{ab}(0)=1$, and they vanish at the threshold $r=1/9$.

If the ALP mass is in the perturbative regime (i.e., for $m_a\gg\Lambda_{\rm QCD}$), its inclusive decay rate into hadrons can be calculated under the assumption of quark-hadron duality \cite{Poggio:1975af,Shifman:2000jv}. Setting the light quark masses to zero (since here by assumptions $m_a\gg m_q$ for all light quarks) and including the one-loop QCD corrections to the decay rate as calculated in \cite{Spira:1995rr}, we obtain 
\begin{equation}\label{Gamma_had}
\begin{aligned}
   \Gamma(a\to\mbox{hadrons}) 
   &= \frac{32\pi\,\alpha_s^2(m_a)\,m_a^3}{\Lambda^2}
    \left[ 1 + \left( \frac{97}{4} - \frac{7n_q}{6} \right) \frac{\alpha_s(m_a)}{\pi} \right]
    \bigg| C_{GG} + \sum_{q=1}^{n_q} \frac{c_{qq}}{32\pi^2} \bigg|^2 \\
   &\equiv \frac{32\pi\,\alpha_s^2(m_a)\,m_a^3}{\Lambda^2}
    \left[ 1 + \frac{83}{4}\,\frac{\alpha_s(m_a)}{\pi} \right] \left| C_{GG}^{\rm eff} \right|^2 ,
\end{aligned}
\end{equation}
where $n_q=3$ is the number of light quark flavors. To good approximation this rate scales with the third power of the ALP mass. Decays into heavy quarks, if kinematically allowed, can be reconstructed by heavy-flavor tagging. The corresponding rates are (with $Q=b$ or $c$)
\begin{equation}
   \Gamma(a\to Q\bar Q) 
   = \frac{3m_a\,\overline{m}_Q^2(m_a)}{8\pi\Lambda^2} \left| c_{QQ}^{\rm eff} \right|^2
    \sqrt{1-\frac{4m_Q^2}{m_a^2}} \,,
\end{equation}
where at leading order in perturbation theory $c_{QQ}^{\rm eff}=c_{QQ}$.

One-loop corrections to the ALP--quark couplings $c_{qq}$ for both light and heavy quarks can be calculated in analogy with those to the ALP--lepton couplings discussed in Section~\ref{sec:cll}. The obvious replacements to be applied to relation (\ref{clleff}) are $Q_\ell\to Q_q$ and $T_3^\ell\to T_3^q$. In addition, the $W$-boson contribution picks up a factor $V_{ik} V_{jk}^*$ or $V_{ki}^* V_{kj}$ (summed over $k$) for external up-type or down-type quarks with generation indices $i$ and $j$, respectively. If the internal quark with index $k$ is heavy, a non-trivial loop function arises. Note that these contributions can be off-diagonal in generation space. Finally, there is a new one-loop contribution involving the ALP--gluon coupling, whose form is
\begin{equation}\label{eq17}
   \delta c_{qq}^{\rm eff} 
   = - 12 C_F\,\alpha_s^2\,C_{GG} 
    \left[ \ln\frac{\mu^2}{m_q^2} + \delta_1 + g(\tau_q) \right] ,
\end{equation}
with $C_F=4/3$. The perturbative calculation of this expression can be trusted as long as $m_a\gg\Lambda_{\rm QCD}$ and $m_q\gg\Lambda_{\rm QCD}$. For the light quarks, the appropriate infrared scale is not the quark mass but a typical hadronic scale such as $m_\pi$. We have derived the estimate (\ref{deltac}) by using the above result for the gluon contribution to $c_{qq}$ in (\ref{di-photonrate}).

\subsection{Summary of ALP decay modes}

\begin{figure}[t]
\begin{center}
\includegraphics[width=0.84\textwidth]{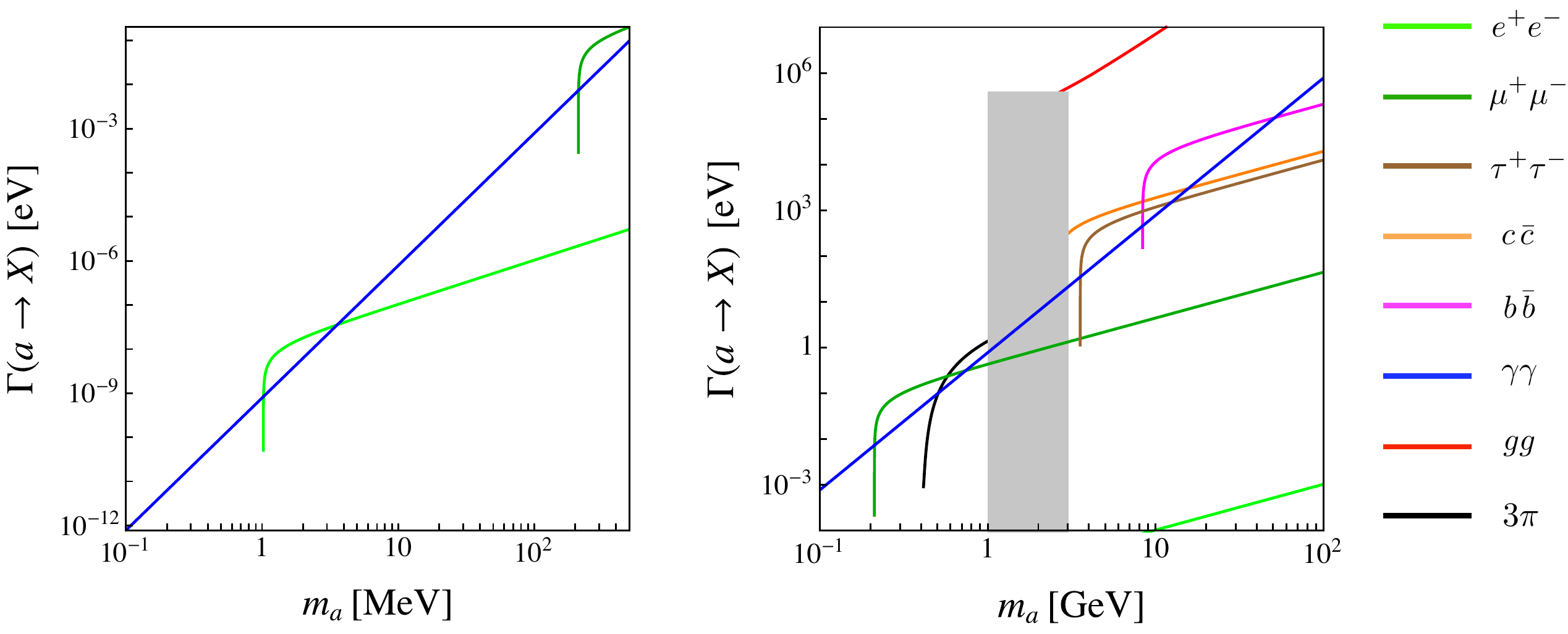}
\includegraphics[width=0.84\textwidth]{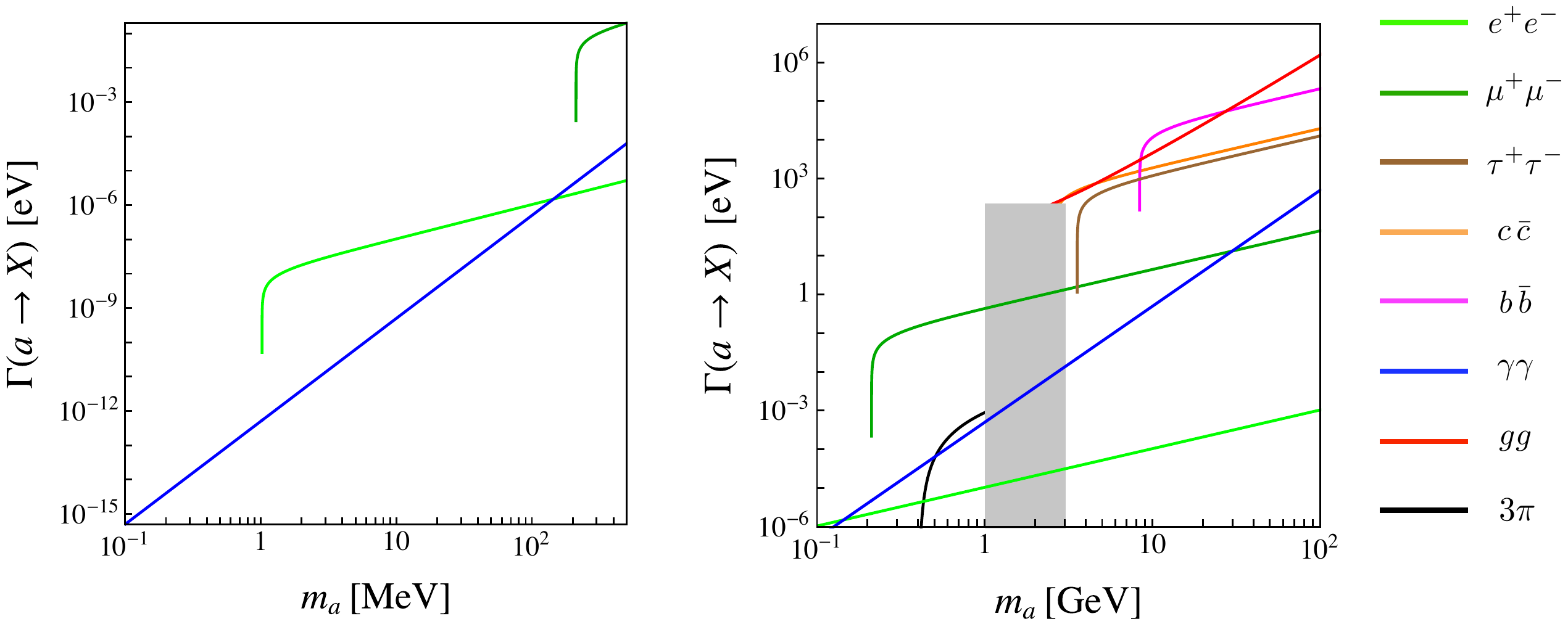}
\end{center}
\vspace{-4mm}
\caption{\label{fig:decayrates} ALP decay rates into pairs of SM particles obtained by setting the relevant effective Wilson coefficients to~1 (top), or by setting the ALP--fermion couplings to~1 and the ALP--boson couplings to $1/(4\pi^2)$ (bottom). The gray area between 1 and 3\,GeV shows the region in which various exclusive hadronic (and difficult to calculate) decay channels such as $a\to\rho\rho$ open up. In this interval the rate $\Gamma(a\to\mbox{hadrons})$ is expected to interpolate between the black and red lines. The rates for decays into heavy-flavor jets are shown separately.}
\end{figure}

Above we have presented an overview of possible ALP decay modes into SM particles. The upper panel in Figure~\ref{fig:decayrates} shows the various decay rates for a new-physics scale $\Lambda=1$\,TeV as a function of the ALP mass, under the assumption that the relevant coefficients $|C_{\gamma\gamma}^{\rm eff}|$, $|C_{GG}^{\rm eff}|$ and $|c_{ff}^{\rm eff}|$ are all equal to~1. For different values of these parameters, the rates need to be rescaled by factors $(|C_{ii}^{\rm eff}|/\Lambda)^2$. For example, in the lower panel we assume that the ALP--boson couplings are loop suppressed. If all Wilson coefficients are of the same magnitude and the ALP is lighter than the pion (or if it does not couple to colored particles at all), the dominant decay mode is $a\to\gamma\gamma$. The leptonic modes $a\to\ell^+\ell^-$ are only significant near the thresholds $m_a\gtrsim 2m_\ell$, where they can be dominant. If the ALP--boson couplings are loop suppressed, the leptonic decays can be dominant for ALP masses exceeding $2m_e$. The picture changes significantly if the ALP is heavy enough to decay hadronically, i.e.\ for $m_a>3m_{\pi^0}\approx 405$\,MeV. If the coupling to gluons is unsuppressed, the ALP then decays predominantly into hadronic final states. For $m_a>\mbox{few GeV}$, the inclusive hadronic rate is approximately given by (\ref{Gamma_had}). If, on the other hand, the ALP--gluon coupling is suppressed, there can be a potpourri of decay modes ($a\to\mbox{hadrons}$, $a\to b\bar b$, $a\to c\bar c$, $a\to\tau^+\tau^-$, $a\to\gamma\gamma$) with potentially similar rates. Which of these modes dominates depends on the details of the model.  

If the total decay rate of the ALP is too small, the ALP leaves the detector before it decays. For example, a total rate of $10^{-9}$\,eV corresponds to a lifetime of $6.6\cdot 10^{-7}$\,s. If the ALP is produced in decays of heavier particles, the Lorentz boost can increase its lifetime significantly. It is also a possibility that the ALP decays invisibly into light particles of a hidden sector. In this case the decay products cannot be reconstructed, and hence the ALP signature would be that of missing energy and momentum.

\subsection{Constraints on ALP couplings to photons and electrons}
\label{sec:ALPphoton}

The couplings of ALPs to photons and electrons have been constrained over vast regions of parameter space using a variety of experiments in particle physics, astro-particle physics and cosmology. Since our work is motivated by the idea that ALPs could interact with SM particles with couplings of order $(1\,\mbox{TeV})^{-1}$ to $(100\,\mbox{TeV})^{-1}$, such that these interactions can be probed at the LHC, we need to address the question of how the existing bounds can be satisfied. In Figure~\ref{fig:existingbounds} we show a compilation of existing exclusion regions for the ALP--photon and ALP--electron couplings. Before addressing these bounds in more detail, let us add an important remark concerning the ALP--lepton couplings. In the  absence of a flavor symmetry, under which the three lepton flavors carry different charges (but which must be broken in order to explain neutrino oscillations), the matrices $\bm{C}_L$ and $\bm{C}_e$ entering the ALP--lepton couplings in (\ref{Leff}) must, to an excellent approximation, be proportional to the unit matrix. Otherwise it is impossible to avoid flavor-changing neutral currents in the charged lepton sector, which are generated after electroweak symmetry breaking, see (\ref{KFdef}). The relevant couplings $c_{e\mu}$, $c_{\mu\tau}$ and $c_{e\tau}$ must satisfy very strong constraints from processes such as $\mu\to e\gamma$ and $\mu^-\to e^- e^+ e^-$, and analogous ones involving heavier leptons (see \cite{Calibbi:2017uvl} for a recent review). As a result, one expects that
\begin{equation}\label{LFuniv}
   c_{ee}\simeq c_{\mu\mu}\simeq c_{\tau\tau}
\end{equation}
to very good accuracy. Below we will sometimes make use of this relation.
  
\subsubsection{Constraints on the ALP--photon coupling} 

Consider first the exclusion regions in the $m_a-|C_{\gamma\gamma}^{\rm eff}|$ plane shown in the left panel. The parameter space excluded from cosmological constraints is shaded gray. This includes constraints from measurements of the number of effective degrees of freedom, modifications to big-bang nucleosynthesis, distortions of the cosmic microwave-background spectrum and extragalactic background-light measurements \cite{Cadamuro:2011fd,Millea:2015qra}. Energy loss of stars through radiation of ALPs is constrained by the ratio of red giants to younger stars of the so-called horizontal branch~(HB) \cite{Raffelt:1985nk,Raffelt:1987yu,Raffelt:2006cw} (shaded purple). Another strong constraint arises from the measurement of the length of the neutrino burst from Supernova SN1987a, which would have been shorter in the presence of an energy loss from ALP emission \cite{Payez:2014xsa} (shaded yellow), as well as from the non-observation of a photon burst from SN1987a due to the decay of emitted ALPs  \cite{Jaeckel:2017tud} (shaded orange). These constraints require an extremely tight bound $|C_{\gamma\gamma}^{\rm eff}|/\Lambda\ll 10^{-15}\,\mbox{TeV}^{-1}$ in the mass window between 150\,eV and about 1\,MeV. For smaller ALP masses the bounds are weaker, ranging from $|C_{\gamma\gamma}^{\rm eff}|/\Lambda<10^{-9}\,\mbox{TeV}^{-1}$ for $m_a=150$\,eV to $|C_{\gamma\gamma}^{\rm eff}|/\Lambda<3\cdot 10^{-7}\,\mbox{TeV}^{-1}$ for $m_a<4$\,eV. Below 4\,eV the tightest bounds come from HB stars and axion helioscopes like the Tokyo Axion Helioscope (SUMICO) and the CERN Axion Solar Telescope (CAST), which search for ALPs produced in the Sun and exclude the blue parameter space \cite{Inoue:2008zp,Arik:2008mq,Graham:2015ouw}. Above the threshold $m_a=2m_e\approx 1$\,MeV, decays of the ALPs into electron--positron pairs may affect the assumptions of some of these constraints in a non-trivial way. In the sub-eV mass range, light-shining-through-a-wall experiments (LSW) also provide interesting constraints.

Beam-dump searches are sensitive to ALPs radiated off photons, which are exchanged between the incoming beam and the target nuclei (Primakoff effect) and decay back to photons outside the target. The orange area is a compilation of different runs performed at SLAC \cite{Riordan:1987aw,Bjorken:1988as}. Radiative decays $\Upsilon\to\gamma a$ of Upsilon mesons have been searched for at CLEO and BaBar \cite{Balest:1994ch,delAmoSanchez:2010ac}, and yield the excluded area shaded light green. Bounds from collider searches for ALPs include searches for mono-photons with missing energy ($e^+ e^-\to\gamma a$) at LEP (dark orange), tri-photon searches on and off the $Z$-pole ($e^+ e^-\to 3\gamma$) at LEP (light blue), and searches for the same final state at CDF (purple) and LHC (dark orange). A detailed discussion of these searches can be found in \cite{Mimasu:2014nea,Jaeckel:2015jla,Knapen:2016moh}. For ALP masses in the multi-GeV range, alternative searches for ALP production in ultra-peripheral heavy-ion collisions have the potential to improve the current bounds by up to two orders of magnitude, provided the $a\to\gamma\gamma$ branching ratio is close to 100\% \cite{Knapen:2016moh}. First evidence for light-by-light scattering in 480\,$\mu\mbox{b}^{-1}$ of Pb--Pb collision data has recently been reported by ATLAS \cite{Aaboud:2017bwk}. While the derivation of the precise bound on the ALP--photon coupling is beyond the scope of this work, the green area labeled ``Pb'' shows an estimate obtained based on a rescaling of the projected limit presented in \cite{Knapen:2016moh} to the luminosity used in the ATLAS analysis. Beam-dump experiments and collider searches are directly sensitive to the presence of additional ALP couplings for masses $m_a>2m_e$. The reach of beam-dump experiments, for example, would be strongly reduced if ALPs would decay into electrons before they leave the beam dump. The limits from collider searches and those derived fro heavy-ion collisions shown in the plot assume $\mbox{Br}(a\to\gamma\gamma)=1$. The corresponding exclusion regions would move upwards if this assumption was relaxed. Also, in some cases specific assumptions about the relation between $C_{\gamma\gamma}$ and $C_{\gamma Z}$ were made, which have an influence on the results.

\begin{figure}
\begin{center}
\includegraphics[width=\textwidth]{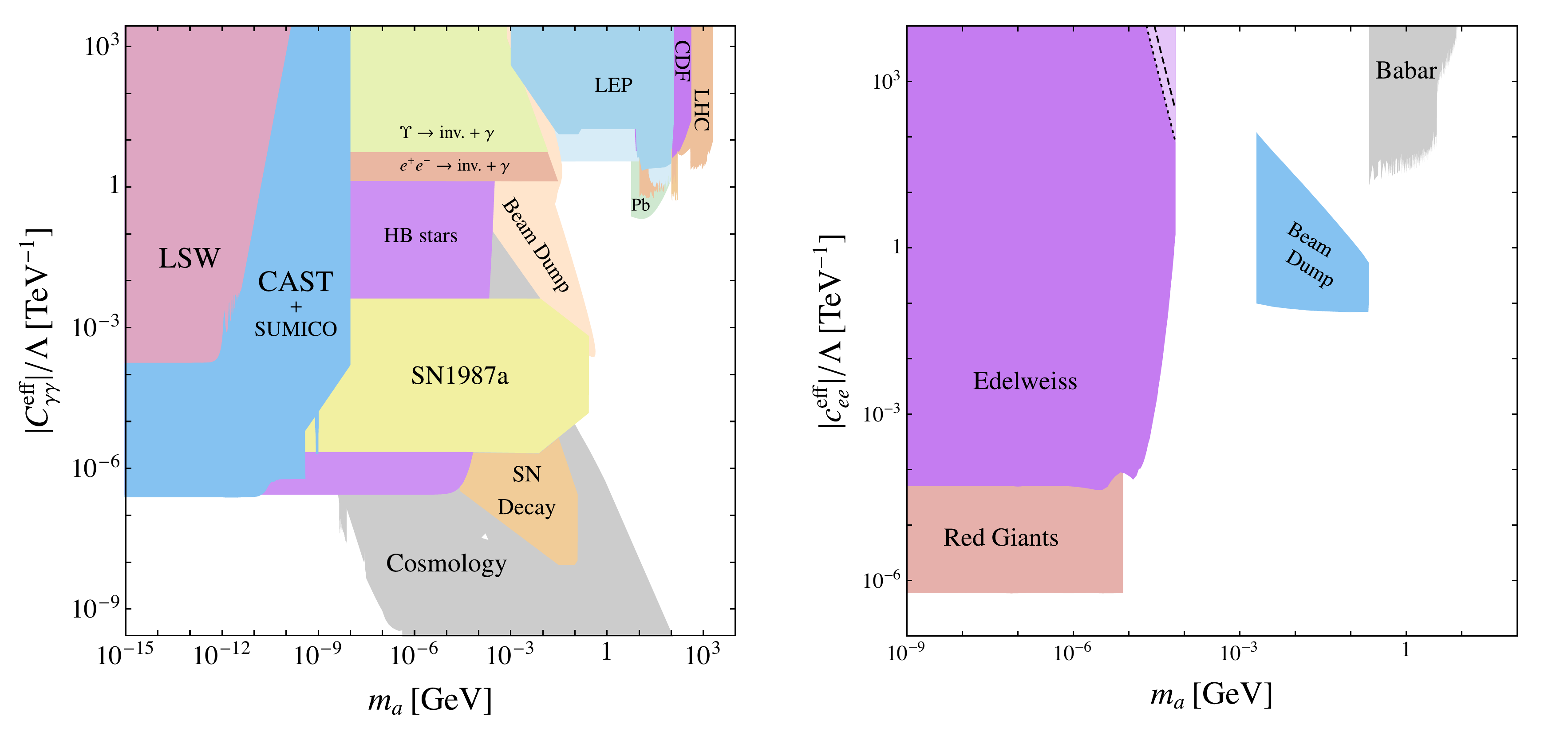}
\end{center}
\vspace{-3mm}
\caption{\label{fig:existingbounds} Existing constraints on the ALP--photon (left) and ALP--electron coupling (right) derived from a variety of particle physics, astro-particle physics and cosmological observations. Several of these bounds are model dependent. The BaBar constraint in the right-hand plot assumes $c_{\mu\mu}\approx c_{ee}$, see (\ref{LFuniv}); otherwise, this is a bound on $|c_{\mu\mu}^{\rm eff}|$. See the text for more details.}
\end{figure}

It follows from this discussion that the ALP--photon coupling is most severely constrained for all ALP masses below about 1\,MeV. At tree-level, this requires that the combination $C_{\gamma\gamma}-1.92\,C_{GG}=C_{WW}+C_{BB}-1.92\,C_{GG}$ of the Wilson coefficients of the operators in which the ALP couples to gauge fields in (\ref{Leff}) must be extremely small, of order $(10^{-9}-10^{-7})\,(\Lambda/\mbox{TeV})$ for $m_a<150$\,eV, and less than $10^{-15}\,(\Lambda/\mbox{TeV})$ for $150\,\mbox{eV}<m_a<1$\,MeV. If we assume that $\Lambda$ lies within a few orders of magnitude of the TeV scale, these constraints would either require an extreme fine tuning or (better) a  mechanism which enforces that $C_{BB}=-C_{WW}$ and $C_{GG}=0$. (However, integrating out a single, complete electroweak multiplet will always generate contributions to $C_{WW}$ and $C_{BB}$ with same sign.) The assumption that such a cancellation can be engineered was made in the recent analysis in \cite{Brivio:2017ije}. Moreover, relation (\ref{di-photonrate}) shows that even in this case an effective coupling $C_{\gamma\gamma}^{\rm eff}\ne 0$ will inevitably be generated at one-loop (and higher-loop) order as long as some couplings in the effective Lagrangian are set by the TeV scale. To see this, consider the following numerical results in the relevant mass window: 
\begin{equation}\label{Cgagaeffvalues}
\begin{aligned}   
   C_{\gamma\gamma}^{\rm eff}(1\,\mbox{MeV}) 
   &\approx C_{\gamma\gamma} - 1.92\,C_{GG} + 5\cdot 10^{-13}\,C_{WW} 
    - 6\cdot 10^{-3}\,c_{ee} - 5\cdot 10^{-8}\,c_{\mu\mu} - 2\cdot 10^{-10}\,c_{\tau\tau} \\
   &\mbox{}- 2\cdot 10^{-7}\,(c_{uu}-c_{dd}) - {\cal O}(10^{-8})\,c_{ss}
    - 4\cdot 10^{-10}\,c_{cc} - 1\cdot 10^{-11}\,c_{bb} - 3\cdot 10^{-14}\,c_{tt} \,, \\
   C_{\gamma\gamma}^{\rm eff}(100\,\mbox{keV}) 
   &\approx C_{\gamma\gamma} - 1.92\,C_{GG} + 5\cdot 10^{-15}\,C_{WW} 
    - 2\cdot 10^{-5}\,c_{ee} - 5\cdot 10^{-10}\,c_{\mu\mu} - 2\cdot 10^{-12}\,c_{\tau\tau} \\
   &\mbox{}- 2\cdot 10^{-9}\,(c_{uu}-c_{dd}) - {\cal O}(10^{-10})\,c_{ss} 
    - 4\cdot 10^{-12}\,c_{cc} - 1\cdot 10^{-13}\,c_{bb} - 3\cdot 10^{-16}\,c_{tt} \,.
\end{aligned}
\end{equation}
For ALP masses below 100\,keV each loop contribution scales with $m_a^2$. We observe that reaching $|C_{\gamma\gamma}^{\rm eff}|/\Lambda<10^{-15}\,\mbox{TeV}^{-1}$ requires a significant fine-tuning of essentially all Wilson coefficients in the effective Lagrangian (\ref{Leff}). This includes the coefficient $C_{WW}$, even though its one-loop contribution is very small. As we will show below, the one-loop radiative corrections to the ALP--electron coupling induce a contribution $\delta c_{ee}\approx -0.8\cdot 10^{-2}\,C_{WW}$ independently of the ALP mass, which adds the terms $5\cdot 10^{-5}\,C_{WW}$ and $2\cdot 10^{-7}\,C_{WW}$ to the two values shown in (\ref{Cgagaeffvalues}). It follows that ALPs with masses in the range between 150\,eV and 1\,MeV are incompatible with the assumption of couplings to SM particles that could be probed at high-energy particle colliders. For masses below 150\,eV, on the other hand, a mechanism which sets $C_{\gamma\gamma}=0$ and $C_{GG}=0$ at tree level would be sufficient to satisfy the relevant constraints irrespective of the values of the remaining ALP couplings.

The left panel in Figure~\ref{fig:existingbounds} shows that above 30\,MeV a window opens for $|C_{\gamma\gamma}^{\rm eff}|/\Lambda\sim 1\,\mbox{TeV}^{-1}$, and above 400\,MeV the ALP--photon coupling is essentially unconstrained as long as it falls between $(10^{-6}-10^{+1})\,\mbox{TeV}^{-1}$. The mass range $m_a>30$\,MeV is thus the best motivated region to search for ALPs at high-energy particle colliders. It is interesting to study loop corrections also for this high-mass region. They can be sizable for all particles lighter than the ALP. For example, at $m_a=10$\,GeV we find
\begin{equation}
\begin{aligned}
   C_{\gamma\gamma}^{\rm eff}(10\,\mbox{GeV})
   &\approx C_{\gamma\gamma} + 10^{-2}\,\Big[ - (12.0-0.3\,i)\,C_{GG} 
    + 0.6\,c_{ee} + 0.6\,c_{\mu\mu} + (0.7-0.4\,i)\,c_{\tau\tau} \\
   &\hspace{2.83cm}\mbox{}+ 0.8\,c_{uu} + 0.2\,c_{dd} + 0.2\,c_{ss} + (0.9-0.4\,i)\,c_{cc} \\[-1mm]
   &\hspace{2.83cm}\mbox{}- (0.1+0.3\,i)\,c_{bb} - 3\cdot 10^{-4}\,c_{tt} + 5\cdot 10^{-3}\,C_{WW} \Big] \,,
\end{aligned}
\end{equation}
where we have included the two-loop estimate (\ref{deltac}) for the contribution from $C_{GG}$. In addition, we expect a two-loop contribution proportional to $C_{WW}$ of order $10^{-2}$ inside the square bracket. Sizable loop corrections can be generated either if $C_{GG}={\cal O}(1)$, or if some of the fermion couplings are of ${\cal O}(10)$. For example, setting $c_{ee}=c_{\mu\mu}=c_{\tau\tau}=10$ for the charged leptons would not lead to any tensions with perturbativity.

\subsubsection{Constraints on the ALP--electron coupling} 
\label{subsec:aeecoupl}

Exclusion limits on the ALP--electron coupling are shown on the right panel of Figure~\ref{fig:existingbounds}. They include searches by the Edelweiss collaboration (shaded purple) \cite{Armengaud:2013rta} for ALPs produced in the Sun by the Compton process $\gamma e^-\to e^- a$, by bremsstrahlung $e^- X\to X a$ off electrons or hydrogen and helium nuclei in the plasma, and by ALP radiation from excited ions. Even stronger limits for $m_a<10^{-5}$\,GeV are derived from observations of Red Giants (shaded red). ALP radiation can lead to the cooling of the cores of these stars, which leads to delayed Helium ignition and modifies the brightness-temperature relation \cite{Raffelt:2006cw}. Axion radiation from electron beams is further constrained by beam-dump experiments performed at SLAC (shaded blue) \cite{Essig:2010gu}. The presence of a sizable ALP--photon coupling would reduce the reach of beam-dump experiments and could affect the astrophysical constraints in a non-trivial way. In particular, the Edelweiss bounds assume that ALPs produced in the Sun do not decay on their way to Earth, which would require that the ALP--photon coupling is tuned to zero with high precision, which is rather implausible in view of our discussion in the previous section. We note, however, that a viable scenario can be obtained by setting the tree-level ALP couplings to quarks and gauge bosons to zero. An ALP--photon coupling is then still induced at one-loop order, see (\ref{di-photonrate2}), and for $m_a<m_e$ it is to good approximation given by $C_{\gamma\gamma}^{\rm eff}\approx-\frac{c_{ee}\,m_a^2}{192\pi^2 m_e^2}$. Requiring that the average decay length of the ALP is larger than the Earth's distance to the Sun, we then obtain the bound
\begin{equation}
   \frac{|c_{ee}^{\rm eff}|}{\Lambda} < \frac{0.022}{\mbox{TeV}}\,\sqrt{\frac{E_a^2}{m_a^2}-1} 
    \left[ \frac{m_a}{\mbox{MeV}} \right]^{-7/2} ; \quad m_a<m_e \,.
\end{equation}
The dashed and dotted lines intersecting the Edelweiss constraint in Figure~\ref{fig:existingbounds} indicate this bound for $E_a=14$\,keV and 1\,keV, respectively. Below these lines, ALPs with the corresponding minimum energies are sufficiently long-lived to travel from the Sun to the Earth before decaying. We also note that limits on the ALP--electron coupling in the mass range between 20\,MeV and 10\,GeV can be derived from dark-photon searches performed at MAMI \cite{Merkel:2014avp} and BaBar \cite{Lees:2014xha}. While a proper conversion of these limits is non-trivial \cite{Liu:2017htz} and beyond the scope of this work, the bounds one obtains are typically rather weak, of order $|c_{ee}^{\rm eff}|/\Lambda\gtrsim 10^3\,\mbox{TeV}^{-1}$. Assuming the approximate universality of the ALP--lepton couplings shown in (\ref{LFuniv}), a stronger constraint can be derived from a dark-photon search in the channel $e^+ e^-\to\mu^+\mu^- Z'$ performed by BaBar \cite{TheBABAR:2016rlg}, which we will reanalyze in the context of our model in the next section. For $C_{\gamma\gamma}=0$, this gives rise to the bound shaded in gray in Figure~\ref{fig:existingbounds}.

Of the one-loop contributions to the effective ALP--electron coupling in (\ref{clleff}), only the photon term shows a sizable sensitivity to the ALP mass, and only in the region where $m_a\gtrsim m_e$. We find (with $\mu=\Lambda=1$\,TeV in the argument of the logarithms)
\begin{equation}
\begin{aligned}   
   c_{ee}^{\rm eff}(m_a=1\,\mbox{GeV}) 
   &\approx c_{ee} \left[ 1 + {\cal O}\big(\alpha\big) \right]
    - 0.8\cdot 10^{-2}\,C_{WW} + (0.7-1.1\,i)\cdot 10^{-2}\,C_{\gamma\gamma} \,, \\
   c_{ee}^{\rm eff}(m_a=1\,\mbox{keV})
   &\approx c_{ee} \left[ 1 + {\cal O}\big(\alpha\big) \right]
    - 0.8\cdot 10^{-2}\,C_{WW} - 1.4\cdot 10^{-2}\,C_{\gamma\gamma} \,.
\end{aligned}
\end{equation}
To satisfy the model-independent bound $|c_{ee}^{\rm eff}|/\Lambda<10^{-6}\,\mbox{TeV}^{-1}$ in the mass range $m_a<10$\,keV would require that $|C_{\gamma\gamma}|$ and $|C_{WW}|$ (and hence both $|C_{WW}|$ and $|C_{BB}|$) must be smaller than approximately $10^{-4}\,(\Lambda/\mbox{TeV})$ in this low-mass region.

\section{Anomalous magnetic moment of the muon}
\label{sec:amu}

\begin{figure}
\begin{center}
\includegraphics[width=0.48\textwidth]{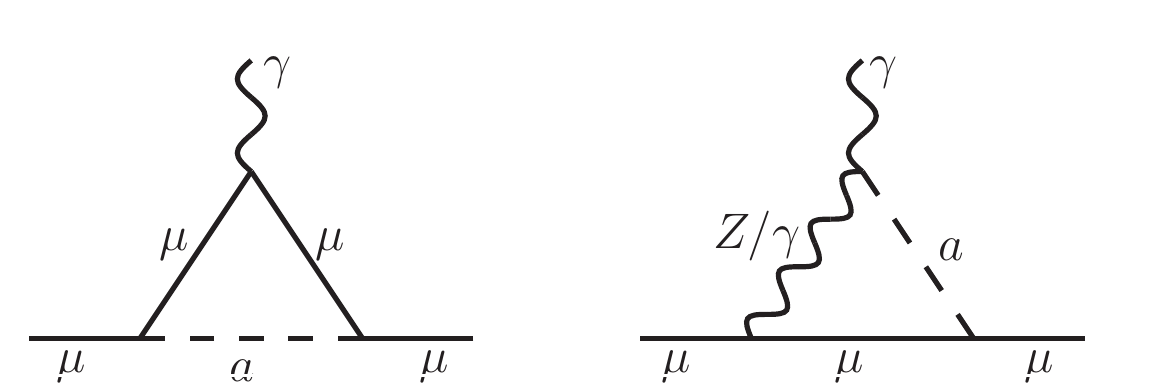}
\end{center}
\vspace{-2mm}
\caption{\label{fig:amugraphs} One-loop diagrams contributing to the anomalous magnetic moment of the muon.}
\end{figure}

The persistent deviation of the measured value of the muon anomalous magnetic moment $a_\mu=(g-2)_\mu/2$ \cite{Bennett:2006fi} from its SM value provides one of the most compelling hints for new physics. The difference $a_\mu^{\rm exp}-a_\mu^{\rm SM}=(29.3\pm 7.6)\cdot 10^{-10}$, where we have taken an average of two recent determinations \cite{Davier:2016iru,Jegerlehner:2017lbd}, differs from zero by about~4 standard deviations. It has been emphasized recently that this discrepancy can be accounted for by an ALP with an enhanced coupling to photons \cite{Marciano:2016yhf}. At one-loop order, the effective Lagrangian gives rise to the contributions to $a_\mu$ shown in Figure~\ref{fig:amugraphs}. The first graph, in which the ALP couples to the muon line, gives a contribution of the wrong sign \cite{Leveille:1977rc,Haber:1978jt}; however, its effect may be overcome by the second diagram, which involves the ALP coupling to photons (or to $\gamma Z$), if the Wilson coefficient $C_{\gamma\gamma}$ in (\ref{Leff}) is sufficiently large \cite{Chang:2000ii,Marciano:2016yhf}. Performing a complete one-loop analysis, we find that the effective ALP Lagrangian gives rise to the new-physics contribution 
\begin{equation}\label{deltaamu}
\begin{aligned}
   \delta a_\mu &= \frac{m_\mu^2}{\Lambda^2}\,\bigg\{ K_{a_\mu}(\mu) 
    - \frac{(c_{\mu\mu})^2}{16\pi^2}\,h_1\bigg(\frac{m_a^2}{m_\mu^2}\bigg) 
    - \frac{2\alpha}{\pi}\,c_{\mu\mu}\,C_{\gamma\gamma} \left[ \ln\frac{\mu^2}{m_\mu^2} + \delta_2 + 3
    - h_2\bigg(\frac{m_a^2}{m_\mu^2}\bigg) \right] \\
   &\hspace{1.5cm}\mbox{}- \frac{\alpha}{2\pi}\,\frac{1-4s_w^2}{s_w c_w}\,c_{\mu\mu}\,C_{\gamma Z}
    \left( \ln\frac{\mu^2}{m_Z^2} + \delta_2 + \frac32 \right) \!\bigg\} \,.
\end{aligned}
\end{equation}
The loop functions read (with $x=m_a^2/m_\mu^2+i0$)
\begin{equation}
\begin{aligned}
   h_1(x) &= 1 + 2x + x (1-x) \ln x - 2x(3-x)\,\sqrt{\frac{x}{4-x}}\,\arccos\frac{\sqrt{x}}{2} \,, \\
   h_2(x) &= 1 - \frac{x}{3} + \frac{x^2}{6} \ln x
    + \frac{2+x}{3}\,\sqrt{x(4-x)}\,\arccos\frac{\sqrt{x}}{2} \,.
\end{aligned}
\end{equation}    
They are positive and satisfy $h_{1,2}(0)=1$ as well as $h_1(x)\approx (2/x)(\ln x-\frac{11}{6})$ and $h_2(x)\approx(\ln x+\frac32)$ for $x\gg 1$. The scheme-dependent constant $\delta_2=-3$ is again related to the treatment of the Levi--Civita symbol in $d$ dimensions, see Appendix~\ref{app:A}. 

Note that in processes in which the ALP only appears in loops but not as an external particle, the scale dependence arising from the UV divergences of the ALP-induced loop contributions are canceled by the scale dependence of a Wilson coefficient in the $D=6$ effective Lagrangian of the SM. In the present case the relevant term yielding a tree-level contribution to $a_\mu$ reads (written in the broken phase of the electroweak theory)
\begin{equation}
   {\cal L}_{\rm eff}^{D=6} 
   \ni - K_{a_\mu}\,\frac{e m_\mu}{4\Lambda^2}\,\bar\mu\,\sigma_{\mu\nu} F^{\mu\nu}\mu \,.
\end{equation}
In order to calculate the Wilson coefficient $K_{a_\mu}$ one would need to consider a specific UV completion of the effective Lagrangian (\ref{Leff}). The large logarithm in the term proportional to $C_{\gamma\gamma}$ in (\ref{deltaamu}) is, however, unaffected by this consideration. The coefficient we obtain for this logarithm agrees with \cite{Marciano:2016yhf} (the remaining finite terms were not displayed in this reference). Two-loop light-by-light contributions proportional to $(C_{\gamma\gamma}/\Lambda)^2$ have been estimated in \cite{Marciano:2016yhf} and were found to be approximately given by 
\begin{equation}
   \delta a_\mu\big|_{\rm LbL} 
   \approx \frac{m_\mu^2}{\Lambda^2}\,\frac{12\alpha^3}{\pi}\,C_{\gamma\gamma}^2\,\ln^2\frac{\mu^2}{m_\mu^2} \,.
\end{equation}
For $\mu=\Lambda=1$\,TeV this evaluates to $\delta a_\mu|_{\rm LbL}\approx 5.6\cdot 10^{-12}\,C_{\gamma\gamma}^2$. 
In the region of parameter space we consider, where $|C_{\gamma\gamma}|/\Lambda\lesssim 2\,\mbox{TeV}^{-1}$ (see below), the impact of this effect is tiny. 

\begin{figure}[t]
\begin{center}
\includegraphics[width=0.88\textwidth]{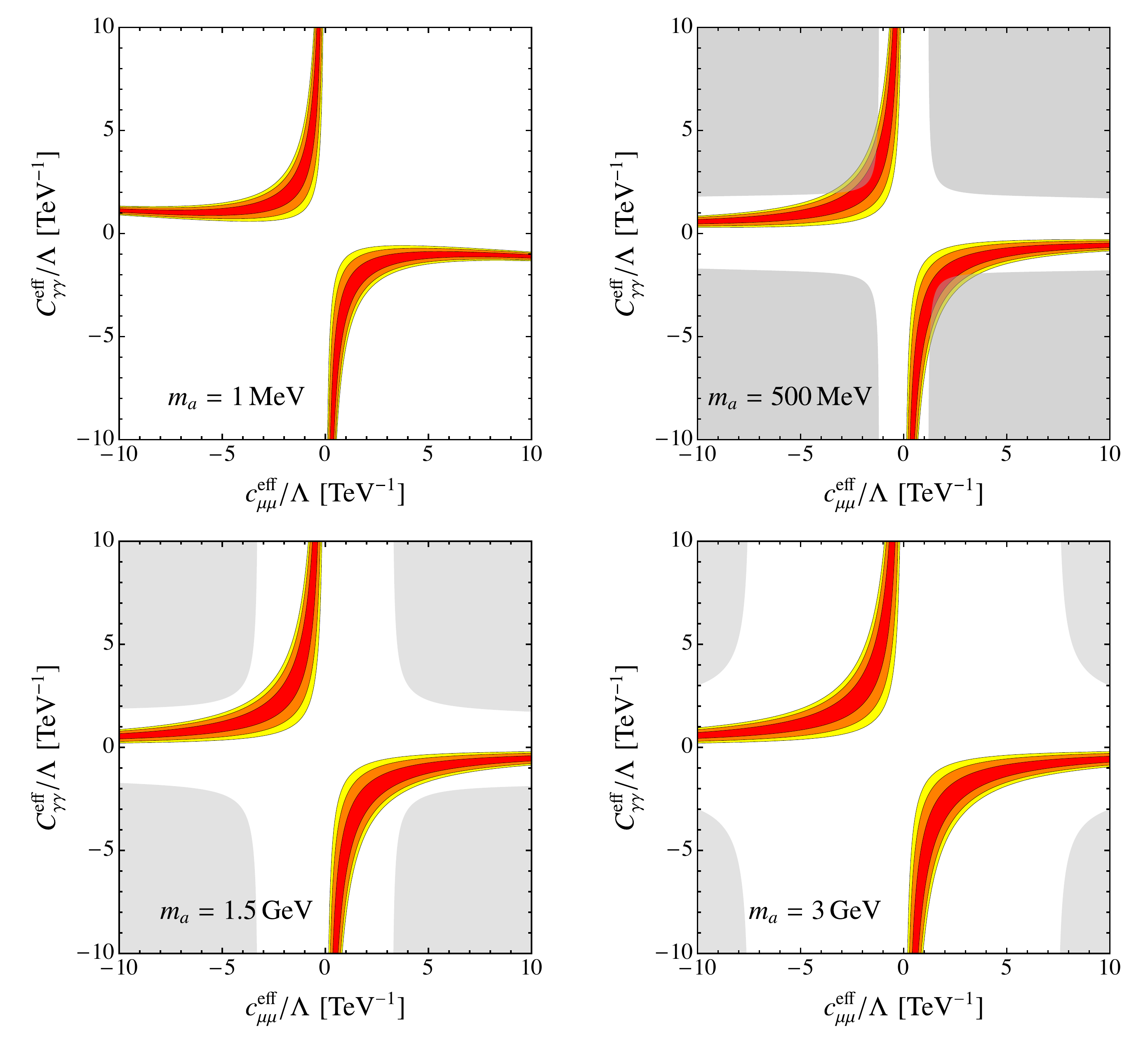}
\end{center}
\vspace{-3mm}
\caption{\label{fig:g-2} Regions in ALP coupling space where the experimental value of $(g-2)_\mu$ is reproduced at 68\% (red), 95\% (orange) and 99\% (yellow) confidence level (CL), for different values of $m_a$. We assume $K_{a_\mu}(\Lambda)=0$ at $\Lambda=1$\,TeV and neglect the tiny contribution proportional to $C_{\gamma Z}$. For $m_a>2m_\mu$, the gray regions are excluded by a dark-photon search in the $e^+ e^-\to\mu^+\mu^-+\mu^+\mu^-$ channel performed by BaBar \cite{TheBABAR:2016rlg}.}
\end{figure}

In our numerical analysis, we will assume that the contribution of $K_{a_\mu}(\mu)$ is subleading at the high scale $\mu=\Lambda$. If the Wilson coefficients $c_{\mu\mu}$ and $C_{\gamma\gamma}$ are of similar magnitude, the logarithmically enhanced contribution is the parametrically largest one-loop correction. It gives a positive shift of $a_\mu$ provided the product $c_{\mu\mu}\,C_{\gamma\gamma}$ is negative. The correction proportional to $C_{\gamma Z}$ is suppressed by $(1-4s_w^2)$ and hence is numerically subdominant. Note also that the contribution proportional to $(c_{\mu\mu})^2$ is suppressed in the limit where $m_a^2\gg m_\mu^2$, while the remaining terms remain unsuppressed. 

Figure~\ref{fig:g-2} shows the regions in the parameter space of the couplings $c_{\mu\mu}$ and $C_{\gamma\gamma}$ in which the experimental value of the muon anomalous magnetic moment can be explained in terms of the ALP-induced loop corrections shown in Figure~\ref{fig:amugraphs}, without invoking a large contribution from the unknown short-distance coefficient $K_{a_\mu}(\Lambda)$. There is a weak dependence on the ALP mass, such that the allowed parameter space increases for $m_a^2\gg m_\mu^2$. Interestingly, we find that an explanation of the anomaly is possible without much tuning as long as one coefficients is of order $\Lambda/\mbox{TeV}$, while the other one can be of similar order or larger. Since $c_{\mu\mu}$ enters observables always in combination with $m_\mu$, it is less constrained by perturbativity than $C_{\gamma\gamma}$. 

\begin{figure}
\begin{center}
\includegraphics[width=0.95\textwidth]{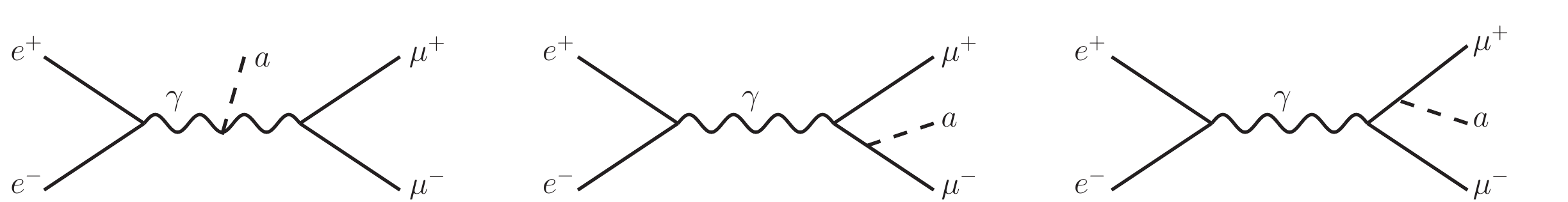}
\end{center}
\vspace{-2mm}
\caption{\label{fig:BaBargraphs} Tree-level Feynman diagrams contributing to the process $e^+ e^-\to\mu^+\mu^- a$.}
\end{figure}

An important constraint on the ALP--photon and ALP--muon couplings, $C_{\gamma\gamma}$ and $c_{\mu\mu}$, can be derived from a search for light $Z'$ bosons performed by BaBar, which constrains the resonant production of muon pairs in the process $e^+e^-\to\mu^+\mu^- +Z'\to\mu^+\mu^- +\mu^+\mu^-$ \cite{TheBABAR:2016rlg}. The Feynman diagrams contributing to this process at tree level (and for $m_e=0$) are shown in Figure~\ref{fig:BaBargraphs}. Neglecting the electron mass and averaging over the initial-state polarizations, we obtain for the cross section
\begin{equation}\label{sigmafin}
   \sigma(e^+ e^-\to\mu^+\mu^- a) 
   = \frac{\alpha^2}{3\pi\Lambda^2} \left[ I_{\gamma\gamma}(r,\epsilon) \left| e^2\,C_{\gamma\gamma} \right|^2 
    + \epsilon\,I_{\gamma\mu}(r,\epsilon)\,\mbox{Re}\left( e^2\,C_{\gamma\gamma}\,c_{\mu\mu}^* \right)
    + \epsilon\,I_{\mu\mu}(r,\epsilon) \left| c_{\mu\mu} \right|^2 \right] ,
\end{equation}
where $r=m_a^2/s$ and $\epsilon=m_\mu^2/s$ are dimensionless ratios, and $\sqrt{s}\approx 10.58$\,GeV is the center-of-mass energy. Note that the contributions involving the ALP--muon coupling are chirally suppressed by a factor $\epsilon=m_\mu^2/s$ and hence are numerically very small in the region where $C_{\gamma\gamma}$ and $c_{\mu\mu}$ take values of similar magnitude. The contributions involving the ALP--photon coupling are logarithmically divergent in the limit $m_\mu\to 0$. Neglecting terms of ${\cal O}(\epsilon)$ and higher in the coefficient functions, which is an excellent approximation numerically, we find
\begin{equation}
\begin{aligned}
   I_{\gamma\gamma}(r,\epsilon) 
   &= \frac23\,(1-r)^3 \ln\frac{(1-r)^2}{\epsilon} 
    - \frac23\,(3-r)\,r^2 \ln r - \frac{7-17r+17r^2-7r^3}{3} \,, \\
   I_{\gamma\mu}(r,\epsilon) 
   &= (1-r)^2 \left[ 8\,\mbox{Li}_2(1-r) + 2\ln r \ln\frac{(1-r)^2}{\epsilon} + \ln^2 r \right]
    - (3+4r+3r^2) \ln r - 5 (1-r^2) \,, \\
   I_{\mu\mu}(r,\epsilon) 
   &= r^2 \left[ \,\frac14 \ln^2 r - \ln r \ln(1+r) - \mbox{Li}_2(-r) - \frac{\pi^2}{12} \right] 
    - \frac{1-2r-3r^2}{4} \ln r - \frac{1-4r+3r^2}{2} \,.
\end{aligned}
\end{equation}
In order to compute the resonant $e^+ e^-\to\mu^+\mu^- a\to\mu^+\mu^-\mu^+\mu^-$ cross section, we need to multiply expression (\ref{sigmafin}) with the $a\to\mu^+\mu^-$ branching ratio. Assuming that only the Wilson coefficients $C_{\gamma\gamma}$ and $c_{\mu\mu}$ are non-zero, and that the ALP couplings to charged leptons are flavor universal, we obtain (for $m_a>2m_\mu$)
\begin{equation}\label{Br}
   \mbox{Br}(a\to\mu^+\mu^-) 
   = \frac{\frac{m_\mu^2}{2m_a^2} \sqrt{1-\frac{4m_\mu^2}{m_a^2}} \left| c_{\mu\mu} \right|^2}%
    {\left| e^2\,C_{\gamma\gamma}\right|^2 + \sum_\ell\,\frac{m_\ell^2}{2m_a^2} \sqrt{1-\frac{4m_\ell^2}{m_a^2}}
     \left| c_{\mu\mu} \right|^2} \,,
\end{equation}
where the sum in the denominator extends over all lepton flavors with $2m_\ell<m_a$. If additional decay channels were present, the bounds derived below would become weaker.

At one-loop order, the effective ALP--photon coupling receives contributions proportional to $c_{\mu\mu}$, which have been shown in (\ref{di-photonrate}) and (\ref{di-photonrate2}). These loop-induced effects contribute to (\ref{sigmafin}) at a level comparable to the chirally-suppressed tree-level contributions involving $c_{\mu\mu}$. In order to properly account for the full dependence on $c_{\mu\mu}$, one should thus use the effective ALP--photon coupling
\begin{equation}\label{Cgagaeff}
   C_{\gamma\gamma}^{\rm eff} 
   = C_{\gamma\gamma} + c_{\mu\mu} \sum_{\ell=e,\mu,\tau}\,\frac{B_1(\tau_\ell)}{16\pi^2}
\end{equation}
instead of $C_{\gamma\gamma}$ in (\ref{sigmafin}) and (\ref{Br}).

For a given value of the ALP mass in the range $2m_\mu<m_a<\sqrt{s}-2m_\mu$ the product
\begin{equation}
   \sigma(e^+ e^-\to\mu^+\mu^- a\to\mu^+\mu^- +\mu^+\mu^-) 
   = \sigma(e^+ e^-\to\mu^+\mu^- a)\,\mbox{Br}(a\to\mu^+\mu^-) 
\end{equation}
is bounded from above by the values shown in Figure~4 of \cite{TheBABAR:2016rlg}. In applying these bounds, we perform an average over the mass range $[m_a-0.5\,\mbox{GeV},m_a+0.5\,\mbox{GeV}]$ to smooth out the spiky structures seen in the figure. The resulting exclusion regions in the $c_{\mu\mu}-C_{\gamma\gamma}$ plane arising at 90\% CL are shown by the gray regions in Figure~\ref{fig:g-2}. In the mass range just above the di-muon threshold, the exclusion region derived from the BaBar analysis lies close to the region where $(g-2)_\mu$ can be explained and indeed excludes a small portion of this region. On the other hand, for ALP masses below $2m_\mu$ no constraints arise, and for $m_a>1.5$\,GeV the constraints quickly become rather weak. We emphasize, however, that ALP searches at the upcoming Belle~II super flavor factory, both in the $a\to\mu^+\mu^-$ and $a\to\gamma\gamma$ channels, have the potential to significantly tighten these constraints and exclude an ALP-based explanation of the muon anomaly in the mass range from $2m_\mu$ up to a few GeV.

\section{Exotic decays of the Higgs boson into ALPs}
\label{sec:Higgs}

The presence of ALP couplings to SM particles gives rise to the possibility of various exotic decay modes of the Higgs boson, which might be discoverable during the high-luminosity run of the LHC. The relevant decay modes are $h\to Za$ and $h\to aa$. These offer a variety of interesting search channels for ALPs, depending on how the ALP and the $Z$ boson decay. In some regions of parameter space, the decay $h\to Za$ may be reconstructed in the $h\to Z\gamma$ search channel and appear as a new-physics contribution to this decay mode. The present experimental upper limits on the $pp\to h\to Z\gamma$ rates reported by CMS \cite{Chatrchyan:2013vaa} and ATLAS \cite{Aad:2014fia} (both at 95\% confidence level (CL)) are 9 and 11 times above the SM value, respectively, thus leaving plenty of room for new-physics effects. A discovery of the $h\to Z\gamma$ decay mode and an accurate measurement of its rate are among the most pressing targets for the high-luminosity LHC run. Very importantly, we will show that ALP searches in the $h\to Za$ and $h\to aa$ channels with subsequent $a\to \gamma\gamma$ or $a\to e^+e^-$ decays can potentially probe regions in the $m_a$\,--\,$C_{\gamma\gamma}^\text{eff}$ and $m_a$\,--\,$c_{ee}^{\rm eff}$ parameter spaces that are inaccessible to any other searches.

\begin{figure}
\begin{center}
\includegraphics[width=0.45\textwidth]{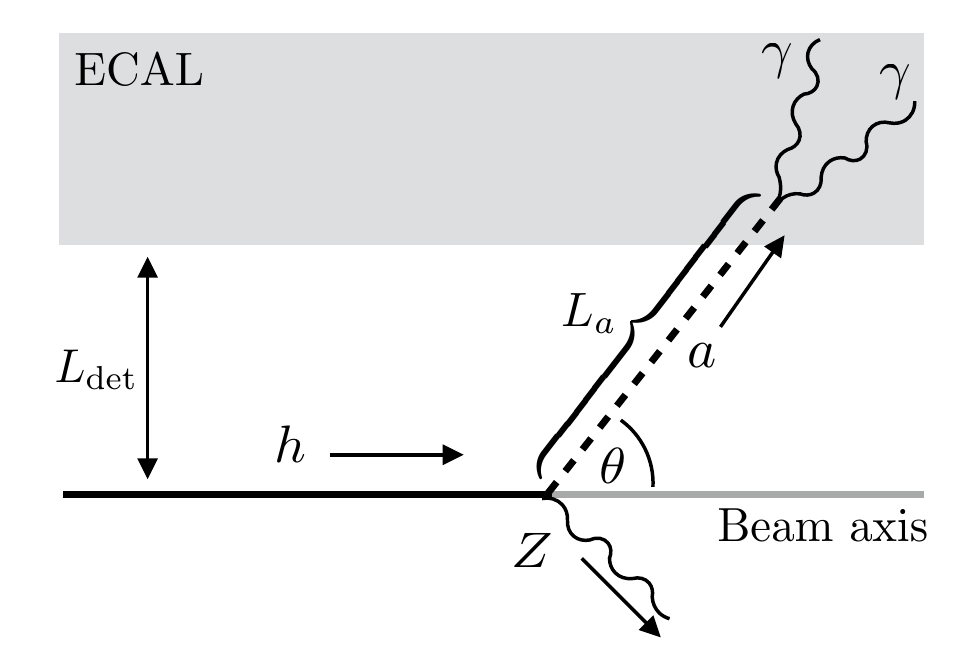}
\end{center}
\vspace{-3mm}
\caption{\label{fig:geo} Sketch of the decay $h\to Za\to Z\gamma\gamma$ in a vertical cross section of the detector. The gray shaded area represents the position of the electromagnetic calorimeter (ECAL).}
\end{figure}

The lifetime of ALPs and their boost factor have important consequences for their detectability. For very light ALPs or very weak couplings, the decay length can become macroscopic and hence only a small fraction of ALPs decay inside the detector. Since to good approximation Higgs bosons at the LHC are produced along the beam direction, the average decay length of the ALP perpendicular to the beam axis is 
\begin{equation}\label{eq:Lperp}
   L_a^\perp(\theta) = \frac{\beta_a\gamma_a}{\Gamma_a}\,\sin\theta \equiv L_a\,\sin\theta \,, 
\end{equation}
where $\theta$ is the angle of the ALP with respect to the beam axis, $\beta_a$ and $\gamma_a$ are the usual relativistic factors, and $\Gamma_a$ is the total decay width of the ALP. For the example of $h\to Za$ decay followed by $a\to\gamma\gamma$, the geometry is sketched in Figure~\ref{fig:geo}. Note that the quantity $L_a^\perp(\theta)$ (but not $L_a$) is invariant under longitudinal boosts along the beam axis, and we are thus free to define $L_a$ and the angle $\theta$ in the Higgs-boson rest frame. If the ALP is observed in the decay mode $a\to X\bar X$, we can express its total width in terms of the branching fraction and partial width for this decay, yielding
\begin{equation}
   L_a = \sqrt{\gamma_a^2-1}\,\,\frac{\mbox{Br}(a\to X\bar X)}{\Gamma(a\to X\bar X)} \,,
\end{equation}
irrespective of the choice of the final state $X\bar X$. The relevant boost factors in the Higgs-boson rest frame are $\gamma_a=(m_h^2-m_Z^2+m_a^2)/(2m_a m_h)$ for $h\to Za$ and $\gamma_a=m_h/(2m_a)$ for $h\to aa$.

We call $f_{\rm dec}^{Za}$ and $f_{\rm dec}^{aa}$ the fraction of all $h\to Za$ and $h\to aa$ events where the ALPs decay before they have traveled a perpendicular distance $L_{\rm det}$ set by the relevant detector components needed for the reconstruction of the particles $X$ (i.e., the electromagnetic calorimeter if $X$ is a photon, and the inner tracker if $X$ is an electron). Since two-body decays of the Higgs boson are isotropic in the Higgs rest frame, it follows that
\begin{equation}\label{eq:29}
\begin{aligned}
   f_{\rm dec}^{Za} &= \int_0^{\pi/2}\!d\theta\sin\theta
    \left( 1 - e^{-L_{\rm det}/L_a^\perp(\theta)} \right) , \\
   f_{\rm dec}^{aa} &= \int_0^{\pi/2}\!d\theta\sin\theta
    \left( 1 - e^{-L_{\rm det}/L_a^\perp(\theta)} \right)^2 . 
\end{aligned}
\end{equation} 
These integrals are discussed in more detail in Appendix~\ref{app:B}. Both event fractions are exponentially close to~1 if $L_a\ll L_{\rm det}$. Numerically, one finds that $f_{\rm dec}^{aa}\approx(f_{\rm dec}^{Za})^2$ to very good approximation, unless the ratio $L_{\rm det}/L_a\ll 1$. In the latter case one obtains
\begin{equation}\label{eq:fdecasy}
   f_{\rm dec}^{Za} \approx \frac{\pi}{2}\,\frac{L_{\rm det}}{L_a} \,, \qquad
   f_{\rm dec}^{aa} \approx \left( \frac{L_{\rm det}}{L_a} \right)^2 \ln\frac{1.258 L_a}{L_{\rm det}} \,.
\end{equation}
We now define the effective branching ratios
\begin{equation}\label{eq:effBR}
\begin{aligned}
   \mbox{Br}(h\to Za\to\ell^+\ell^- + X\bar X) \big|_{\rm eff} 
   &= \mbox{Br}(h\to Za)\,\mbox{Br}(a\to X\bar X)\,f_{\rm dec}^{Za}\,\,\mbox{Br}(Z\to \ell^+\ell^-) \,, \\
   \mbox{Br}(h\to aa\to X\bar X + X\bar X) \big|_{\rm eff} 
   &= \mbox{Br}(h\to aa)\,\mbox{Br}(a\to X\bar X)^2\,f_{\rm dec}^{aa} \,, 
\end{aligned}
\end{equation}
where $\mbox{Br}(Z\to \ell^+\ell^-)=0.0673$ for $\ell=e,\mu$. If the decay length $L_a\ll L_\text{det}$, the effective branching ratios are just the products of the relevant branching fractions for the individual decays. They depend on the squares of the Wilson coefficients $C_{Zh}^{\rm eff}$ and $C_{ah}^{\rm eff}$, which govern the Higgs decay rates into ALPs, and on the branching ratio $\mbox{Br}(a\to X\bar X)$ for the decay mode in which the ALP is reconstructed. In the opposite case, where the ALP decay length is larger than the detector scale $L_{\rm det}$, the dependence on the $a\to X\bar X$ branching ratio drops out to good approximation, because the relevant product $\mbox{Br}(a\to X\bar X)/L_a\propto\Gamma(a\to X\bar X)$ is governed by the $a\to X\bar X$ partial decay rate. Via this rate enters a dependence on the Wilson coefficient $C_{XX}^{\rm eff}$ responsible for the decay $a\to X\bar X$. 

\begin{figure}[t]
\begin{center}
\includegraphics[width=0.55\textwidth]{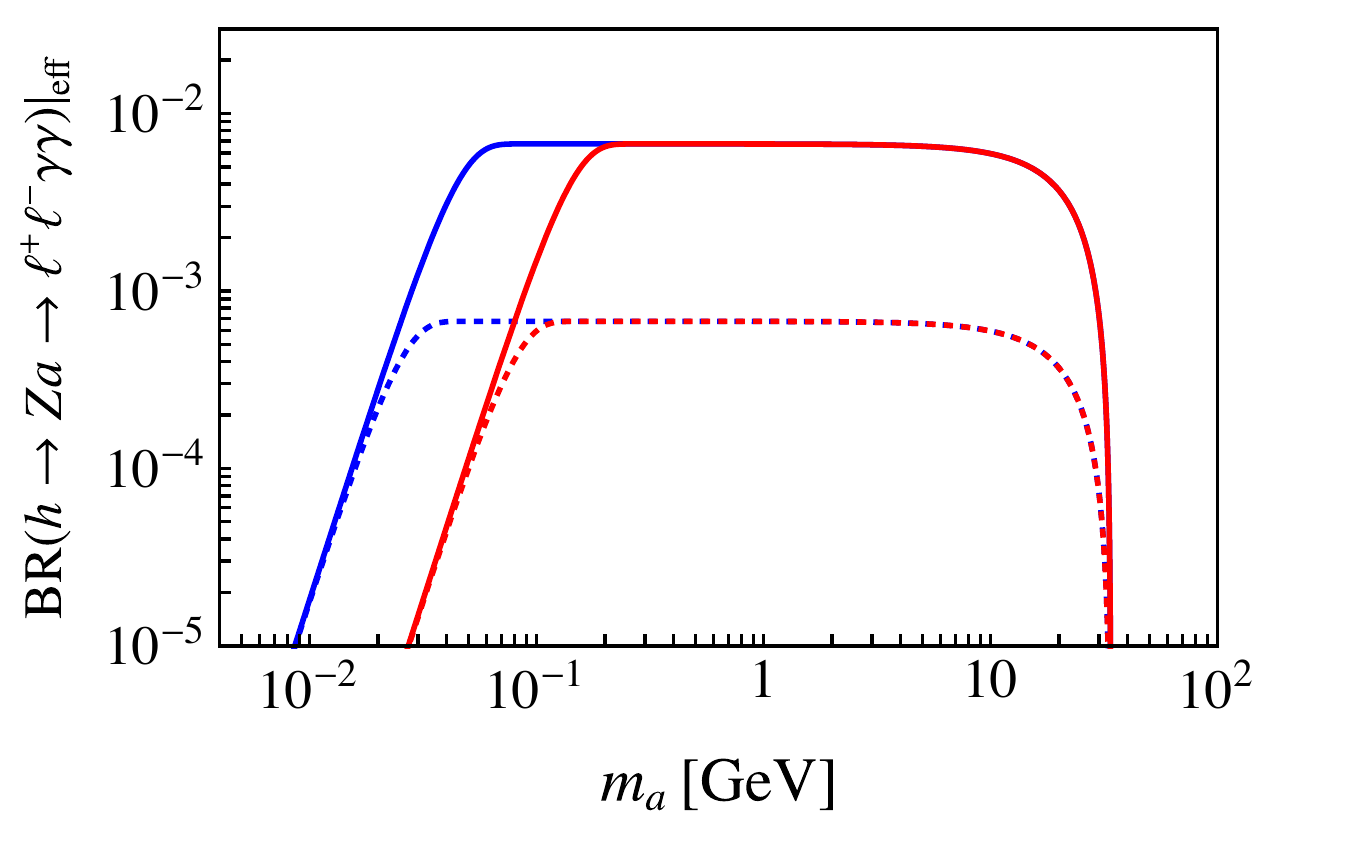}
\end{center}
\vspace{-3mm}
\caption{\label{fig:BRs} Effective $h\to Za\to\ell^+\ell^-\gamma\gamma$ branching ratio as a function of the ALP mass for a fixed value $\text{Br}(h\to Za)=0.1$ at $m_a=1$\,GeV. The solid lines refer to a 100\% $a\to\gamma\gamma$ branching ratio along with $|C_{\gamma\gamma}^{\rm eff}|/\Lambda=1/\mbox{TeV}$ (blue) and $|C_{\gamma\gamma}^{\rm eff}|/\Lambda=0.1/\mbox{TeV}$ (red). The dotted lines are obtained by lowering the $a\to\gamma\gamma$ branching ratio to 10\%.}
\end{figure}

This behavior is illustrated in Figure~\ref{fig:BRs}, which shows the effective branching ratio $\mbox{Br}(h\to Za\to\ell^+\ell^-\gamma\gamma) \big|_{\rm eff}$ for different values of the $a\to\gamma\gamma$ branching ratio and the relevant coefficient $C_{\gamma\gamma}^{\rm eff}$ mediating the di-photon decay. We keep the $h\to Za$ branching fraction fixed at 10\% for $m_a=1$\,GeV. The two solid curves correspond to fixed $\mbox{Br}(a\to\gamma\gamma)=1$ along with $|C_{\gamma\gamma}^{\rm eff}|/\Lambda=1/\mbox{TeV}$ (blue) and $|C_{\gamma\gamma}^{\rm eff}|/\Lambda=0.1/\mbox{TeV}$ (red). For sufficiently large ALP mass the same asymptotic value for the effective branching ratio is obtained, but the reach towards low masses depends sensitively on the value of $C_{\gamma\gamma}^{\rm eff}$. The two dotted lines are obtained in the same way, but with $\mbox{Br}(a\to\gamma\gamma)=0.1$. In this case the asymptotic value for the effective branching ratio is reduced by a factor 10, but the behavior in the low-mass region is the same as before. As explained above, for low masses the effective branching ratio becomes independent of $\mbox{Br}(a\to\gamma\gamma)$, while for large masses it becomes independent of $C_{\gamma\gamma}^{\rm eff}$.

\subsection[ALP searches in $h\to Za$ decay]{\boldmath ALP searches in $h\to Za$ decay}
\label{sec:hZa}

\begin{figure}
\begin{center}
\includegraphics[width=0.67\textwidth]{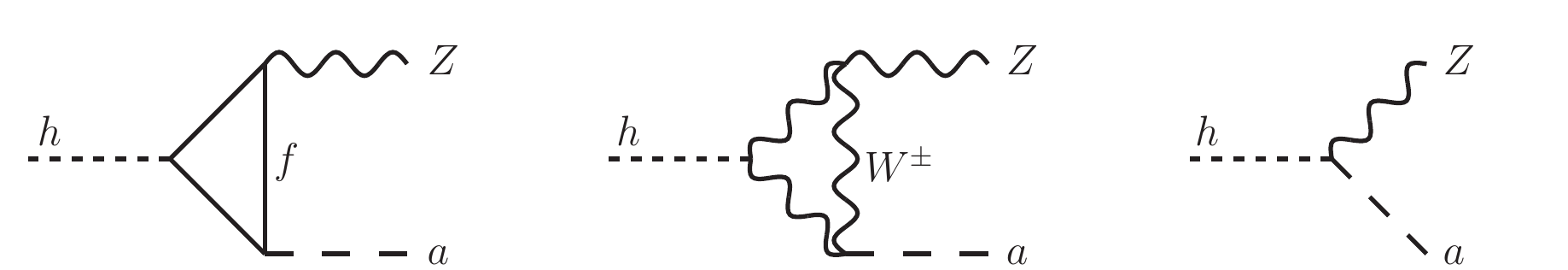}
\end{center}
\vspace{-2mm}
\caption{\label{fig:hZagraphs} Feynman diagrams contributing to the decay $h\to Za$.}
\end{figure}

\begin{figure}
\begin{center}
\includegraphics[width=\textwidth]{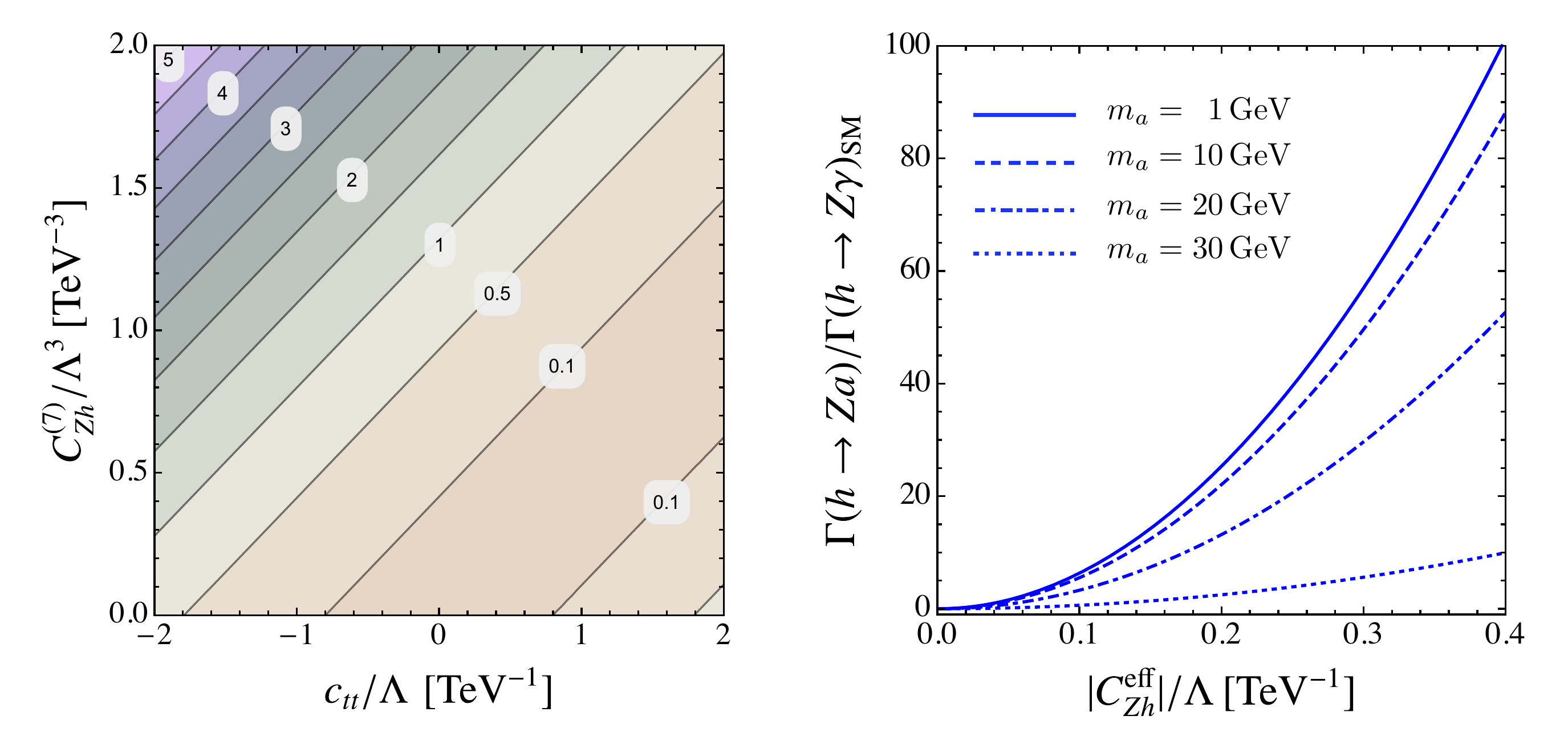}
\end{center}
\vspace{-3mm}
\caption{\label{fig:hZarate} Left: Contours for the ratio $\Gamma(h\to Za)/\Gamma(h\to Z\gamma)_{\rm SM}$ in the plane of the Wilson coefficients $c_{tt}$ and $C_{Zh}^{(7)}$ for $m_a<1$\,GeV and $\Lambda=1$\,TeV. Right: The same rate ratio as a function of the effective Wilson coefficient $C_{Zh}^{\rm eff}$ for different ALP masses.}
\end{figure}

The relevant Feynman diagrams contributing to the $h\to Za$ decay amplitude up to one-loop order are depicted in Figure~\ref{fig:hZagraphs}. The effective Lagrangian (\ref{Leff}) does not contain a dimension-5 operator contributing to the $h\to Za$ decay amplitude at tree level. The only contribution arising at this order is due to fermion loop graphs. Because both the Higgs boson and the ALP couple to fermions proportional to the fermion mass, the only relevant effects comes from the top quark. The $W$-boson loop diagram shown in the second graph vanishes, since there are not enough 4-vectors available to saturate the indices of the Levi--Civita tensor in the $aWW$ vertex. A tree-level contribution to the $h\to Za$ decay amplitude (third graph) arises first at dimension-7 order, from the third operator shown in (\ref{LeffD>5}). Evaluating all contributions, we obtain \cite{Bauer:2016ydr}  
\begin{equation}
   \Gamma(h\to Za) = \frac{m_h^3}{16\pi\Lambda^2} \left| C_{Zh}^{\rm eff} \right|^2 
     \lambda^{3/2}\bigg(\frac{m_Z^2}{m_h^2},\frac{m_a^2}{m_h^2}\bigg) \,,
\end{equation}
where $\lambda(x,y)=(1-x-y)^2-4xy$, and we have defined
\begin{equation}\label{GammaSZh7}
   C_{Zh}^{\rm eff} = C_{Zh}^{(5)} - \frac{N_c\,y_t^2}{8\pi^2}\,T_3^t\,c_{tt}\,F
     + \frac{v^2}{2\Lambda^2}\,C_{Zh}^{(7)} \,.
\end{equation}
Here $y_t$ and $T_3^t=\frac12$ are the top-quark Yukawa coupling and weak isospin, and $C_{Zh}^{(5)}=0$. The top-quark contribution involves the parameter integral
\begin{equation}
   F = \int_0^1\!d[xyz]\,\frac{2m_t^2-x m_h^2-z m_Z^2}{m_t^2-xy m_h^2-yz m_Z^2-xz m_a^2} 
   \approx 0.930 + 2.64\cdot 10^{-6}\,\frac{m_a^2}{\mbox{GeV}^2} \,,
\end{equation}
where $d[xyz]\equiv dx\,dy\,dz\,\delta(1-x-y-z)$. Numerically, we obtain
\begin{equation}\label{C5eff}
   C_{Zh}^{\rm eff}\approx C_{Zh}^{(5)} - 0.016\,c_{tt}
    + 0.030\,C_{Zh}^{(7)} \left[ \frac{\mbox{1\,TeV}}{\Lambda} \right]^2 .
\end{equation}
The left plot in Figure~\ref{fig:hZarate} shows our predictions for the $h\to Za$ decay rate normalized to the SM rate $\Gamma(h\to Z\gamma)_{\rm SM}=6.32\cdot 10^{-6}$\,GeV \cite{Dittmaier:2011ti}. We set $C_{Zh}^{(5)}=0$ and display the rate ratio in the plane of the Wilson coefficients $c_{tt}$ and $C_{Zh}^{(7)}$. Since only the relative sign of the two coefficients matters, we take $C_{Zh}^{(7)}$ to be positive without loss of generality. We find that, in a large portion of parameter space, the exotic $h\to Za$ mode can naturally have a similar decay rate as the $h\to Z\gamma$ mode in the SM, especially if the top-quark contribution interferes constructively with the dimension-7 contribution proportional to $C_{Zh}^{(7)}$. 

The argument for the absence of a tree-level dimension-5 contribution to the $h\to Za$ decay amplitude holds in all new-physics models, in which the operators in the effective Lagrangian arise from integrating out heavy particles whose mass remains large in the limit of unbroken electroweak symmetry \cite{Bauer:2016ydr}. However, this argument does not apply for the class of models featuring new heavy particles which receive all or most of their mass from electroweak symmetry breaking. Concrete examples of such models include little-Higgs models, in which fermionic top partners can have very large Higgs couplings \cite{Han:2003wu,Perelstein:2003wd}, and triplet--doublet dark matter models with vector-like leptons \cite{Dedes:2014hga,Freitas:2015hsa}, which are generalizations of the Wino--Higgsino dark matter scenario in the minimal supersymmetric standard model. The effective Lagrangian for such models generically contains operators which are non-polynomial in the Higgs field (see e.g.\ \cite{Pierce:2006dh}). At dimension-5 order, there is a unique such operator relevant to the decay $h\to Za$. It is given by \cite{Bauer:2016ydr} 
\begin{equation}\label{Leffnonpol} 
   {\cal L}_{\rm eff}^{\rm non-pol}\ni \frac{C_{Zh}^{(5)}}{\Lambda} \left( \partial^\mu a\right)
    \left( \phi^\dagger\,iD_\mu\,\phi + \mbox{h.c.} \right) \ln\frac{\phi^\dagger\phi}{\mu^2} + \dots \,.
\end{equation}
Its contribution to the decay amplitude was already included in (\ref{GammaSZh7}) and (\ref{C5eff}). The decay $h\to Za$ is unique in the sense that, at dimension-5 order, a tree-level $hZa$ coupling can only arise in such special models. Note that the non-polynomial operator in (\ref{Leffnonpol}) also arises at one-loop order in the SM. Integrating out the top-quark from the effective Lagrangian generates a contribution to $C_{Zh}^{(5)}$ given by the second term in (\ref{GammaSZh7}) evaluated with $F=1$.

In the right plot in Figure~\ref{fig:hZarate}, we allow for non-zero $C_{Zh}^{(5)}$ and display the rate ratio as a function of the effective Wilson coefficient $C_{Zh}^{\rm eff}$ defined in (\ref{C5eff}) for different ALP masses. In models where a tree-level dimension-5 contribution is present, one can naturally obtain $h\to Za$ rates exceeding the SM $h\to Z\gamma$ rate by orders of magnitude. For example, with $|C_{Zh}^{\rm eff}|/\Lambda=0.3\,\mbox{TeV}^{-1}$ and for a light ALP ($m_a<1$\,GeV) one finds a ratio of about 60, corresponding to a 9\% $h\to Za$ branching ratio. This would be a spectacular new-physics effect. We find that the decay rate is approximately independent of the ALP mass as long as $m_a$ is below a few GeV. The decay $h\to Za$ is kinematically allowed as long as $m_a<m_h-m_Z\approx 33.9$\,GeV. Figure~\ref{fig:hZarate} shows that significant decay rates can be found even close to the kinematic limit. 

The LHC collaborations have reported the 95\% CL upper limit $\mbox{Br}(h\to\mbox{BSM})<0.34$ on decays of the Higgs boson into non-SM final states, obtained from a combined analysis of the Higgs-boson production and decays rates \cite{Khachatryan:2016vau}. This implies the bound $\Gamma(h\to\mbox{BSM})<2.1$\,MeV on any decay rate involving new particles. For the special case of $h\to Za$ decay, we thus obtain 
\begin{equation}\label{eq:cZhbound}
   \big| C_{Zh}^{\rm eff} \big| < 0.72 \left[ \frac{\Lambda}{1\,\mbox{TeV}} \right] .
\end{equation}
This bound is obtained by neglecting the ALP mass and gets weaker if $m_a$ approaches the kinematic limit $m_a=m_h-m_Z$. Using the projected bound $\text{Br}(h\to\text{BSM})<0.1$ that can be obtained with an integrated luminosity of 3000\,fb$^{-1}$ at $\sqrt{s}=14$\,TeV \cite{ATL-PHYS-PUB-2014-016} (assuming no new physics) one would find $|C_{Zh}^{\rm eff}|<0.34\,(\Lambda/\mbox{TeV})$. The existing upper bounds for Higgs-boson decays into invisible particles, which are $\mbox{Br}(h\to\mbox{invisible})<0.23$ from ATLAS \cite{Aad:2015pla} and $\mbox{Br}(h\to\mbox{invisible})<0.24$ from CMS \cite{Khachatryan:2016whc}, do not currently constrain the $h\to Za$ decay rate, even if $\text{Br}(a\to\mbox{invisible})=1$.

\begin{figure}[t]
\begin{center}
\includegraphics[width=\textwidth]{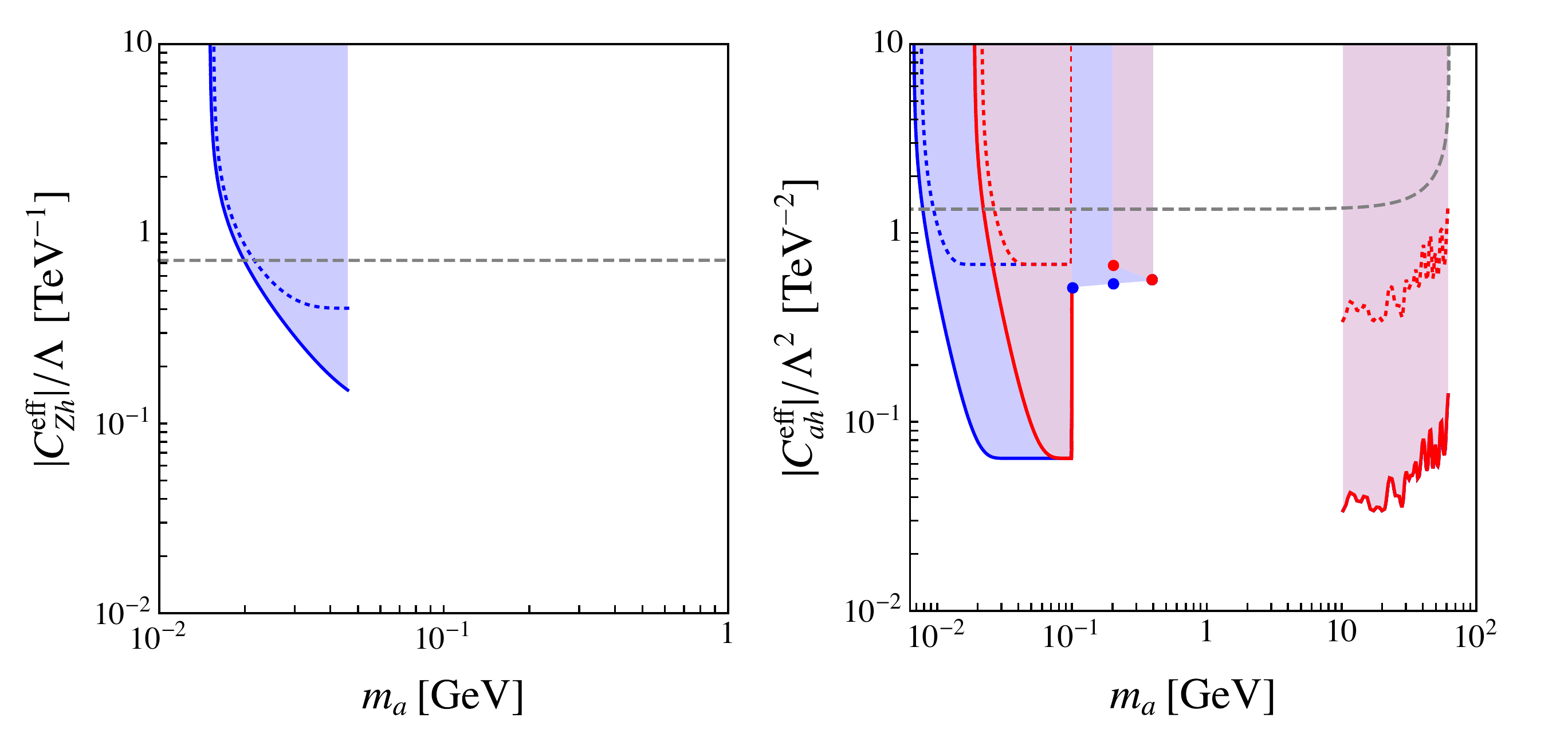}
\end{center}
\vspace{-3mm}
\caption{\label{fig:haa} Parameter space excluded by existing searches for $h\to Z\gamma$ decays (left panel), by the measurements of the $pp\to h\to\gamma\gamma$ rate (low-mass region in the right panel), and by dedicated searches for $h\to\gamma\gamma+\gamma\gamma$ in the mass range between 100 and 400\,MeV (three points) and in the region $m_a=(10-62.5)$\,GeV (right panel). Solid and dotted curves are obtained for $\text{Br}(a\to\gamma\gamma)=1$ and 0.1, respectively, while red and blue lines (and points) refer to $|C_{\gamma\gamma}^{\rm eff}|/\Lambda=1/\mbox{TeV}$ and $0.1/\mbox{TeV}$. The gray dashed lines indicate the model-independent bounds (\ref{eq:cZhbound}) and (\ref{haaBSM}).}
\end{figure}

Depending on the dominant branching ratio of the ALP, the decay $h\to Za$ can give rise to various interesting experimental signatures. ALP decays into photons can be searched for in the $h\to Za\to\ell^+\ell^-\gamma\gamma$ final state. No dedicated searches have been performed in this channel yet. However, for strongly boosted ALPs the two photons would be reconstructed as a single photon jet, and the decays $h\to Za$ would then lead to a modification of the observed $pp\to h\to Z\gamma$ rate. Since there is no interference term, this rate would necessarily be enhanced in this case. From Figure~\ref{fig:hZarate} it follows that this enhancement can easily be of ${\cal O}(1)$ and stronger. We estimate the mass below which a di-photon decay of the ALP will mimic a single photon in the detector to be about 47\,MeV by following the analysis for $h\to aa$ decay of \cite{ATLAS:2012soa} and accounting for the different Lorentz boost factors (see the discussion in Section~\ref{subsec:htoaa}). The current best limit on the cross section of $\sigma(pp\to h\to Z\gamma)<9\,\sigma(pp\to h\to Z\gamma)_\text{SM}$ \cite{Chatrchyan:2013vaa} then rules out the shaded area above the solid and dotted blue lines in the left panel of Figure~\ref{fig:haa}. The lines in this figure have the same meaning as in Figure~\ref{fig:BRs}. Solid and dotted lines refer to $\text{Br}(a\to\gamma\gamma)=1$ and $\text{Br}(a\to \gamma\gamma)=0.1$, respectively. Blue lines are obtained with $|C_{\gamma\gamma}^{\rm eff}|/\Lambda=1/\mbox{TeV}$, while red lines correspond to $|C_{\gamma\gamma}^{\rm eff}|/\Lambda=0.1/\mbox{TeV}$. With present luminosity, only the former choice gives rise to non-trivial bounds. As explained above, for low ALP masses the constraints become independent of the $a\to\gamma\gamma$ branching ratio. For very low ALP masses sensitivity is lost, because most of the ALPs decay outside the detector.

\begin{figure}
\begin{center}
\includegraphics[width=\textwidth]{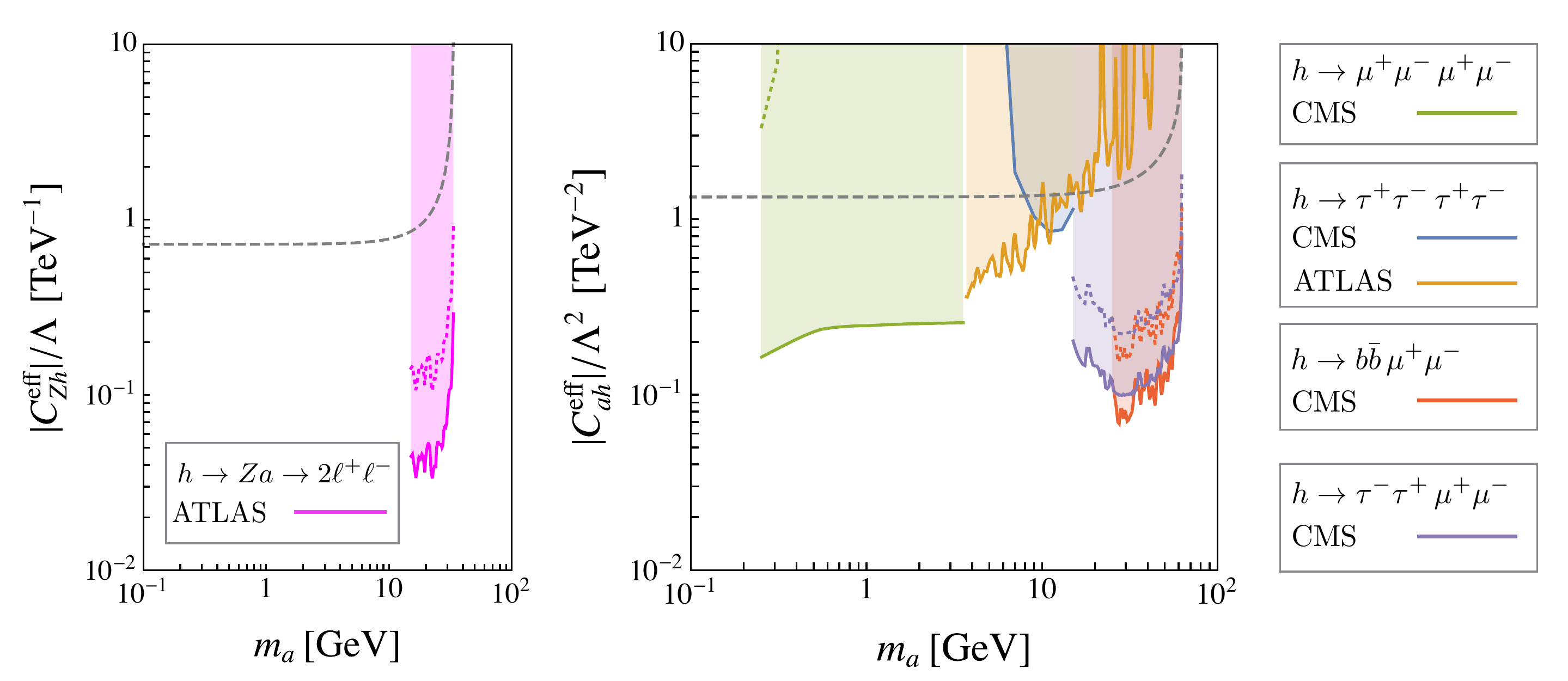}
\end{center}
\vspace{-3mm}
\caption{\label{fig:otherbounds} Left: Parameter space excluded by a search for $h\to ZZ_d\to 2\ell^+\ell^-$, assuming $\text{Br}(a\to\ell^+\ell^-)=1$ (solid line) and $\text{Br}(a\to\ell^+\ell^-)=0.1$ (dotted line). Right: Constraints from dedicated LHC searches for $h\to aa$ with subsequent ALP decays into fermion pairs. The solid contours assume $\text{Br}(a\to\ell^+\ell^-)=1$ if all fermions have the same flavor, and $\text{Br}(a\to\mu^+\mu^-)=\text{Br}(a\to\tau^+\tau^-)=\text{Br}(a\to b\bar b)=0.5$ otherwise. The dotted contours correspond to $\text{Br}(a\to\ell^+\ell^-)=0.1$ if all fermions have the same flavor, and $\text{Br}(a\to\mu^+\mu^-)=0.1$, $\text{Br}(a\to\tau^+\tau^-)=\text{Br}(a\to b\bar b)=0.9$ otherwise. The gray dashed lines indicate the model-independent bounds (\ref{eq:cZhbound}) and (\ref{haaBSM}).}
\end{figure}

If the leptonic decay modes are relevant, ALPs can be searched for in $h\to Za\to 4\ell$ decays. An analysis by ATLAS searching for new ``dark'' bosons $Z_d$ produced in Higgs decays $h\to Z Z_d$ with subsequent decays $Z Z_d\to 4\ell$, where $\ell=e$ or $\mu$, can be reinterpreted to constrain $C_{Zh}^\text{eff}$ in the considered mass window $m_{Z_d}=(15-35)$\,\text{GeV} \cite{Aad:2015sva}. We show the excluded region in the left panel of Figure~\ref{fig:otherbounds}, in which the solid and dotted contours correspond to $\text{Br}(a\to\ell^+\ell^-)=1$ and 0.1, respectively. For these high ALP masses, the $h\to Za\to 4\ell$ rate is essentially independent of the values of the Wilson coefficients $|c_{\ell\ell}^{\rm eff}|$. We strongly encourage our experimental colleagues to extend these searches to lower masses and to separate the final-state lepton flavors. The expected asymmetry between electron, muon and tau final states from ALP decays would be a striking signature of a light pseudoscalar boson. The possibility to observe light new particles in Higgs decays with this final state has also been pointed out in \cite{Gonzalez-Alonso:2014rla}. A heavier ALP can also decay into heavy-quark pairs, which would provide spectacular signatures such as $h\to Za\to\ell^+\ell^- b\bar b$, or into di-jets, i.e.\ $h\to Za\to\ell^+\ell^- j(j)$, where a single jet would be observed in the case of two strongly collimated jets. Very light or weakly coupled ALPs can remain stable on detector scales. In this case, a Higgs produced in vector-boson fusion or in association with a $Z$-boson or a top-quark pair can lead to interesting signatures of the type $pp\to hjj\to Z+\rlap{\,/}E_T+jj$, $pp\to hZ\to ZZ+\rlap{\,/}E_T$, or $pp\to ht\bar t\to Z+\rlap{\,/}E_T+t\bar t$. 

\subsection[ALP searches in $h\to aa$ decay]{\boldmath ALP searches in $h\to aa$ decay}
\label{subsec:htoaa}

\begin{figure}
\begin{center}
\includegraphics[width=0.7\textwidth]{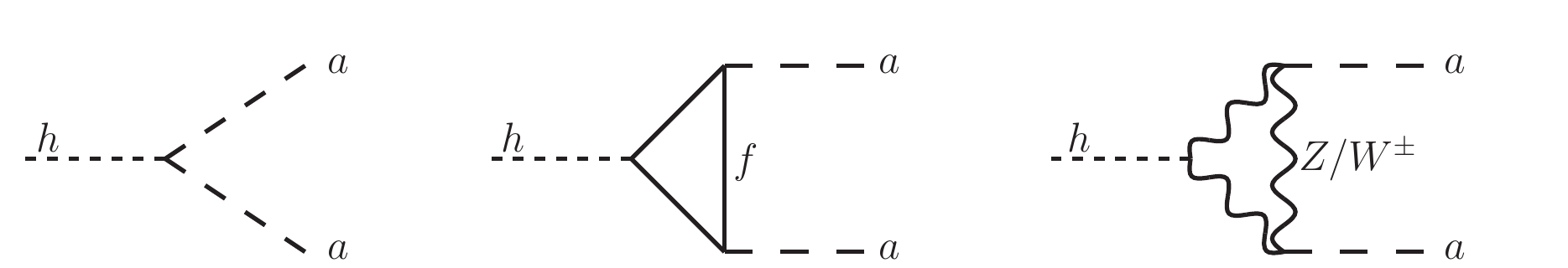}
\end{center}
\vspace{-2mm}
\caption{\label{fig:haadiagrams} Feynman diagrams contributing to the decay $h\to aa$. The last diagram involves the Higgs couplings to $W$ and $Z$ bosons.}
\end{figure}

By means of the Higgs portal interactions in the dimension-6 effective Lagrangian (\ref{LeffD>5}), as well as by loop-mediated dimension-6 processes, a Higgs boson can decay into a pair of ALPs. We have calculated the $h\to aa$ decay rate including the tree-level Higgs-portal interactions as well as all one-loop corrections arising from two insertions of operators from the dimension-5 effective Lagrangian~(\ref{Leff}). The relevant diagrams are shown in Figure~\ref{fig:haadiagrams}. Since both the Higgs boson and the ALP couple to fermions proportional to their mass, only the top-quark contribution needs to be retained in the second diagram. Keeping $m_a$ only in the phase space and neglecting it everywhere else, we find
\begin{equation}
   \Gamma(h\to aa) 
   = \frac{v^2 m_h^3}{32\pi\Lambda^4} \left|C_{ah}^{\rm eff}\right|^2
    \left( 1 - \frac{2m_a^2}{m_h^2} \right)^2 \sqrt{1 - \frac{4m_a^2}{m_h^2}} \,,
\end{equation}
where the effective coupling is given by 
\begin{equation}\label{Cahres}
\begin{aligned}
   C_{ah}^{\rm eff} &= C_{ah}(\mu)
    + \frac{N_c\,y_t^2}{4\pi^2}\,c_{tt}^2 \left[ \ln\frac{\mu^2}{m_t^2} - g_1(\tau_{t/h}) \right]
    - \frac{3\alpha}{2\pi s_w^2} \left( g^2 C_{WW} \right)^2
    \left[ \ln\frac{\mu^2}{m_W^2} + \delta_1 - g_2(\tau_{W/h}) \right] \\
   &\quad\mbox{}- \frac{3\alpha}{4\pi s_w^2 c_w^2} \left( \frac{g^2}{c_w^2}\,C_{ZZ} \right)^2
    \left[ \ln\frac{\mu^2}{m_Z^2} + \delta_1 - g_2(\tau_{Z/h}) \right] ,
\end{aligned}
\end{equation}
with $\tau_{i/h}\equiv 4m_i^2/m_h^2$ and $\delta_1=-\frac{11}{3}$. The relevant loop functions read
\begin{equation}
   g_1(\tau) = \tau\,f^2(\tau) + 2\sqrt{\tau-1}\,f(\tau) - 2 \,, \qquad
   g_2(\tau) = \frac{2\tau}{3}\,f^2(\tau) + 2\sqrt{\tau-1}\,f(\tau) - \frac83 \,.
\end{equation}
Note that the second Higgs-portal interaction in (\ref{LeffD>5}) does not contribute in this approximation, because its effect is suppressed by $m_a^2/m_h^2$. Numerically, we obtain for $\Lambda=1$\,TeV
\begin{equation}
   C_{ah}^{\rm eff}\approx C_{ah}(\Lambda)
    + 0.173\,c_{tt}^2 - 0.0025 \left( C_{WW}^2 + C_{ZZ}^2 \right) ,
\end{equation}
indicating that the top-quark contribution, in particular, can be sizable. Relation (\ref{Cahres}) shows that even if the portal coupling $C_{ah}$ vanishes at some scale, an effective coupling is induced at one-loop order if the ALP couples to at least one of the heavy SM particles ($t$, $Z$ or $W$). Also, because of the presence of UV divergences in the various terms, the coupling $C_{ah}(\mu)$ must cancel the scale dependence of the various other terms, and hence it is not consistent to set it to zero in general. For a light ALP ($m_a<1$\,GeV) a 10\% $h\to aa$ branching ratio is obtained for $|C_{ah}^{\rm eff}|/\Lambda^2=0.62\,\mbox{TeV}^{-2}$. Note that a Wilson coefficient of this size could even be due to a loop-induced contribution from the top quark, if $|c_{tt}|/\Lambda\approx 1.9\,\mbox{TeV}^{-1}$.

\begin{figure}
\begin{center}
\includegraphics[width=0.45\textwidth]{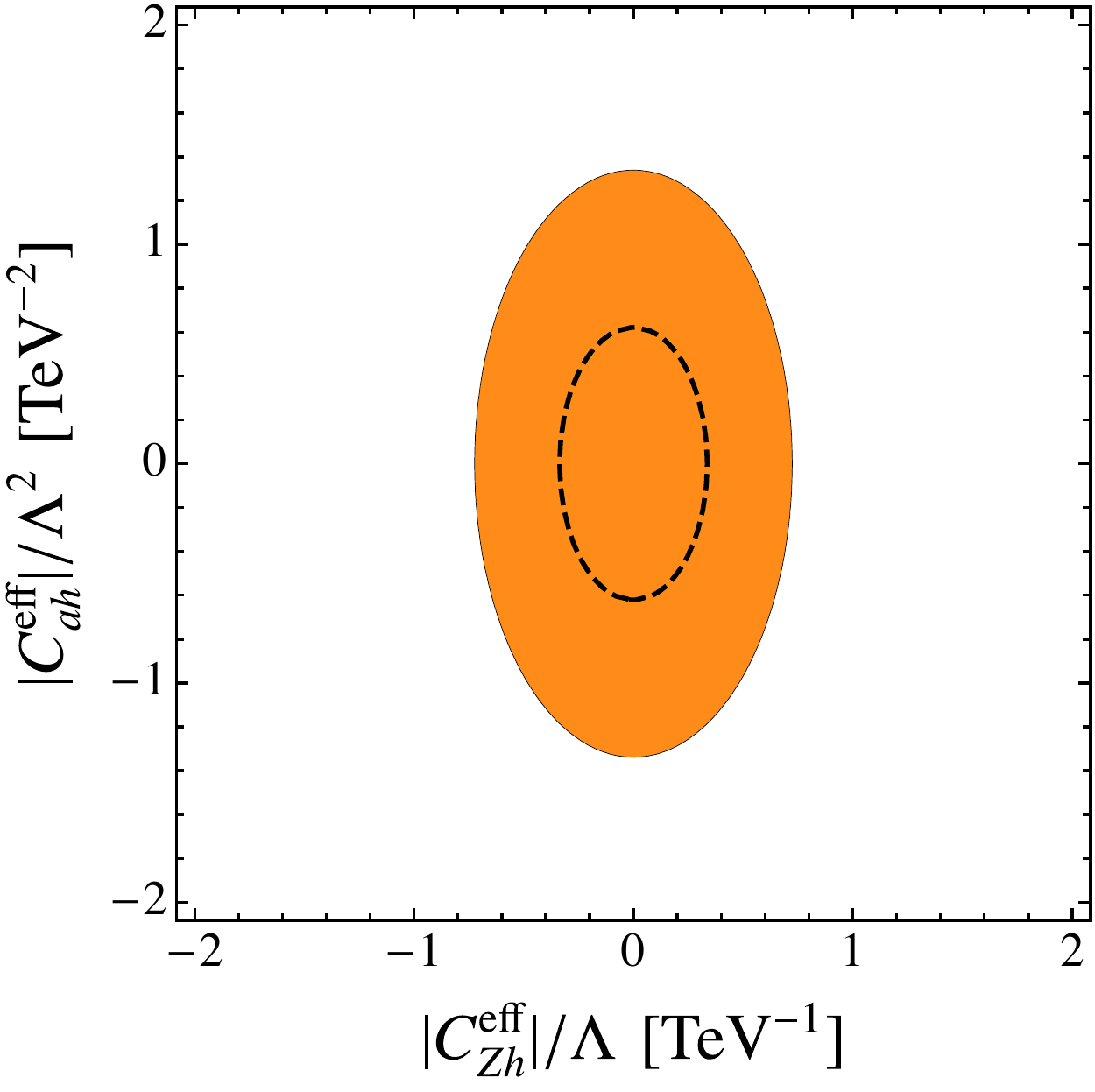}
\end{center}
\vspace{-3mm}
\caption{\label{fig:CZhvsCah} Allowed region for the Wilson coefficients $C_{Zh}^{\rm eff}$ and $C_{ah}^{\rm eff}$ obtained from the present bound $\text{Br}(h\to\text{BSM})<0.34$ (orange) derived from the global analysis of Higgs decays \cite{Khachatryan:2016vau}. The black dashed line shows the projected bound one would obtain for $\text{Br}(h\to\text{BSM})<0.1$, as expected for 3000\,fb$^{-1}$ of integrated luminosity at $\sqrt{s}=14$ TeV.}
\end{figure}

Imposing the current upper limit $\mbox{Br}(h\to\mbox{BSM})<0.34$ (at 95\% CL) \cite{Khachatryan:2016vau}, we obtain
\begin{equation}\label{haaBSM}
   \big| C_{ah}^{\rm eff} \big| < 1.34\,\bigg[ \frac{\Lambda}{1\,\mbox{TeV}} \bigg]^2 \,.
\end{equation}
More generally, if both coefficients are non-zero, the allowed values for $C_{Zh}^\text{eff}$ and $C_{ah}^\text{eff}$ are constrained to lie within the orange region in Figure~\ref{fig:CZhvsCah}. At the end of LHC operation, with a projected integrated luminosity of 3000\,fb$^{-1}$ at $\sqrt{s}=14$\,TeV, one expects the improved bound $\text{Br}(h \to \text{BSM})<0.1$ \cite{ATL-PHYS-PUB-2014-016}, which would imply that the two coefficients must be inside the dashed black contour in the figure. The constraint on $C_{ah}^{\rm eff}$ alone would then be $|C_{ah}^{\rm eff}|<0.62\,(\Lambda/\mbox{TeV})^2$. Invisible ALP decays would lead to invisible Higgs-boson decays, for which the bounds $\mbox{Br}(h\to\mbox{invisible})<0.23$ from ATLAS \cite{Aad:2015pla} and $\mbox{Br}(h\to\mbox{invisible})<0.24$ from CMS \cite{Khachatryan:2016whc} imply the constraint $|C_{ah}^{\rm eff}|<1.02\,(\Lambda/\mbox{TeV})^2$ for $\text{Br}(a\to\text{invisible})=1$.

Depending on the pattern of ALP decay modes, promising signals arise from multi-photon and multi-lepton final states, but also from ALP decays into jets or $b$ quarks. Very light ALPs can only decay into photons and are boosted along the beam direction with a boost factor $\gamma_a=m_h/(2 m_a)\gg 1$, for which the photons are highly collimated. For masses $m_a<625$\,MeV, the opening angle between the final state photons $\Delta\phi=\arccos(1 - 2/\gamma_a^2)\approx 2/\gamma_a$ is smaller than the angular resolution of the ATLAS and CMS electromagnetic calorimeters (ECALs) of $\sim 20$\,mrad, and hence the photons enter the same calorimeter cell \cite{Dobrescu:2000jt,Chang:2006bw,Chala:2015cev}. However, shower-shape analyses allow one to differentiate between single and multiple photons even if the opening angle is below the angular resolution. To be conservative, and based on the analysis in \cite{ATLAS:2012soa}, we therefore assume that ALP masses below 100\,MeV cannot be distinguished from $h\to\gamma\gamma$ decays. In this case, we can turn the limit on the signal strength parameter $\mu_{\rm exp}^{h\to\gamma\gamma}=1.14\,^{+\,0.19}_{-\,0.18}$ \cite{Khachatryan:2016vau} into a constraint on the $h\to aa\to\gamma\gamma+\gamma\gamma$ branching ratio,
\begin{equation}
   \mu^{h\to\gamma\gamma}
   = \frac{\sigma(pp\to h\to \gamma\gamma)}{\sigma(pp\to h\to\gamma\gamma)_\text{SM}}
   = 1 + \frac{\text{Br}(h\to aa\to \gamma\gamma+\gamma\gamma)\big\vert_\text{eff}}{\text{Br}(h\to \gamma\gamma)_\text{SM}}\,,
\end{equation}
where the effective Higgs branching ratio $\text{Br}(h\to aa\to \gamma\gamma+\gamma\gamma)\big\vert_\text{eff}$ is defined as in (\ref{eq:effBR}) and takes into account the lifetime of the ALPs. This constraint is shown by the contours in the low-mass region of the right panel of Figure~\ref{fig:haa}, where the meaning of the curves is the same as in Figure~\ref{fig:BRs}. The solid and dotted curves correspond to $\text{Br}(a\to\gamma\gamma)=1$ and $\text{Br}(a\to\gamma\gamma)=0.1$, respectively, while the blue and red curves refer to $|C_{\gamma\gamma}^{\rm eff}|/\Lambda=1\,\mbox{TeV}^{-1}$ and $|C_{\gamma\gamma}^{\rm eff}|/\Lambda=0.1\,\mbox{TeV}^{-1}$. ATLAS further provides limits on $\text{Br}(h\to aa\to\gamma\gamma+\gamma\gamma)$ for the three mass values $m_a=100$\,MeV, $m_a=200$\,MeV and $m_a=400$\,MeV, based on the $\sqrt{s}=7$\,TeV dataset \cite{ATLAS:2012soa}. The corresponding limits are indicated by the three blue or red points in the figure. For ALP masses in the range $m_a=(10-62.5)$\,GeV, ATLAS has performed a dedicated search for $h\to aa\to 4\gamma$ \cite{Aad:2015bua}. We show the corresponding bounds in the right panel of Figure~\ref{fig:haa}. In this case the red contours overlap with the blue ones, since the value of $C_{\gamma\gamma}^{\rm eff}$ becomes irrelevant as long as the $a\to\gamma\gamma$ branching ratio takes a fixed value. It is apparent that the limits for very light ALP masses are independent of the choice of $\text{Br}(a\to\gamma\gamma)$, while the limits for heavy ALPs are unchanged for smaller Wilson coefficients $C_{\gamma\gamma}^{\rm eff}$, as expected from the discussion of Figure~\ref{fig:BRs}.

Various searches for $h\to aa$ decays with subsequent ALP decays into heavy fermion pairs have been performed. This includes $h\to aa\to\tau^+\tau^-\tau^+\tau^- $, $h\to aa\to\tau^+\tau^-\mu^+\mu^-$ \cite{Aad:2015oqa,Khachatryan:2015nba,Khachatryan:2017mnf}, $h\to aa\to b\bar b\mu^+\mu^-$, and $h\to aa\to b\bar b b\bar b$ \cite{Khachatryan:2017mnf,Aaboud:2016oyb}. Constraints from the latter are not yet sensitive to SM-like Higgs production cross sections. The other constraints are shown in the right panel of Figure~\ref{fig:otherbounds}. The solid contours assume $\text{Br}(a\to\ell^+\ell^-)=1$ for decays probing a single leptonic decay mode, and $\text{Br}(a\to\mu^+\mu^-)=\text{Br}(a\to\tau^+\tau^-)=\text{Br}(a\to b\bar b)=0.5$ if two different fermion species are considered. The dotted contours correspond to $\text{Br}(a\to\ell^+\ell^-)=0.1$ for decays probing a single leptonic decay mode, and $\text{Br}(a\to\mu^+\mu^-)=0.1$, $\text{Br}(a\to\tau^+\tau^-)=\text{Br}(a\to b\bar b)=0.9$ otherwise. 

\subsection{Probing the parameter space of ALPs}

Given the rich phenomenology of ALP decays, there is a plethora of promising searches at the LHC for both $h\to Za$ and $h\to aa$ decays. If the $a\to\gamma\gamma$ branching ratio is sufficiently large, these exotic Higgs decays with subsequent ALP decays into photons would give rise to very clean signatures, which can be used to discover or constrain the ALP--photon coupling in a vast region of so far unexplored parameter space \cite{Bauer:2017nlg}. Equally interesting are ALP decays into lepton pairs, which would also lead to clean final states. We now discuss the prospects for searches in these two channels and present projections for the reach of Run-2 of the LHC. ALP decays into hadronic final states, such as di-jets or heavy $Q\bar Q$ pairs, are experimentally more challenging and would require dedicated analyses. We emphasize that our focus in this work is on visibly decaying ALPs, which can be reconstructed in the detector. Searches for invisibly decaying ALPs can be performed using the missing-energy signature in mono-$X$ final states such as $pp\to Z^*\to ha\to h+\rlap{\,/}E_T$ or $pp\to Z^*\to Za\to Z+\rlap{\,/}E_T$ \cite{Brivio:2017ije}. 

\subsubsection{Constraining the ALP--photon coupling}

Present and future searches for $h\to\gamma\gamma+\gamma\gamma$ and $h\to\ell^+\ell^- +\gamma\gamma$ decays at the LHC can probe a large range of ALP--photon couplings. In our estimates below we focus on Run-2 of the LHC, which will provide an integrated luminosity of 300\,fb$^{-1}$ at $\sqrt{s}=13$\,TeV. We require 100 signal events in each search channel and require that the ALPs decay before the electromagnetic calorimeter, which is typically located at a distance of approximately 1.5\,m from the beam axis. We assume that the Higgs bosons are produced in gluon fusion with a cross section of $\sigma_{13\,\text{TeV}}(gg\to h)=48.52$\,pb \cite{Anastasiou:2016cez}. Projections for higher luminosity (3\,ab$^{-1}$ at $\sqrt{s}=14$\,TeV) and for a 100\,TeV proton--proton collider will be presented elsewhere \cite{inprep}.

In our analysis we consider very different experimental searches. Light ALPs can effectively enhance the $h\to\gamma\gamma$ branching ratio, heavier ALPs produce clearly separated di-photon resonances in $h\to aa\to\gamma\gamma+\gamma\gamma$ decays, and ALPs with very small couplings can lead to displaced vertices. Experimental strategies to isolate the signal and suppress the background differ significantly for these searches. We are not in a position to provide detailed estimates of detector and reconstruction efficiencies, or to perform solid background estimates. Nevertheless, we believe that our requirement of 100 signal events in the respective search channels is realistic. For comparison, we note that the current precision of the $h\to\gamma\gamma$ rate measurements excludes more than 340 new-physics events in this channel \cite{Khachatryan:2016vau}, the upper limit on $h\to Z\gamma$ decay allows for 400 new-physics events \cite{Chatrchyan:2013vaa}, and the search for $h\to aa$ for heavy ALPs \cite{Aad:2015bua} is sensitive to 120--390 events depending on the ALP mass (all at 95\% CL).\footnote{In $Z\to\gamma a\to 3\gamma$ decay discussed in Section~\ref{sec:Zga}, the experimental analysis can reject 273 new-physics events at 95\% CL.} 
Note that in the present work we do not make use of displaced-vertex signatures, which will help to greatly reduce the background in the region of parameter space where only a small fraction of the ALPs decays inside the detector. We hope that our analysis will trigger sufficient interest in the experimental community that dedicated analysis strategies will be developed by the experimental collaborations themselves.

\begin{figure}[t]
\begin{center}
\includegraphics[width=0.48\textwidth]{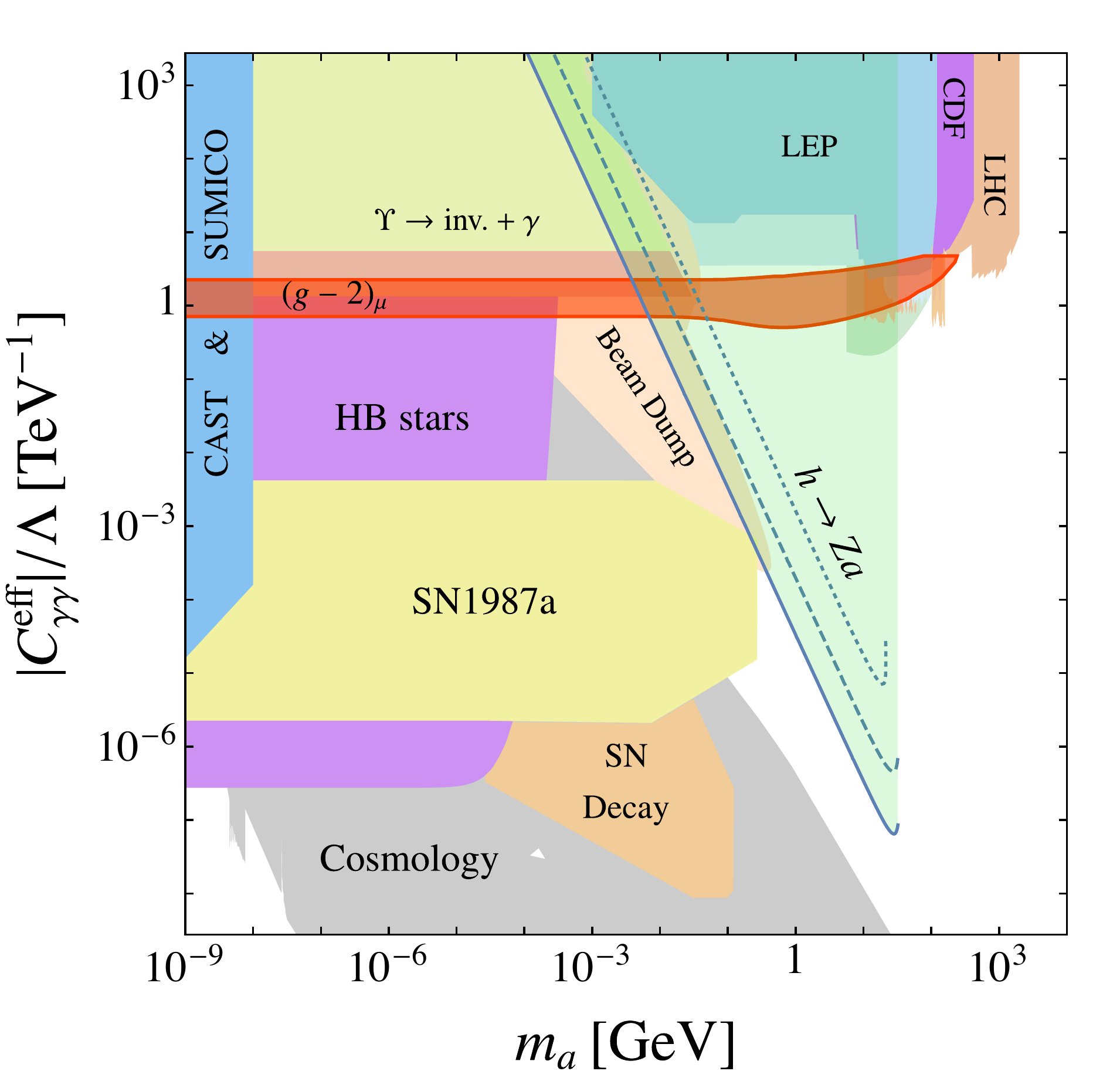}\quad 
\includegraphics[width=0.48\textwidth]{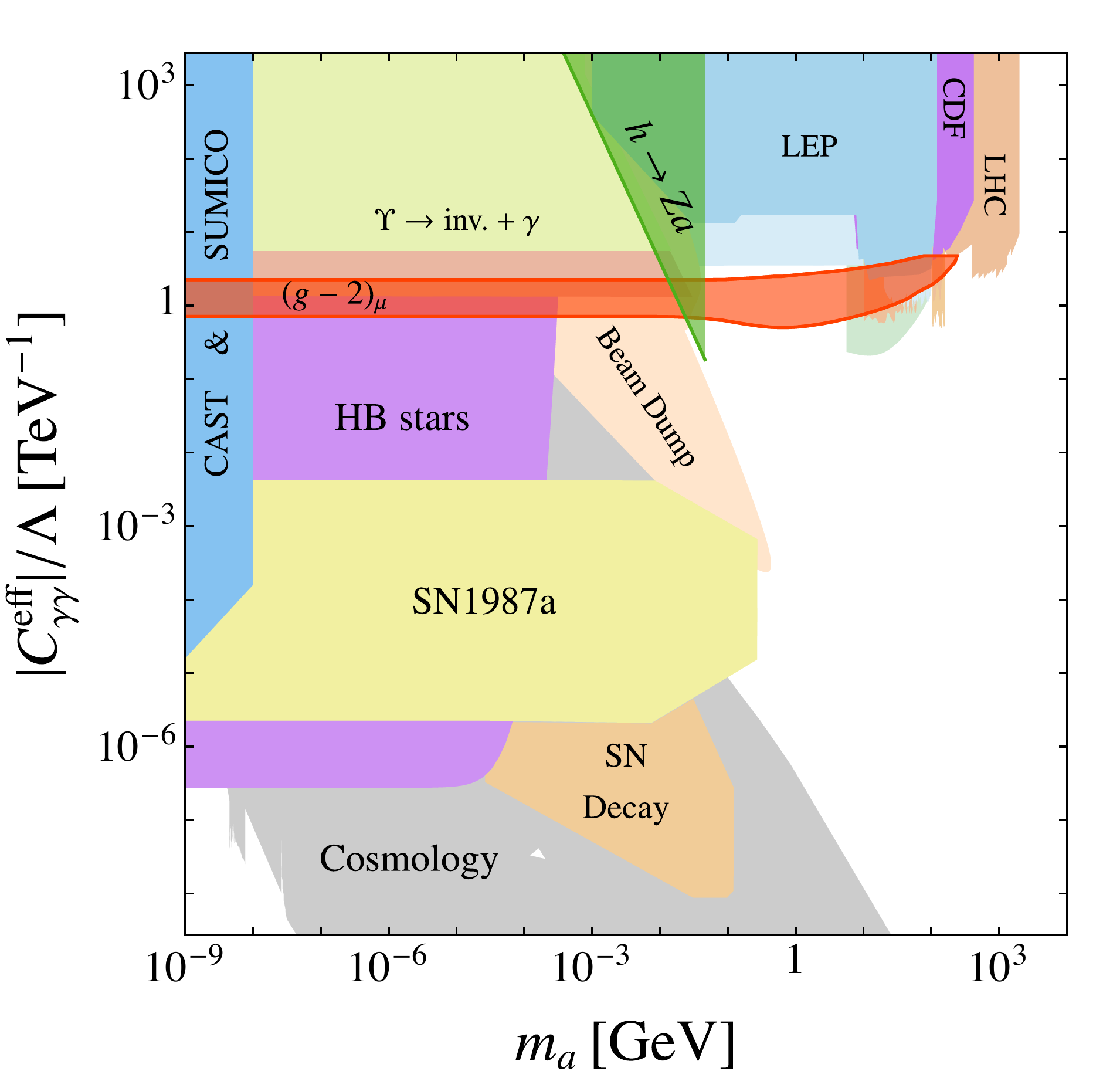}
\end{center}
\vspace{-3mm}
\caption{\label{fig:CggexclusionZA} Constraints on the ALP mass and coupling to photons derived from various experiments (colored areas without boundaries, adapted from \cite{Jaeckel:2015jla}) along with the parameter regions that can be probed using the Higgs decays $h\to Za\to\ell^+\ell^-\gamma\gamma$. The left panel shows the reach of LHC Run-2 with 300\,fb$^{-1}$ of integrated luminosity (shaded in light green). We require at least 100 signal events. The contours correspond to $|C_{Zh}^{\rm eff}|/\Lambda=0.72\,\mbox{TeV}^{-1}$ (solid), $0.1\,\mbox{TeV}^{-1}$ (dashed) and $0.015\,\mbox{TeV}^{-1}$ (dotted). The red band shows the preferred parameter space where the $(g-2)_\mu$ anomaly can be explained at 95\% CL. The right panel shows the regions excluded by existing searches for $h\to Z\gamma$ (shaded in dark green), where we assume $|C_{Zh}^{\rm eff}|/\Lambda=0.72\,\mbox{TeV}^{-1}$.}
\end{figure}

We begin by presenting the projected reach of searches for the decay $h\to Za\to\ell^+\ell^-+\gamma\gamma$, for which the effective branching ratio has been defined in the first line of (\ref{eq:effBR}). In this case we require that 
\begin{equation}
   N_{\rm signal} = \mathcal{L}_\text{LHC} \times \sigma_{13\,\text{TeV}}(gg\to h)
    \times \text{Br}(h\to Za\to\ell^+\ell^- +\gamma\gamma) \Big|_{\text{eff}} > 100 \,.
\end{equation}
The green shaded regions in the left panel of Figure~\ref{fig:CggexclusionZA} show the parameter space which can be probed in Run-2 for different values of the relevant Wilson ALP--Higgs coupling. The three lines limiting these regions correspond to $|C_{Zh}^{\rm eff}|/\Lambda=0.72\,\mbox{TeV}^{-1}$ (solid contour), $0.1\,\mbox{TeV}^{-1}$ (dashed contour) and $0.015\,\mbox{TeV}^{-1}$ (dotted contour), taking into account the model-independent upper bound from $h\to\text{BSM}$ derived in \eqref{eq:cZhbound}. Note that the dotted line roughly corresponds to a TeV-scale coupling suppressed by a loop factor. With 300\,fb$^{-1}$ of luminosity it is possible to extend the search to slightly smaller couplings, but reaching sensitivity to couplings smaller than $|C_{Zh}^{\rm eff}|/\Lambda<0.01\,\mbox{TeV}^{-1}$ would require a larger luminosity. To draw the contours in the figure we have assumed that $\text{Br}(a\to\gamma\gamma)=1$; however, it is important to realize that their shape is essentially independent of the value of the $a\to\gamma\gamma$ branching ratio as long as this quantity is larger than a certain critical value, which is set by the required number of signal events (and as long as the ALP mass is not too close to the kinematic limit). These limiting values are $\text{Br}(a\to\gamma\gamma)>3\cdot 10^{-4}$ (solid), 0.011 (dashed) and 0.46 (solid). Importantly, it is thus possible to probe the ALP--photon coupling even if the ALP predominantly decays into other final states. The triangular shape of the region of the projected reach is a consequence of the fact that ALPs with either small masses or small couplings, which fall beyond the left boundary of the region of sensitivity, live long enough (on average) to leave the detector. As discussed in Section~\ref{sec:Higgs}, the line in the $m_a-|C_{\gamma\gamma}^{\rm eff}|$ plane where this happens only depends on the partial width $\Gamma(a\to\gamma\gamma)\propto m_a^3\,|C_{\gamma\gamma}^{\rm eff}|^2/\Lambda^2$, but not on $\text{Br}(a\to \gamma\gamma)$. This argument only breaks down near the kinematic boundary $m_a=m_h-m_Z$, where the $h\to Za$ decay rate becomes sensitive to the ALP mass. This behavior can also be understood from Figure~\ref{fig:BRs}. Note that the region in parameter space that can be probed using exotic Higgs decays into ALPs almost perfectly complements the regions covered by existing searches. This will also be true for the other search channels discussed below. Whereas existing searches probe signatures of long-lived ALPs, in our case the ALPs are so short lived that their decays can be reconstructed in the detector. The red band in Figure~\ref{fig:CggexclusionZA} shows the parameter space in which the anomalous magnetic moment of the muon can be explained in terms of loop corrections involving a virtual ALP exchange, assuming $|C_{\gamma\gamma}^\text{eff}|/\Lambda\le|c_{\mu\mu}|/\Lambda\le 5\,\text{TeV}^{-1}$. The upper bound on $|c_{\mu\mu}|$ ensures that there is a substantial $a\to\gamma\gamma$ branching ratio everywhere inside the red band. Notice that almost this entire parameter space can be covered by searches for exotic Higgs decays, provided that the Higgs--ALP coupling $C_{Zh}$ is sufficiently large. In the right panel of Figure~\ref{fig:CggexclusionZA} we present the parameter space already excluded by present analyses placing upper bounds on the $h\to Z\gamma$ branching ratio \cite{Chatrchyan:2013vaa,Aad:2014fia}. These bounds apply in the low-mass region, where the two photons produced in the decay of the ALP are seen as a single photon jet in the calorimeter. The excluded parameter space shaded in dark green is obtained assuming $|C_{Zh}^{\rm eff}|/\Lambda=0.72\,\mbox{TeV}^{-1}$ and $\text{Br}(a\to\gamma\gamma)>0.04$.

\begin{figure}[t]
\begin{center}
\includegraphics[width=0.48\textwidth]{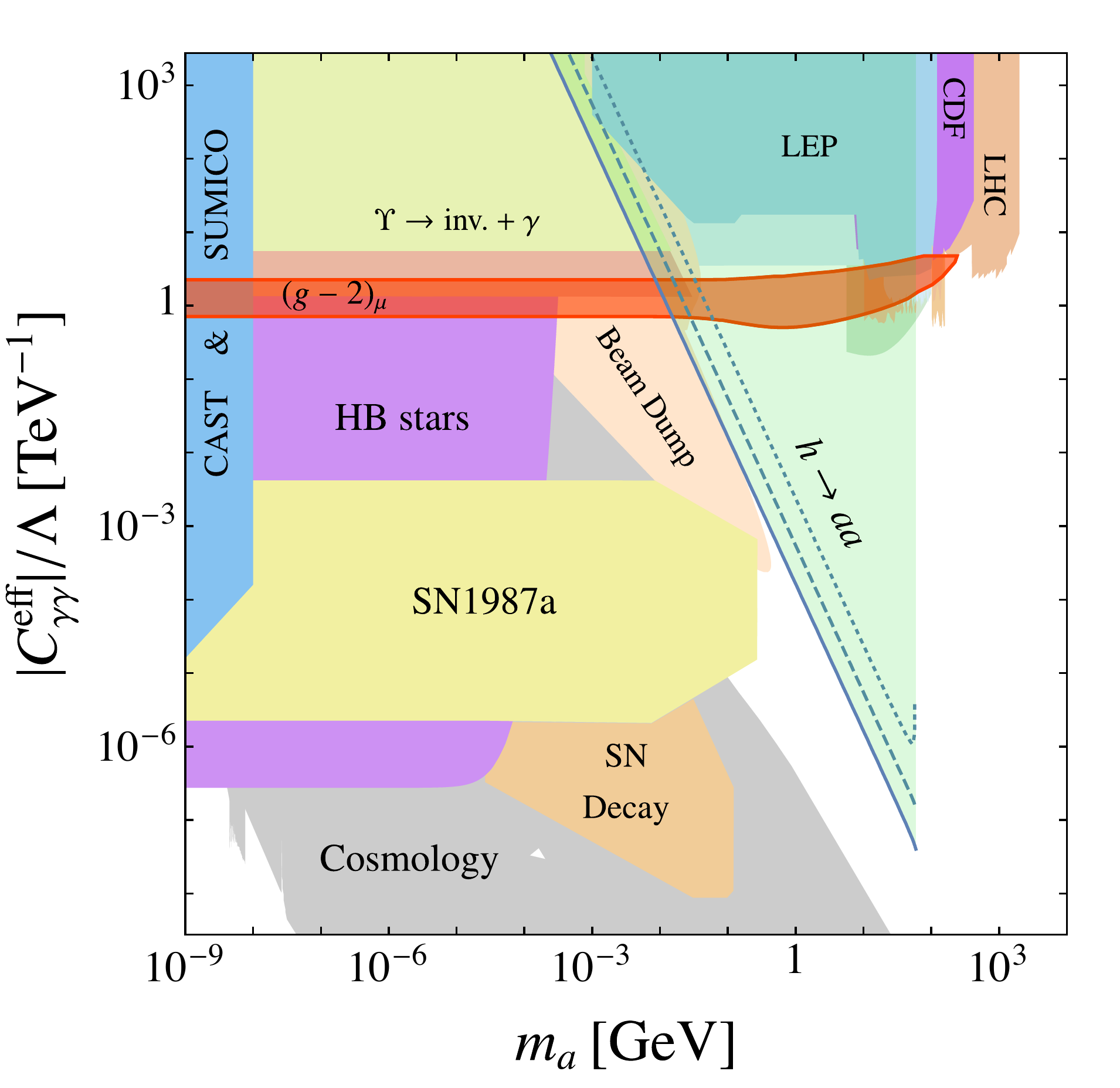}\quad 
\includegraphics[width=0.48\textwidth]{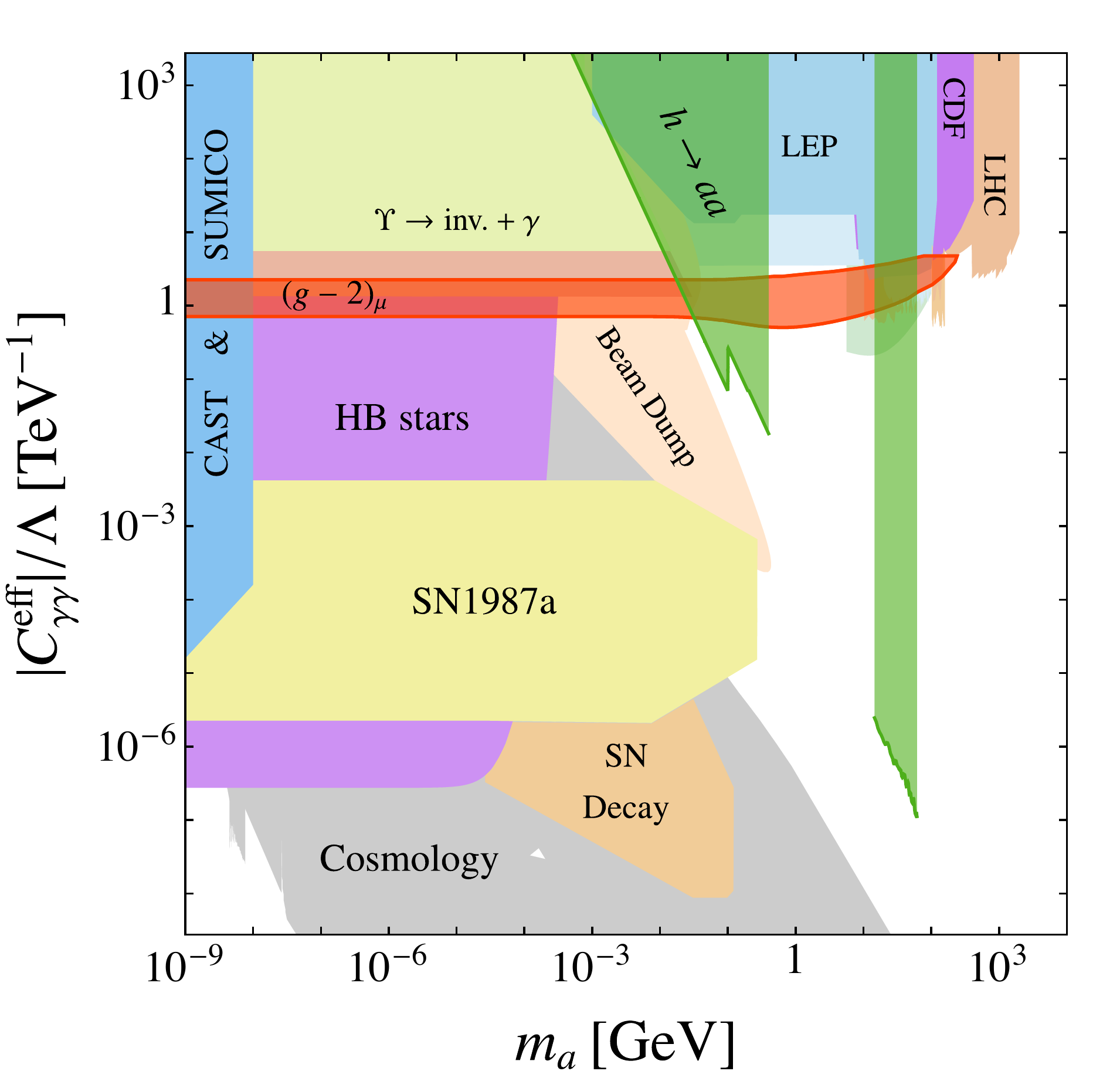}
\end{center}
\vspace{-3mm}
\caption{\label{fig:CggexclusionAA} Constraints on the ALP mass and coupling to photons derived from various experiments (colored areas without boundaries, adapted from \cite{Jaeckel:2015jla}) along with the parameter regions that can be probed using the Higgs decays $h\to aa\to 4\gamma$. The left panel shows the reach of LHC Run-2 with 300\,fb$^{-1}$ of integrated luminosity (shaded in light green). We require at least 100 signal events. The contours correspond to $|C_{ah}^{\rm eff}|/\Lambda^2=1\,\mbox{TeV}^{-2}$ (solid), $0.1\,\mbox{TeV}^{-2}$ (dashed) and $0.01\,\mbox{TeV}^{-2}$ (dotted). The red band shows the preferred parameter space where the $(g-2)_\mu$ anomaly can be explained at 95\% CL. The right panel shows the regions excluded by existing searches for $h\to\gamma\gamma$ and $h\to 4\gamma$ (shaded in dark green), where we assume $|C_{ah}^{\rm eff}|/\Lambda^2=1\,\mbox{TeV}^{-2}$.}
\end{figure}

In Figure~\ref{fig:CggexclusionAA} we present the projected reach of searches for the decay $h\to aa\to\gamma\gamma+\gamma\gamma$, for which the effective branching ratio has been defined in the second line of (\ref{eq:effBR}). As previously, we require that 
\begin{equation}
   N_{\rm signal} = \mathcal{L}_\text{LHC} \times \sigma_{13\,\text{TeV}}(gg\to h)
    \times \text{Br}(h\to aa\to\gamma\gamma+\gamma\gamma) \Big|_{\text{eff}} > 100 \,.
\end{equation}
The lines limiting the green shaded regions in the left panel correspond to $|C_{ah}^\text{eff}|/\Lambda^2=1\,\text{TeV}^{-2}$ (solid), $0.1\,\text{TeV}^{-2}$ (dashed) and $0.01\,\text{TeV}^{-2}$ (dotted), where the last value corresponds to a TeV-scale coefficient times a loop factor. We have used $\mbox{Br}(a\to\gamma\gamma)=1$ in the plot, but once again the contours are essentially independent of the $a\to\gamma\gamma$ branching ratio except for ALP masses close to the kinematic limit $m_a=m_h/2$. The corresponding limiting $a\to\gamma\gamma$ branching ratios are $\text{Br}(a\to\gamma\gamma)>0.006$, 0.049 and 0.49, respectively. With 300\,fb$^{-1}$ of luminosity it is possible to extend the search to slightly smaller couplings, but reaching sensitivity to couplings smaller than $|C_{ah}^{\rm eff}|/\Lambda^2<0.005\,\mbox{TeV}^{-2}$ would require larger luminosity. In the right panel of Figure~\ref{fig:CggexclusionAA} we show the exclusion regions derived from the experimental searches presented in the right panel of Figure~\ref{fig:haa}, now projected into the $m_a-|C_{\gamma\gamma}^\text{eff}|$ plane. We assume $|C_{ah}^{\rm eff}|/\Lambda^2=1\,\text{TeV}^{-2}$. These bounds are valid for branching ratios $\text{Br}(a\to\gamma\gamma)>0.07$, 0.57, and 0.04 for the cases of the low-mass region below 100\,MeV, the mass range between 100 and 400\,MeV, and the high-mass region, respectively. They are obtained from the absence of a significant enhancement of the $h\to\gamma\gamma$ rate \cite{Khachatryan:2016vau}, the search for $h\to\gamma\gamma+\gamma\gamma$ for intermediate masses \cite{ATLAS:2012soa}, and the corresponding search in the high-mass region \cite{Aad:2015bua}. The fact that the exclusion region obtained in the low-mass region with a luminosity of 25\,fb$^{-1}$ per experiment is not much weaker than our projection for 300\,fb$^{-1}$ shown by the solid line in the left panel indicates that our requirement of 100 signal events is not unreasonable.

\begin{figure}[t]
\begin{center}
\includegraphics[width=0.82\textwidth]{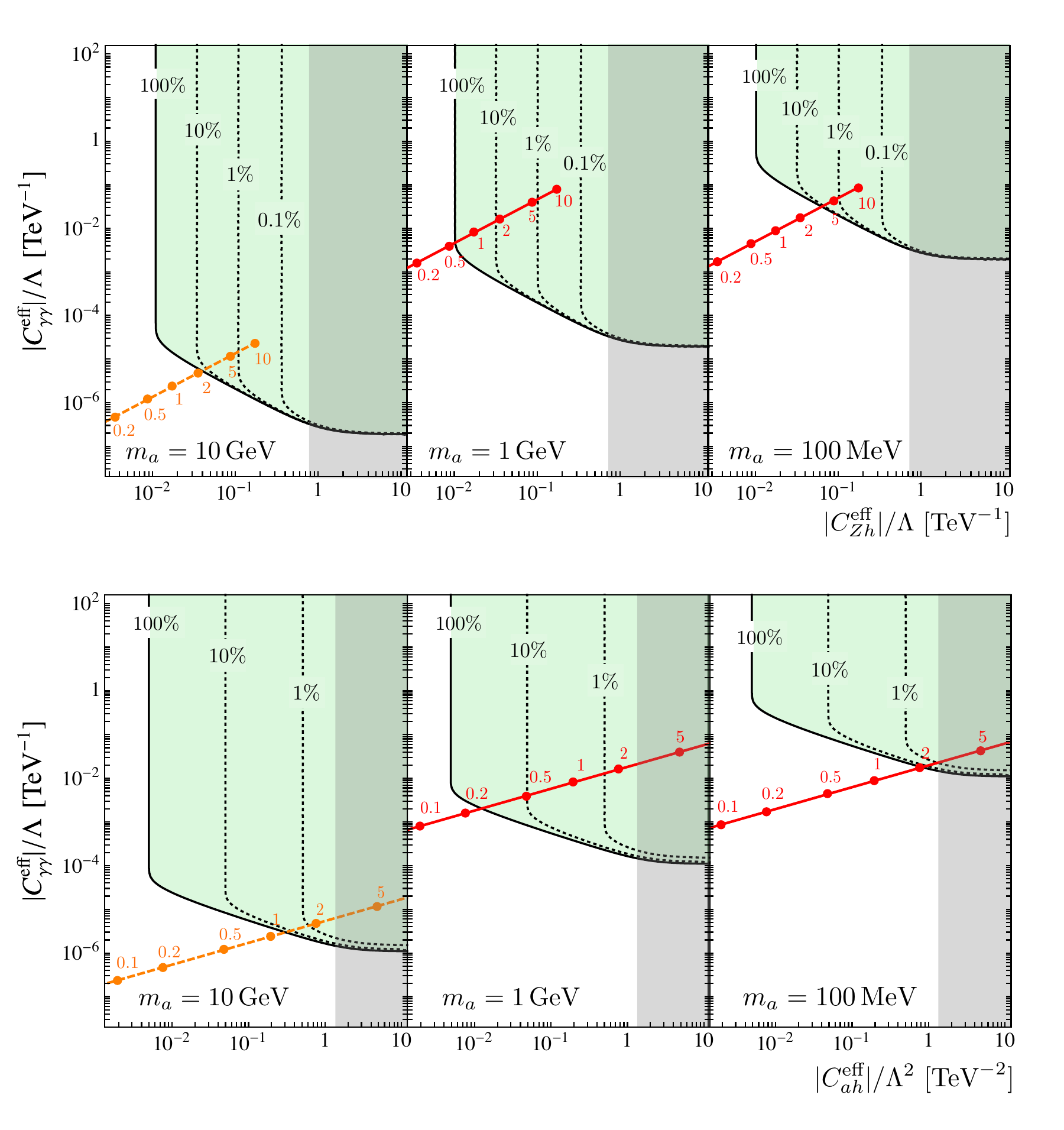}
\end{center}
\vspace{-6mm}
\caption{\label{fig:CZhCgg} 
Parameter space in the plane of the ALP--photon and ALP--Higgs couplings (green regions to the right of the black contours) for which at least 100 events are produced in the $h\to Za\to\ell^+\ell^-\gamma\gamma$ (top) and $h\to aa\to 4\gamma$ (bottom) search channels at the LHC Run-2 with 300\,fb$^{-1}$ and for $m_a=10$\,GeV, 1\,GeV and 100\,MeV. The contours correspond to $\text{Br}(a\to\gamma\gamma)=1$ (solid) and 0.1, 0.01, 0.001 (dotted), as indicated. The gray areas indicate the regions excluded by the bounds (\ref{eq:cZhbound}) and (\ref{haaBSM}). The colored lines show the values of the Wilson coefficients in two specific scenarios, in which the ALP--boson couplings are induced by loops of SM quarks (see text for more details).}
\end{figure}

While the graphical displays in Figures~\ref{fig:CggexclusionZA} and \ref{fig:CggexclusionAA} correctly represent the regions in the $m_a-|C_{\gamma\gamma}^{\rm eff}|$ parameter space which can be probed using exotic Higgs decays, it is important to emphasize that finding a signal in these search regions will require sufficiently large ALP--Higgs couplings, as indicated by the solid, dashed and dotted contour lines in the plots. Consequently, {\em not\/} finding a signal in any of these searches would not necessarily exclude the existence of an ALP in this parameter space. An alternative way to present our results, which makes this fact more explicit, is shown in Figure~\ref{fig:CZhCgg} for $h\to Za$ (upper panel) and $h\to aa$ (lower panel). For three different values of the ALP mass, the green-shaded areas to the right of the solid or dashed contours in the various plots now show the regions in the parameter space of the relevant ALP--Higgs and ALP--photon couplings which can be probed (again requiring at least 100 signal events) for different values of the $a\to\gamma\gamma$ branching ratio. This representation is more faithful in the sense that a negative search result would definitely exclude the corresponding region of parameter space. 

The colored lines overlaid in the plots indicate two interesting yet rather pessimistic scenarios, in which the ALP couplings to bosons are induced via loops of SM quarks only. Of course, larger couplings can be expected if new particles contribute in the loops, or if for some reason the couplings arise at tree level. The red line corresponds to a model in which $C_{\gamma\gamma}^{\rm eff}$, $C_{Zh}^{\rm eff}$ and $C_{ah}^{\rm eff}$ are generated from one-loop diagrams involving the three SM up-type quarks, which are assumed to have equal couplings $c_{uu}=c_{cc}=c_{tt}$. The orange dashed line corresponds to a model in which only the top-quark coupling $c_{tt}$ is non-zero. This provides a concrete example of a scenario in which the loop-induced ALP--Higgs couplings can be sizable, while the induced ALP--photon coupling tends to be very small. In each case, the relevant coupling $|c_{tt}|/\Lambda$ is varied between 0.1\,TeV$^{-1}$ and 10\,TeV$^{-1}$, as indicated by the labels along the line. The $a\to\gamma\gamma$ branching ratios obtained in these scenarios are $7\cdot 10^{-4}$ for $m_a=10$\,GeV, 27\% for $m_a=1$\,GeV, and 100\% for $m_a=100$\,MeV. In the high-mass case ($m_a=10$\,GeV), the di-jet final state $a\to\mbox{2 jets}$ would provide for a more promising search channel.

\subsubsection{Constraining the ALP--lepton couplings}

The analysis of the previous section can be extended to any other decay mode of the ALP. As a second example we consider the decays $a\to\ell^+\ell^-$, which are kinematically accessible if $m_a>2m_\ell$. We stress that analogous analyses to the ones presented here could (and should) be performed for all other possible ALP decay modes. 

\begin{figure}[t]
\begin{center}
\includegraphics[width=0.48\textwidth]{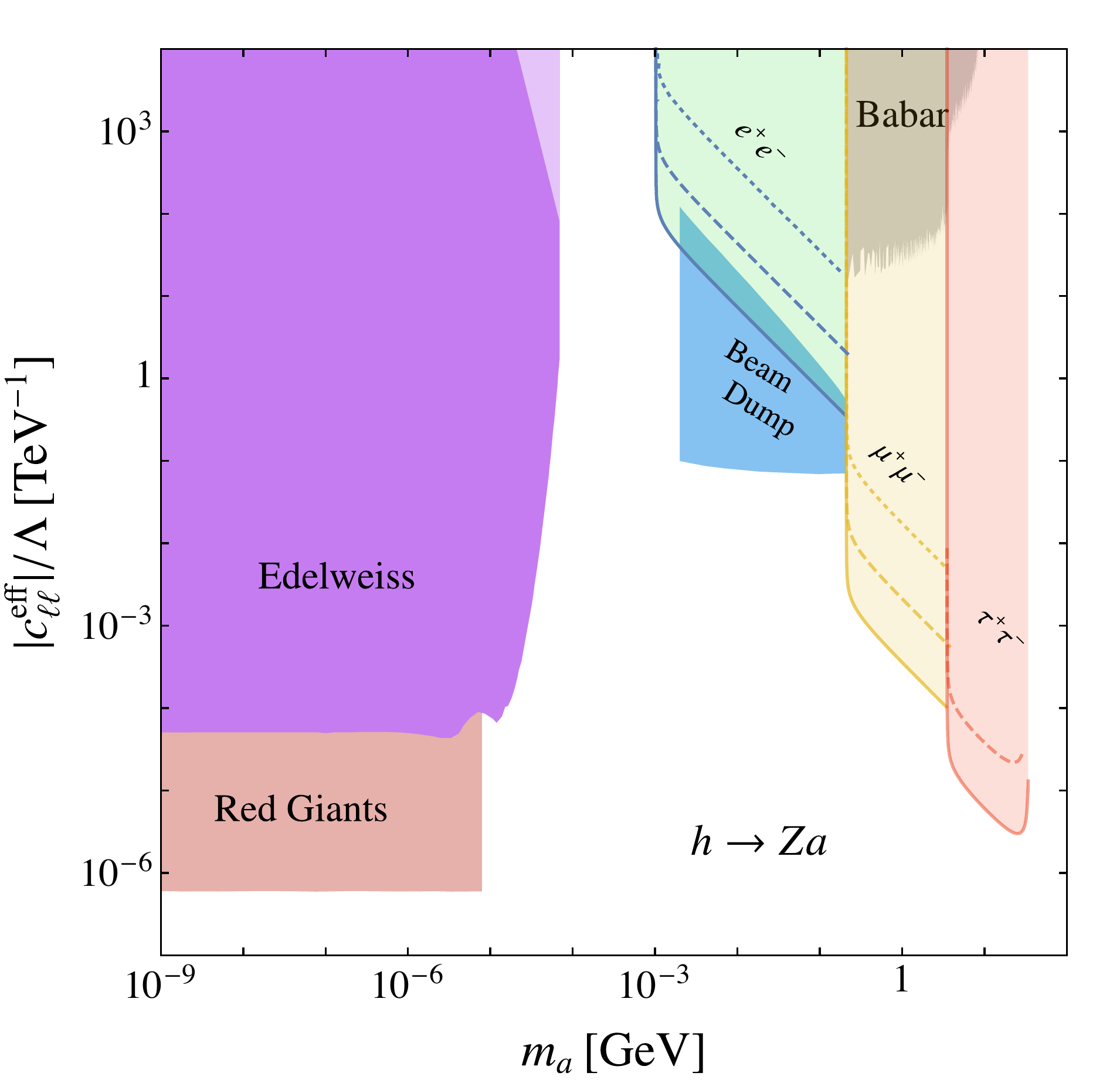}\quad
\includegraphics[width=0.48\textwidth]{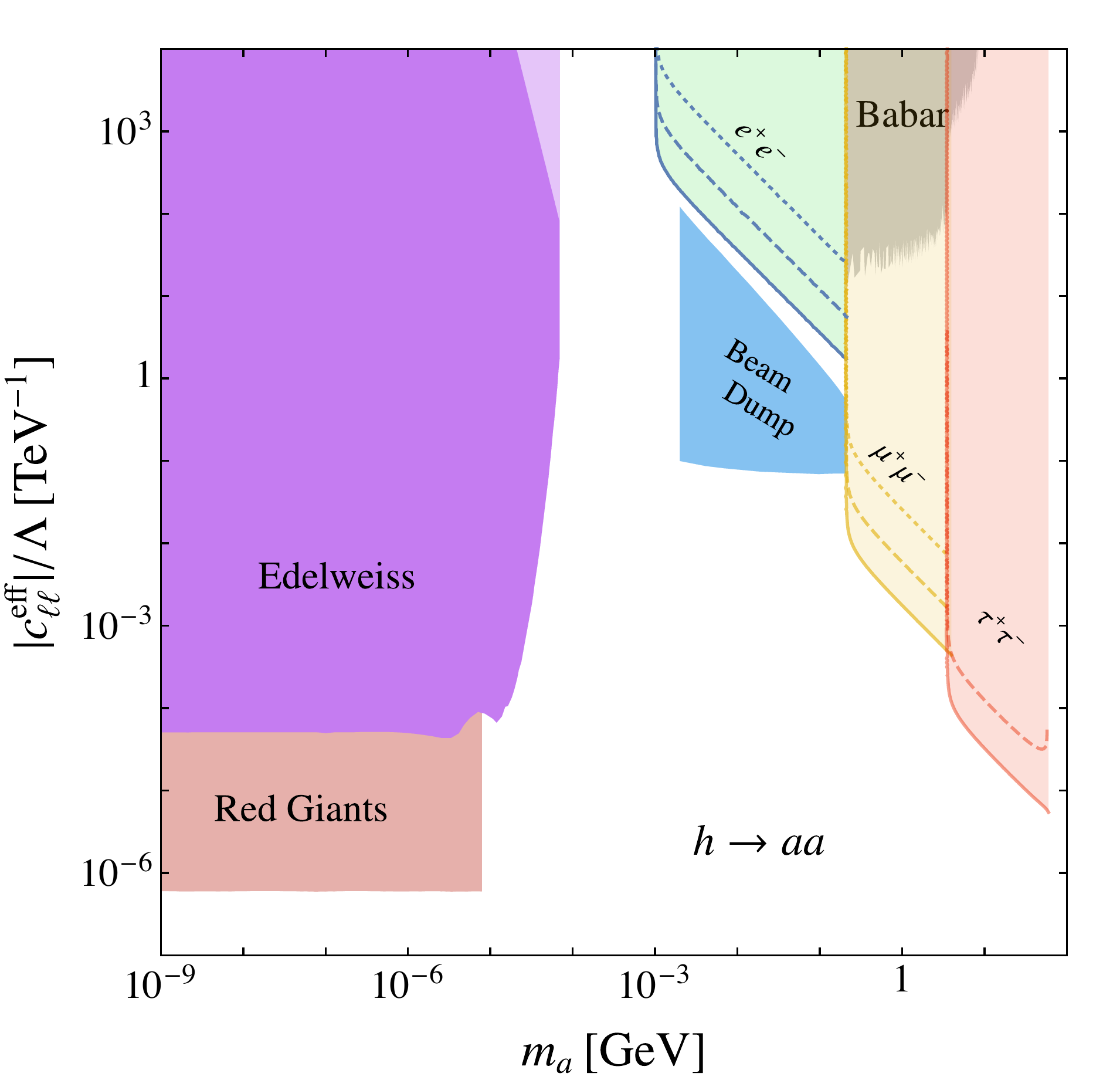}
\end{center}
\vspace{-3mm}
\caption{\label{fig:CEEexclusion} Constraints on the ALP mass and coupling to leptons derived from various experiments (colored areas without boundaries, adapted from \cite{Armengaud:2013rta,Essig:2010gu}) along with the parameter regions that can be probed using the Higgs decays $h\to Za\to\ell_1^+\ell_1^- e^+e^-$ (left) and $h\to aa\to e^+e^- e^+e^-$ (right). The areas shaded in light green show the reach of LHC Run-2 with 300\,fb$^{-1}$ of integrated luminosity. We require at least 100 signal events. The contours in the left panel correspond to $|C_{Zh}^{\rm eff}|/\Lambda=0.72\,\mbox{TeV}^{-1}$ (solid), $0.1\,\mbox{TeV}^{-1}$ (dashed) and $0.015\,\mbox{TeV}^{-1}$ (dotted), while those in the right panel refer to $|C_{ah}^{\rm eff}|/\Lambda^2=1\,\mbox{TeV}^{-2}$ (solid), $0.1\,\mbox{TeV}^{-2}$ (dashed) and $0.01\,\mbox{TeV}^{-2}$ (dotted). The orange and red regions overlaid in the plots show the corresponding parameter space that can be covered in searches for the decay modes $a\to\mu^+\mu^-$ and $a\to\tau^+\tau^-$ (see text for more explanations).}
\end{figure}

The $a\to e^+ e^-$ decay mode is of particular interest, since in the sub-MeV region the ALP--electron coupling has been constrained using a variety of experimental searches, as discussed in Section~\ref{subsec:aeecoupl}. Using exotic Higgs decays, it will be possible to probe the ALP--electron coupling in the largely unexplored region above 1\,MeV. The decay chains $h\to Za\to\ell_1^+\ell_1^- + e^+e^-$and $h\to aa\to e^+e^- + e^+e^-$ provide clean search channels in this parameter space. The corresponding projections are shown by the green shaded regions in Figure~\ref{fig:CEEexclusion}, where we require that (with $\ell_1=e,\mu$)
\begin{equation}
\begin{aligned}
   N_{\rm signal} &= \mathcal{L}_\text{LHC} \times \sigma_{13\,\text{TeV}}(gg\to h)
    \times \text{Br}(h\to Za\to\ell_1^+\ell_1^- + e^+e^-) \Big|_{\text{eff}}>100 \,, \\
   N_{\rm signal} &= \mathcal{L}_\text{LHC} \times \sigma_{13\,\text{TeV}}(gg\to h)
    \times \text{Br}(h\to aa\to e^+e^- + e^+e^-) \Big|_{\text{eff}}>100 \,,
\end{aligned}
\end{equation}
respectively. In contrast to ALP decay into photons, we now set $L_\text{det}=2$\,cm, since the ALP decay into electrons should take place before the inner tracker. The region of sensitivity is limited by contours obtained for different values of the relevant ALP--Higgs couplings. As before, these values are $|C_{Zh}^{\rm eff}|/\Lambda=0.72\,\mbox{TeV}^{-1}$ (solid), $0.1\,\mbox{TeV}^{-1}$ (dashed) and $0.015\,\mbox{TeV}^{-1}$ (dotted) for $h\to Za\to\ell_1^+\ell_1^- + e^+e^-$, and $|C_{ah}^\text{eff}|/\Lambda^2=1\,\text{TeV}^{-2}$ (solid), $0.1\,\text{TeV}^{-2}$ (dashed) and $0.01\,\text{TeV}^{-2}$ (dotted) for $h\to aa\to e^+e^- + e^+e^-$. We have used $\mbox{Br}(a\to e^+e^-)=1$ for the green-shaded region in the plot, but as previously the contours are essentially independent of the $a\to e^+e^-$ branching ratio unless this quantity falls below certain threshold values, which are the same as before. For $h\to Za$, one needs $\mbox{Br}(a\to e^+e^-)>3\cdot 10^{-4}$ (solid), 0.011 (dashed) and 0.46 (dotted). For $h\to aa$, one needs instead $\mbox{Br}(a\to e^+e^-)>0.006$ (solid), 0.049 (dashed) and 0.49 (dotted). Similar to the case of ALP decays into photons, searches for rare Higgs decays have the potential to probe so far unconstrained parameter space. 

The orange and red regions overlaid in the plots show, for comparison, the corresponding parameter space that can be covered in searches for the decay modes $a\to\mu^+\mu^-$ and $a\to\tau^+\tau^-$. For the latter mode, we have adopted the $\tau$ reconstruction efficiencies from the $h\to aa\to\tau^+\tau^- + \tau^+\tau^-$ search performed by CMS in \cite{CMS:2015iga}. For each ALP, they require one tau lepton to decay into a muon and the second one to decay hadronically (with 60\% reconstruction efficiency), leading to a rate reduction by a factor 0.13 for each ALP. The exclusion contours have been computed assuming $\mbox{Br}(a\to\ell^+\ell^-)=1$ for both cases, but as previously the contours are essentially independent of the branching ratio unless this quantity falls below certain threshold values. For $a\to\mu^+\mu^-$ these are the same as for the electron case. For $a\to\tau^+\tau^-$ the limiting branching fractions are larger, due to the lower reconstruction efficiency. For $h\to Za$, one needs $\mbox{Br}(a\to\tau^+\tau^-)>2\cdot 10^{-3}$ (solid) and 0.008 (dashed). For $h\to aa$, one needs instead $\mbox{Br}(a\to\tau^+\tau^-)>0.041$ (solid) and 0.36 (dashed). We observe that the ALP--muon and ALP--tau couplings which can be probed are significantly smaller than the ALP--electron couplings. This simply reflects that the relevant decay rates scale with the square of the charged-lepton mass. 

So far we have discussed searches in the $a\to e^+ e^-$ channel independently of other leptonic ALP decay modes. We emphasize, however, that in many new-physics models one would expect a strong correlation between these modes. Indeed, if the leptonic couplings $c_{\ell\ell}$ are approximately flavor universal, as shown in (\ref{LFuniv}), then the orange and red areas labeled $\mu^+\mu^-$ and $\tau^+\tau^-$ in Figure~\ref{fig:CEEexclusion} can actually be interpreted as parameter regions in which one can probe the ALP--electron coupling. Indeed, if the ALP is heavy enough to decay into muons or taus, the branching ratios for decays into lighter leptons will be tiny, and it will only be possible to reconstruct the decay in the heaviest lepton that is kinematically allowed. Note that the combination of the three different search regions nicely complements the region covered by beam-dump searches.

\begin{figure}[t]
\begin{center}
\includegraphics[width=0.85\textwidth]{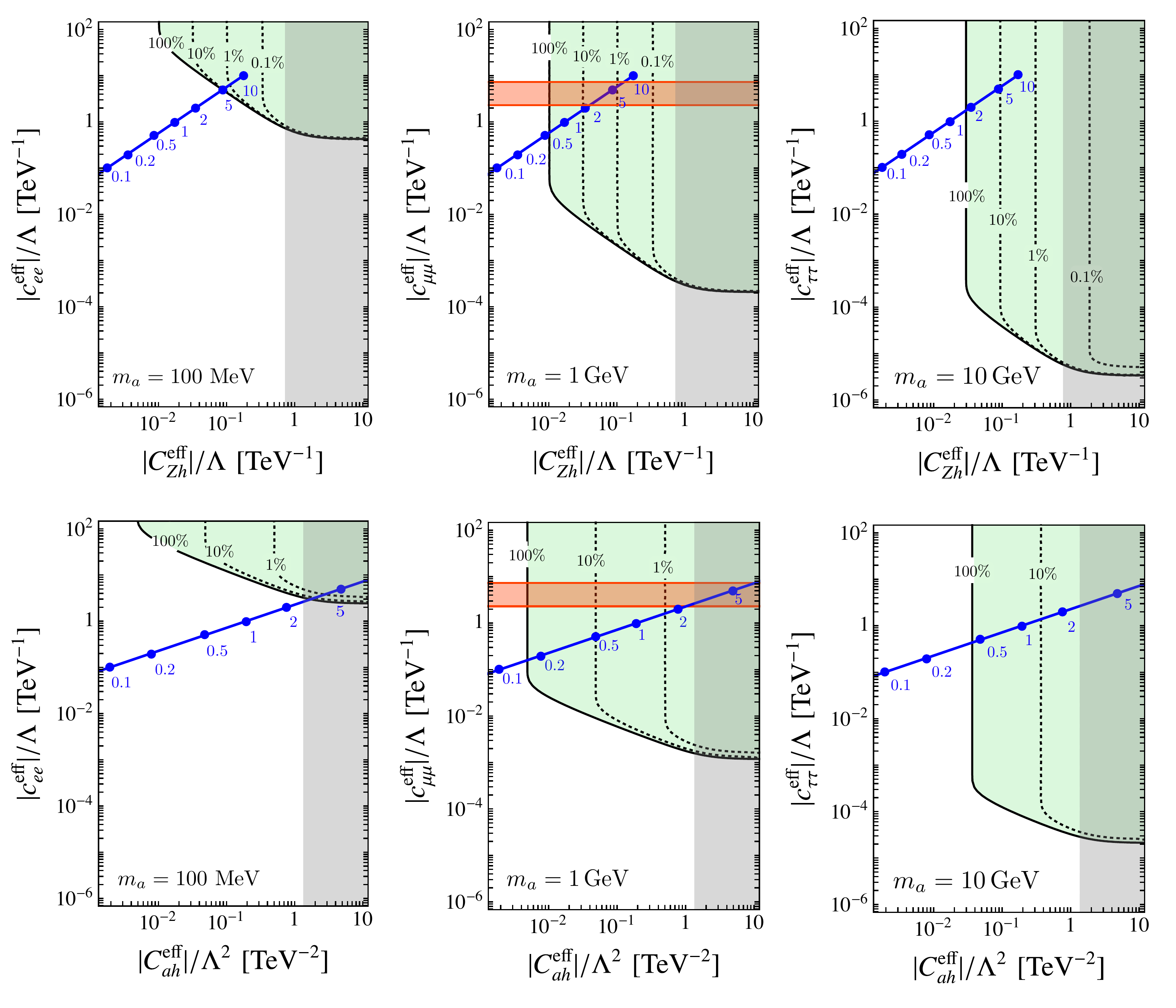} 
\end{center}
\vspace{-6mm}
\caption{\label{fig:CahCZhCee} 
Parameter space in the plane of the ALP--lepton and ALP--Higgs couplings (green regions to the right of the black contours) for which at least 100 events are produced in the $h\to Za\to\ell_1^+\ell_1^-\ell^+\ell^-$ (top) and $h\to aa\to\ell^+\ell^-\ell^+\ell^-$ (bottom) search channels at the LHC Run-2 with 300\,fb$^{-1}$ and for $m_a=10$\,GeV, 1\,GeV and 100\,MeV. The contours correspond to $\text{Br}(a\to\ell^+\ell^-)=1$ (solid) and 0.1, 0.01, 0.001 (dotted), as indicated. The gray area indicates the region excluded by the bounds (\ref{eq:cZhbound}) and (\ref{haaBSM}). The blue line shows the values of the Wilson coefficients in a scenario, in which the ALP couplings to bosons are induced by loops of SM fermions with equal couplings to the ALP (see text for more details). The red band in the center plots shows the parameter space in which $(g-2)_\mu$ can be explained, assuming $|C_{\gamma\gamma}|/\Lambda=1\,\mbox{TeV}^{-1}$.}
\end{figure}

Once again, it is instructive to consider an alternative way of representing the information contained in Figure~\ref{fig:CEEexclusion}. For three different values of the ALP mass, the green-shaded areas to the right of the solid or dashed contours in Figure~\ref{fig:CahCZhCee} show the regions in the parameter space of the relevant ALP--Higgs and ALP--lepton couplings which can be probed in the exotic Higgs decays $h\to Za\to\ell_1^+\ell_1^-+\ell^+\ell^-$ (upper panel) and $h\to aa\to\ell^+\ell^-+\ell^+\ell^-$ (lower panel), again requiring at least 100 signal events, for different values of the $a\to\ell^+\ell^-$ branching ratios. In each case, the decay into the heaviest accessible lepton is shown. The blue line shows the Wilson coefficients in a specific model, in which the ALP couplings to the Higgs boson are generated via loops of SM fermions, assuming that all fermions have equal couplings $c_{ff}$. The relevant leptonic branching ratios is this model are $\mbox{Br}(a\to e^+ e^-)\approx 98\%$ for $m_a=100$\,MeV, $\mbox{Br}(a\to\mu^+\mu^-)\approx 100\%$ for $m_a=1$\,GeV, and $\mbox{Br}(a\to\tau^+\tau^-)\approx 7.5\%$ for $m_a=10$\,GeV.

\section{\boldmath Constraints from $Z$-pole measurements}
\label{sec:Zpole}

\begin{figure}
\begin{center}
\includegraphics[width=0.63\textwidth]{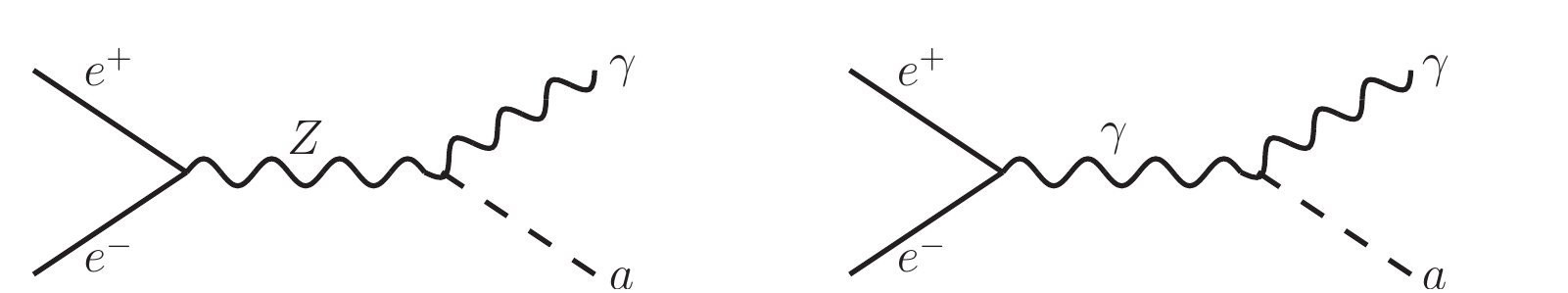}
\end{center}
\vspace{-2mm}
\caption{\label{fig:eegaa} Feynman diagrams contributing to the process $e^+ e^-\to\gamma a$.}
\end{figure}

The ALP couplings to electroweak gauge bosons can also be probed through precision measurements of the properties of $Z$ bosons. As a concrete example, consider the production of a photon in association with an ALP in $e^+ e^-$ collisions. The relevant Born-level diagrams are shown in Figure~\ref{fig:eegaa}. Neglecting the electron mass, we find the cross section 
\begin{equation}
   \frac{d\sigma(e^+ e^-\to\gamma a)}{d\Omega}
   = 2\pi\alpha\,\alpha^2(s)\,\frac{s^2}{\Lambda^2} \left( 1 - \frac{m_a^2}{s} \right)^3 (1+\cos^2\theta)
    \left[ \big|V(s)\big|^2 + \big|A(s)\big|^2 \right] , 
\end{equation}
where $\sqrt{s}$ is the center-of-mass energy and $\theta$ denotes the scattering angle of the photon relative to the beam axis. ALP emission from the initial-state leptons vanishes in the limit $m_e=0$ and is otherwise strongly suppressed. The vector and axial-vector form factors are given by
\begin{equation}\label{VsAs}
   V(s) = \frac{1-4s_w^2}{4s_w^2 c_w^2}\,\frac{C_{\gamma Z}}{s-m_Z^2+i m_Z\Gamma_Z}
    + \frac{C_{\gamma\gamma}}{s} \,, \qquad
   A(s) = \frac{1}{4s_w^2 c_w^2}\,\frac{C_{\gamma Z}}{s-m_Z^2+i m_Z\Gamma_Z} \,,
\end{equation}
where $\Gamma_Z$ is the total width of the $Z$ boson. If one makes the {\it ad hoc\/} assumption that the ALP only couples to photons, while $C_{\gamma Z}=0$, then  measurements of this cross section at LEP can be used to constrain the coupling $C_{\gamma\gamma}$ \cite{Jaeckel:2015jla}. However, in view of the general relations (\ref{Cgagadef}) this assumptions seems very artificial. Let us instead analyze the general structure of the cross section in more detail. At low energy ($s\ll m_Z^2$) the photon contribution dominates and produces a cross section (after integration over  angles)
\begin{equation}
   \sigma(e^+ e^-\to\gamma a) \Big|_{s\ll m_Z^2}
   \approx \frac{32\pi^2\alpha}{3}\,\alpha^2(s) \left( 1 - \frac{m_a^2}{s} \right)^3       
    \frac{|C_{\gamma\gamma}|^2}{\Lambda^2} \,. 
\end{equation}
At high energy ($s\gg m_Z^2$) one finds to good approximation
\begin{equation}\label{sigmahighE}
   \sigma(e^+ e^-\to\gamma a) \Big|_{s\gg m_Z^2}
   \approx \frac{32\pi^2\alpha}{3}\,\alpha^2(s) \left( 1 - \frac{m_a^2}{s} \right)^3       
    \left[ \frac{|C_{\gamma\gamma}|^2}{\Lambda^2} 
    + \frac{|C_{\gamma Z}|^2}{16 s_w^4 c_w^4\,\Lambda^2} \right] , 
\end{equation}
where we have used that $(1-4s_w^2)\approx 0$ in the first term in the expression for $V(s)$ in (\ref{VsAs}). By combining measurements of the cross sections at high and low energies it is thus possible to constraint the two coefficients $C_{\gamma\gamma}$ and $C_{\gamma Z}$ in a model-independent way. A much enhanced sensitivity to the $a\gamma Z$ coupling is obtained on the $Z$ pole, where the cross section is given by
\begin{equation}
   \sigma(e^+ e^-\to\gamma a) \Big|_{s=m_Z^2}
   \approx \frac{32\pi^2\alpha}{3}\,\alpha^2(s) \left( 1 - \frac{m_a^2}{s} \right)^3       
    \left[ \frac{|C_{\gamma\gamma}|^2}{\Lambda^2} 
    + \frac{m_Z^2}{\Gamma_Z^2}\,\frac{|C_{\gamma Z}|^2}{16 s_w^4 c_w^4\,\Lambda^2} \right] .
\end{equation}
Note that the contribution from the $Z$-boson receives an enhancement factor $(m_Z/\Gamma_Z)^2\approx 1336$ relative to (\ref{sigmahighE}). The photon contribution is a background in this case, which can be subtracted by performing a scan about the peak position. In this way one obtains access to $C_{\gamma Z}$ directly. This example nicely illustrates the main idea of our approach. By using on-shell decays of narrow, heavy SM particles into ALPs rather than the production of ALPs via an off-shell particle we obtain a much better sensitivity to the ALP couplings. For the case of on-shell Higgs decays studied in \cite{Bauer:2017nlg} and in Section~\ref{sec:Higgs} of the present work, the relevant enhancement factor is $(m_h/\Gamma_h)^2\approx 9.4\cdot 10^8$ (assuming a SM Higgs width).

It has been pointed out in \cite{Brivio:2017ije} that the Drell--Yan process $pp\to(\gamma/Z)^*\to\gamma a$ at the LHC already provides better constraints on the ALP couplings than the corresponding process $e^+ e^-\to(\gamma/Z)^*\to\gamma a$ at LEP, which we have discussed above. An analogous statement applies for the on-shell decay, which we discuss in Sections~\ref{sec:Zga} and \ref{sec:6.2}. $Z$-pole measurements are also interesting in view of electroweak precision observables placing constraints on the Wilson coefficients $C_{\gamma\gamma}$ and $C_{\gamma Z}$ (or alternatively $C_{WW}$ and $C_{BB}$). These constraints are derived in Section~\ref{sec:EWPT}. Ultra-high precision studies of rare $Z$-boson decays could be performed at a future $e^+ e^-$ collider operating on the $Z$ pole, which could provide samples of almost $10^{12}$ $Z$ bosons per year \cite{Gomez-Ceballos:2013zzn}. Projections for ALP searches at such a facility will be presented elsewhere \cite{inprep}.

\subsection[ALP searches in $Z\to\gamma a$ decay]{\boldmath ALP searches in $Z\to\gamma a$ decay}
\label{sec:Zga}

The second operator in (\ref{gammaZ}) induces the exotic $Z$-boson decay $Z\to\gamma a$ at tree level. Including also the one-loop contributions from fermion loops, we obtain the decay rate
\begin{equation}
   \Gamma(Z\to\gamma a) 
   = \frac{8\pi\alpha\,\alpha(m_Z)\,m_Z^3}{3s_w^2 c_w^2\Lambda^2}\,\big| C_{\gamma Z}^{\rm eff} \big|^2
    \left( 1 - \frac{m_a^2}{m_Z^2} \right)^3 ,
\end{equation}
where the effective Wilson coefficient $C_{\gamma Z}^{\rm eff}$ is given by
\begin{equation}
   C_{\gamma Z}^{\rm eff}
   = C_{\gamma Z} + \sum_f \frac{N_c^f Q_f v_f}{16\pi^2}\,c_{ff}\,B_3(\tau_f,\tau_{f/Z}) \,.
\end{equation}
Here $v_f=\frac12\,T_3^f-s_w^2 Q_f$ is the $Z$-boson vector coupling to fermion $f$, and we have defined the mass ratios $\tau_{f}=4m_f^2/m_a^2$ and $\tau_{f/Z}=4m_f^2/m_Z^2$. The relevant loop function reads
\begin{equation}
   B_3(\tau_1,\tau_2) = 1 - \frac{f^2(\tau_1)-f^2(\tau_2)}{\frac{1}{\tau_1} - \frac{1}{\tau_2}} \,.
\end{equation}
It obeys $B_3(\tau_f,\tau_{f/Z})\approx 1$ for all light fermions other than the top quark, for which $B_3(\tau_t,\tau_{t/Z})\approx B_1(\tau_{t/Z})\approx-0.024$ is very small. As in the case of the $a\to\gamma\gamma$ decay discussed in Section~\ref{sec:agaga}, the main effect of electroweak radiative corrections would be to renormalize the gauge couplings. In the present case the coupling $\alpha$ associated with the photon is evaluated at $q^2=0$, while the coupling $\alpha(m_Z)/(s_w^2 c_w^2)$ associated with the $Z$ boson should be evaluated at $q^2=m_Z^2$ as indicated. The $Z\to\gamma a$ branching fraction is obtained by dividing this partial decay rate by the $Z$-boson total width $\Gamma_Z$. This yields
\begin{equation}\label{ZgaaBR}
   \mbox{Br}(Z\to\gamma a) = 8.17\cdot 10^{-4}\,\big| C_{\gamma Z}^{\rm eff} \big|^2
    \left( 1 - \frac{m_a^2}{m_Z^2} \right)^3 
    \left[ \frac{1\,\mbox{TeV}}{\Lambda} \right]^2 .
\end{equation}
By requiring the $Z$-boson total width to agree with the direct measurement $\Gamma_Z=(2.495\pm 0.0023)$\,GeV performed at LEP \cite{ALEPH:2005ab}, an upper bound on the Wilson coefficient $|C_{\gamma Z}^{\rm eff}|$ can be extracted. At 95\% CL we find $\mbox{Br}(Z\to\mbox{BSM})<0.0018$ and
\begin{equation}\label{eq:cZgammabound}
   \big| C_{\gamma Z}^{\rm eff} \big| < 1.48\,\bigg[ \frac{\Lambda}{1\,\mbox{TeV}} \bigg] \,.
\end{equation}
This bound is obtained by neglecting the ALP mass and gets weaker when $m_a$ approaches the kinematic threshold at $m_a=m_Z$. 

\begin{figure}
\begin{center}
\includegraphics[width=0.95\textwidth]{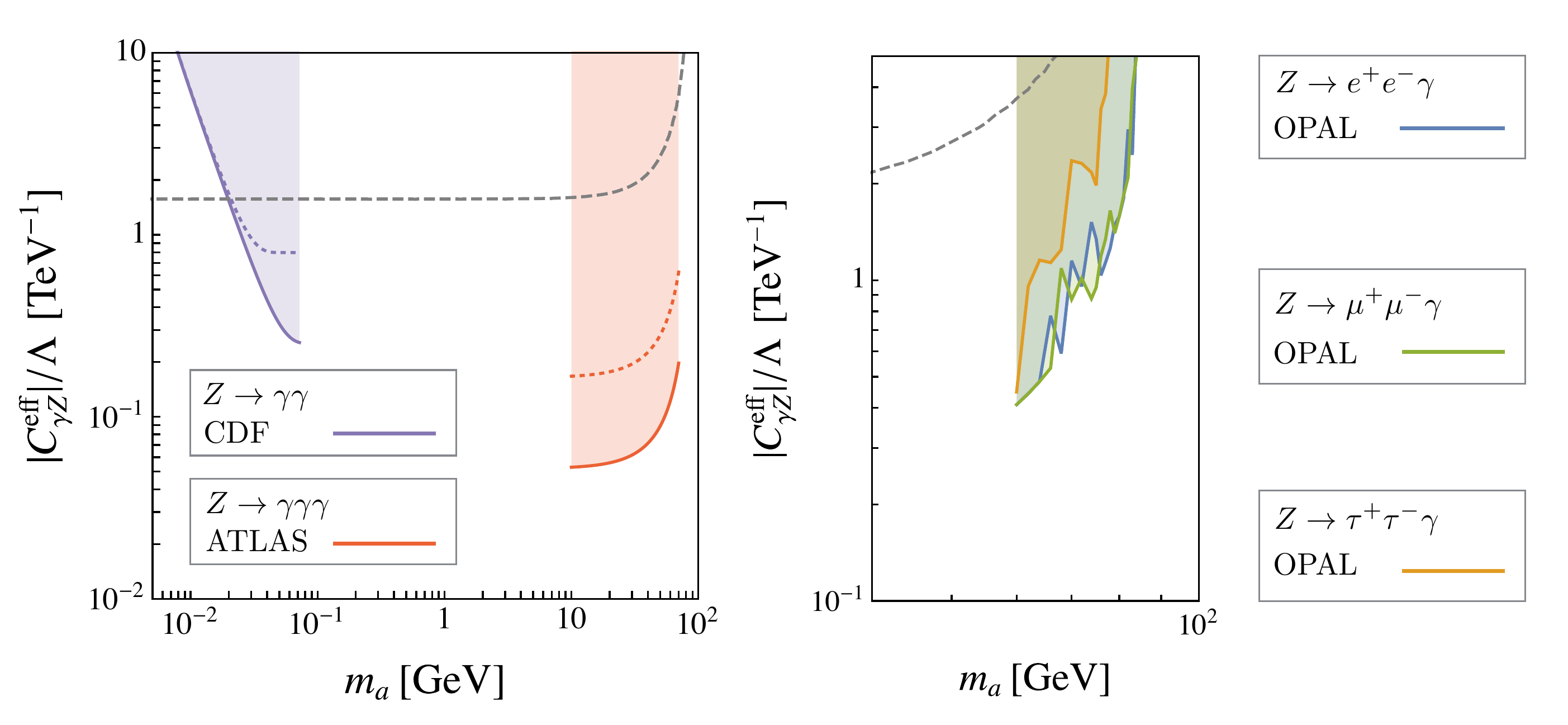}
\end{center}
\vspace{-3mm}
\caption{\label{fig:Zga} Parameter space excluded by measurements of $\mbox{Br}(Z\to\gamma\gamma)$ and $\mbox{Br}(Z\to\gamma\gamma\gamma)$ (left) and measurements of $\mbox{Br}(Z\to\gamma e^+ e^-)$, $\mbox{Br}(Z\to\gamma\mu^+\mu^-)$ and $\mbox{Br}(Z\to\gamma\tau^+\tau^-)$ (right). Regions bounded by solid lines assume $\text{Br}(a\to X\bar X)=1$, those bounded by dashed lines refer to $\text{Br}(a\to X\bar X)=0.1$. The gray dashed line is the bound from \eqref{eq:cZgammabound}.}
\end{figure}

To analyze the reach of this decay mode in probing the ALP--$\gamma Z$, ALP--photon and ALP--electron couplings, we follow a similar strategy as discussed for Higgs decays in Section~\ref{sec:Higgs}. As before, the lifetime of the ALP is taken into account by defining the average decay length of the ALP perpendicular to the beam axis, $L_a^\perp(\theta)$ given in (\ref{eq:Lperp}), where the relevant boost factor in the $Z$-boson rest frame is now $\beta_a\gamma_a=(m_Z^2-m_a^2)/(2m_a m_Z)$. The fraction $f_{\rm dec}^{\gamma a}$ of all $Z\to\gamma a$ events in which $a$ decays before traveling a characteristic distance $L_{\rm det}$ is given by the same expression as in the first line of (\ref{eq:29}). In analogy with (\ref{eq:effBR}), we define the effective branching ratio
\begin{equation}\label{eq:effBR2}
   \mbox{Br}(Z\to \gamma a\to\gamma X\bar X) \big|_{\rm eff} 
   = \mbox{Br}(Z\to \gamma a)\,\mbox{Br}(a\to X\bar X)\,f_{\rm dec}^{\gamma a} \,.
\end{equation}
The ALP branching ratios determine which final states are the most interesting ones. ALP decays into photons lead to the experimental signature $Z\to\gamma a\to\gamma\gamma\gamma$. Bounds on this branching ratio can be derived from precision studies of $Z$-boson decays performed at LEP, the Tevatron and the LHC \cite{Acciarri:1994gb,Abreu:1994du,Aaltonen:2013mfa,Aad:2015bua}. The most stringent constraint is set by a recent ATLAS analysis finding $\mbox{Br}(Z\to\gamma\gamma\gamma)<2.2\cdot 10^{-6}$ at 95\% CL \cite{Aad:2015bua}. Assuming $\text{Br}(a\to\gamma\gamma)=1$ or 0.1, this constraint sets bounds on the Wilson coefficient $|C_{\gamma Z}^{\rm eff}|$, which are depicted by the red solid and dashed lines in the left panel of Figure~\ref{fig:Zga}. The photons have to pass an isolation cut of 4\,GeV in transverse energy. However, to be conservative we take the lower bound at 10\,GeV as in the $h\to\gamma\gamma\gamma\gamma$ search presented in the same paper. The constraint $\mbox{Br}(Z\to\gamma\gamma)<1.46\cdot 10^{-5}$ obtained at 95\% CL by CDF \cite{Aaltonen:2013mfa} becomes relevant below $m_a<73$\,MeV, where the two photons are too collimated to be distinguished in the detector. It implies the exclusion regions shown in violet, which has been derived assuming $|C_{\gamma\gamma}^{\rm eff}|/\Lambda=1\,\mbox{TeV}^{-1}$. ALP decays into lepton pairs give rise to the final states $Z\to\gamma a\to\gamma\ell^+\ell^-$. OPAL sets the most stringent constraints on these processes, namely $\mbox{Br}(Z\to\gamma e^+ e^-)<5.2\cdot 10^{-4}$, $\mbox{Br}(Z\to\gamma\mu^+\mu^-)<5.6\cdot 10^{-4}$ and $\mbox{Br}(Z\to\gamma\tau^+\tau^-)<7.3\cdot 10^{-4}$ at 95\% CL \cite{Acton:1991dq}. The limits on $|C_{\gamma Z}^{\rm eff}|$ derived from these searches are shown in the right panel of Figure~\ref{fig:Zga}, assuming $\mbox{Br}(a\to\ell^+\ell^-)=1$.

\subsection{Probing the ALP--photon and ALP--lepton couplings}
\label{sec:6.2}

\begin{figure}[t]
\begin{center}
\includegraphics[width=\textwidth]{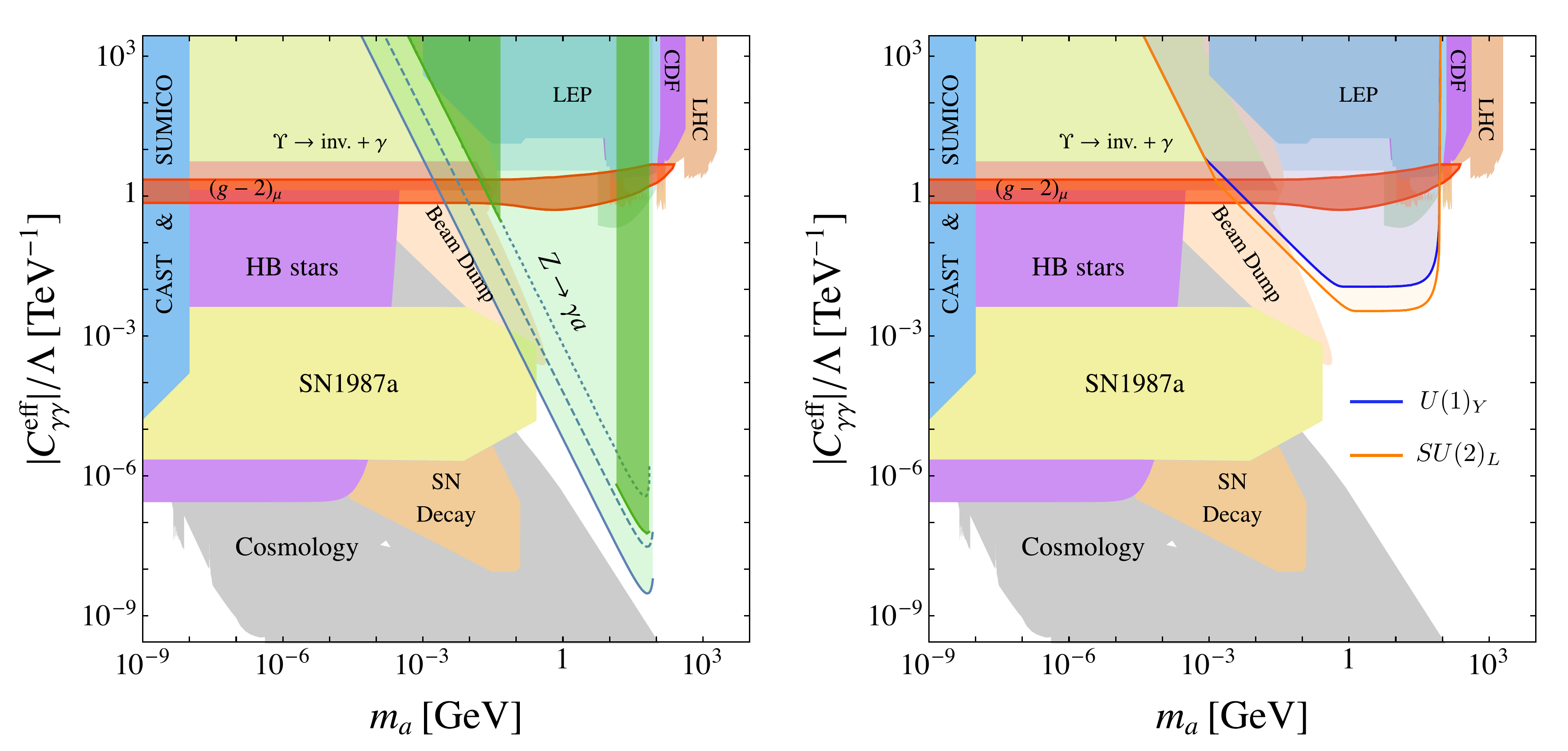}
\end{center}
\vspace{-3mm}
\caption{\label{fig:CggexclusionZagamma} Constraints on the ALP mass and coupling to photons derived from various experiments (colored areas without boundaries, adapted from \cite{Jaeckel:2015jla}) along with the parameter regions that can be probed in LHC Run-2 with 300\,fb$^{-1}$ of integrated luminosity using the decay $Z\to\gamma a\to\gamma\gamma\gamma$. We require at least 100 signal events. Left: Regions that can be probed are shaded in light green. The contours correspond to $|C_{\gamma Z}^{\rm eff}|/\Lambda=1\,\mbox{TeV}^{-1}$ (solid), $0.1\,\mbox{TeV}^{-1}$ (dashed) and $0.01\,\mbox{TeV}^{-1}$ (dotted). The dark green regions are excluded by existing measurements assuming that $|C_{\gamma Z}^{\rm eff}|/\Lambda=1\,\mbox{TeV}^{-1}$. The red band shows the preferred parameter space where the $(g-2)_\mu$ anomaly can be explained at 95\% CL. Right: Regions that can be probed in scenarios where the ALP couples only to hypercharge gauge fields (solid blue) or only to $SU(2)_L$ gauge fields (solid orange). This plot refers to $|C_{\gamma Z}^{\rm eff}|/\Lambda=1\,\mbox{TeV}^{-1}$.}
\end{figure}

Future LHC searches for $Z\to\gamma a\to\gamma\gamma\gamma$ decays can probe a large region in the $m_a-|C_{\gamma\gamma}^{\rm eff}|$ parameter space. The green contours in the left panel in Figure~\ref{fig:CggexclusionZagamma} depict the region where at least 100 signal events are expected at the LHC with $\sqrt{s}=13$\,TeV and 300\,fb$^{-1}$ of integrated luminosity. The $Z$-boson production cross section is $\sigma(pp\to Z)=58.9$\,nb \cite{Aad:2016naf}. The solid, dashed and dotted blue contours correspond to $|C_{\gamma Z}^{\rm eff}|/\Lambda=1\,\mbox{TeV}^{-1}$, $0.1\,\mbox{TeV}^{-1}$ and $0.01\,\mbox{TeV}^{-1}$, respectively. As before, the triangular shape is explained by the fact that ALPs with small masses and couplings are more likely to escape detection. We use $\mbox{Br}(a\to\gamma\gamma)=1$ in the plot, but lowering this branching ratio does not change the contours significantly until a critical value is reached, where less than 100 events are produced for all masses and values of $C_{\gamma Z}^{\rm eff}$. These limiting values are $\text{Br}(a\to\gamma\gamma)>7\cdot 10^{-6}$ (solid), $7\cdot 10^{-4}$ (dashed) and 0.07 (dotted). To reach couplings smaller than $|C_{\gamma Z}^{\rm eff}|/\Lambda=0.0026\,\mbox{TeV}^{-1}$ would require more luminosity. The parameter space shaded in dark green is excluded by present data from CDF \cite{Aaltonen:2013mfa} and ATLAS \cite{Aad:2015bua} (see the left panel of Figure~\ref{fig:Zga}) under the assumption that $|C_{\gamma Z}^{\rm eff}|/\Lambda=1\,\mbox{TeV}^{-1}$ as well as $\text{Br}(a\to\gamma\gamma)>0.065$ (low-mass region) and $\text{Br}(a\to\gamma\gamma)>0.015$ (high-mass region).

\begin{figure}[t]
\begin{center}
\includegraphics[width=0.48\textwidth]{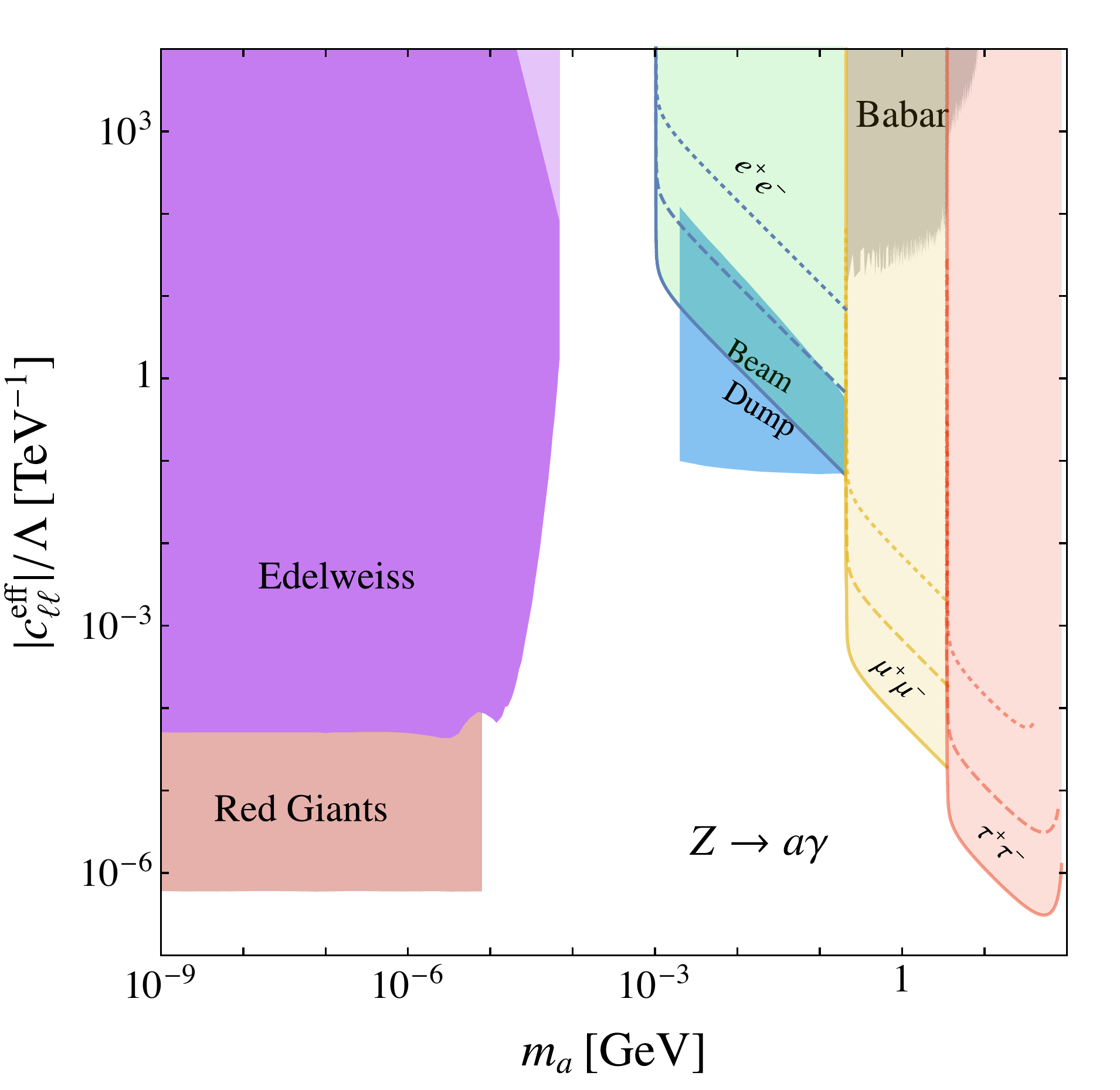}
\end{center}
\vspace{-3mm}
\caption{\label{fig:CEEexclusion2} Constraints on the ALP mass and coupling to leptons derived from various experiments (colored areas without boundaries, adapted from \cite{Armengaud:2013rta,Essig:2010gu}) along with the parameter region that can be probed using the decay $Z\to\gamma a\to\gamma e^+e^-$. The areas shaded in light green show the reach of LHC Run-2 with 300\,fb$^{-1}$ of integrated luminosity. We require at least 100 signal events. The contours correspond to $|C_{\gamma Z}^{\rm eff}|/\Lambda=1\,\mbox{TeV}^{-1}$ (solid), $0.1\,\mbox{TeV}^{-1}$ (dashed) and $0.01\,\mbox{TeV}^{-1}$ (dotted). The orange and red regions overlaid in the plots show the corresponding parameter space that can be covered in searches for the decay modes $a\to\mu^+\mu^-$ and $a\to\tau^+\tau^-$ (see text for more explanations).}
\end{figure}

Comparing the left panel in Figure~\ref{fig:CggexclusionZagamma} with the corresponding plots in Figures~\ref{fig:CggexclusionZA} and \ref{fig:CggexclusionAA} seems to indicate that ALP searches in on-shell $Z\to\gamma a$ decays offer the highest sensitivity to the ALP--photon coupling. This is not necessarily true. The point is that, unlike the case of the Higgs-boson decays considered earlier, in the present case the ALP production process $Z\to\gamma a$ and the ALP decay process $a\to\gamma\gamma$ are governed by Wilson coefficients $C_{\gamma\gamma}$ and $C_{\gamma Z}$, which are correlated via the relations in (\ref{Cgagadef}), since both couplings originate from the gauge-invariant operators with Wilson coefficients $C_{WW}$ and $C_{BB}$ in (\ref{Leff}). It is thus very unlikely that $|C_{\gamma\gamma}^{\rm eff}|$ can take a value that is much smaller than $|C_{\gamma Z}^{\rm eff}|$. In particular, we note that integrating out a single, complete electroweak multiplet will always generate contributions to $C_{WW}$ and $C_{BB}$ with the same sign. If this is the case, then 
\begin{equation}\label{CgaZbound}
   \quad |C_{\gamma Z}|\le c_w^2\,|C_{\gamma\gamma}| \,, \quad \mbox{(single electroweak multiplet)}
\end{equation}
and to very good approximation the same inequality holds for the effective Wilson coefficients including loop corrections. Since $|C_{\gamma Z}^{\rm eff}|/\Lambda>0.0026\,\mbox{TeV}^{-1}$ is required to obtain at least 100 signal events, in the presence of the bound (\ref{CgaZbound}) one cannot probe smaller values of $|C_{\gamma\gamma}^{\rm eff}|$. To illustrate this point, we show in the right panel of Figure~\ref{fig:CggexclusionZagamma} the sensitivity regions obtained for the two cases where the ALP coupling to photons originates only from a coupling to hypercharge (blue line) or only from a coupling to $SU(2)_L$ gauge bosons (orange line). In the first case $C_{\gamma Z}=-s_w^2\,C_{\gamma \gamma}$, while in the second one $C_{\gamma Z}=c_w^2\,C_{\gamma \gamma}$. In both cases we have assumed $\mbox{Br}(a\to\gamma\gamma)=1$, but the contours are essentially independent of this branching ratio as long as $\mbox{Br}(a\to\gamma\gamma)>1.3\cdot 10^{-4}$ for $U(1)_Y$ and $\mbox{Br}(a\to\gamma\gamma)>1.2\cdot 10^{-5}$ for $SU(2)_L$. The sensitivity regions are now significantly reduced, but they still cover the parameter space relevant for an explanation of $(g-2)_\mu$. 

In the leptonic decay channels, future LHC analyses can search for $Z\to\gamma a\to\gamma\ell^+\ell^-$ decays with $\ell=e,\mu,\tau$. Figure~\ref{fig:CEEexclusion2} shows the regions where at least 100 events are expected in the electron (green), muon (orange) and tau (red) channels (red) for $|C_{\gamma Z}^{\rm eff}|/\Lambda=1\,\mbox{TeV}^{-1}$ (solid), $0.1\,\mbox{TeV}^{-1}$ (dashed) and $0.01\,\mbox{TeV}^{-1}$ (dotted). We have used $\mbox{Br}(a\to\ell^+\ell^-)=1$ in each case, but as previously the contours are essentially independent of the $a\to e^+e^-$ branching ratio unless this quantity falls below certain threshold values. For the electron and muon channels the limiting branching ratios are $\text{Br}(a\to\ell^+\ell^-)>7\cdot 10^{-6}$ (solid), $7\cdot 10^{-4}$ (dashed) and 0.07 (dotted). For the tau case, they are instead $\text{Br}(a\to\tau^+\tau^-)>5\cdot 10^{-5}$ (solid), $5\cdot 10^{-3}$ (dashed) and 0.5 (dotted).

\subsection{Electroweak precision tests}
\label{sec:EWPT}

Since we consider ALPs whose mass is significantly lighter than the electroweak scale, loop corrections to electroweak precision observables can in general not simply be described in terms of the usual oblique parameters $S$, $T$ and $U$. Instead, one needs to evaluate the relevant electroweak observables at one-loop order explicitly. Following Peskin and Takeuchi \cite{Peskin:1991sw}, we thus consider the ALP-induced one-loop corrections to three different definitions of the sine squared of the weak mixing angle $s_w^2$, namely $s_*^2$ defined in terms of the neutral-current couplings $\sim (T_3^f-Q_f\,s_*^2)$ of the $Z$ boson to fermions on the $Z$ pole, $s_W^2=1-m_W^2/m_Z^2$ defined in terms of the $W$- and $Z$-boson masses, and $s_0^2$ defined via $\sin2\theta_0=\sqrt{\frac{4\pi\alpha(m_Z)}{\sqrt 2 G_F m_Z^2}}$. 
\begin{figure}
\begin{center}
\includegraphics[width=0.69\textwidth]{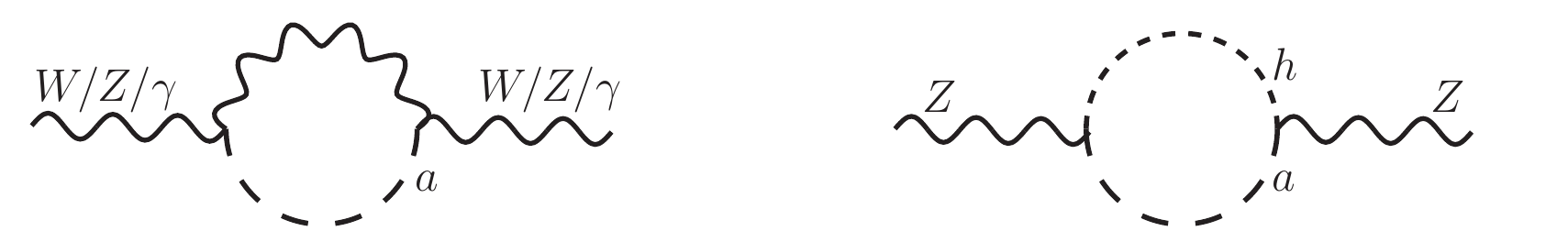}
\end{center}
\vspace{-2mm}
\caption{\label{fig:graphs_STU} One-loop Feynman diagrams contributing to electroweak precision observables.}
\end{figure}
We also consider the $\rho_*$ parameter defined by the low-energy ratio of charged- to neutral-current amplitudes. In terms of vacuum-polarization functions defined by the decomposition $\Pi_{AB}^{\mu\nu}(q)=\Pi_{AB}(q^2)\,g^{\mu\nu}+{\cal O}(q^\mu q^\nu)$, and working to one-loop order, these quantities can be expressed as 
\begin{equation}
\begin{aligned}
   s_*^2 &= \frac{g^{\prime\,2}}{g^2+g^{\prime\,2}} - s_w c_w\,\frac{\Pi_{\gamma Z}(m_Z^2)}{m_Z^2} \,, \\
   s_W^2 &= \frac{g^{\prime\,2}}{g^2+g^{\prime\,2}} 
    - c_w^2 \left[ \frac{\Pi_{WW}(m_W^2)}{m_W^2} - \frac{\Pi_{ZZ}(m_Z^2)}{m_Z^2} \right] , \\
   s_0^2 &= \frac{g^{\prime\,2}}{g^2+g^{\prime\,2}} 
    + \frac{s_w^2 c_w^2}{c_w^2-s_w^2} \left[  \frac{\Pi_{\gamma\gamma}(m_Z^2)}{m_Z^2}
    + \frac{\Pi_{WW}(0)}{m_W^2} - \frac{\Pi_{ZZ}(m_Z^2)}{m_Z^2} \right] , \\
   \rho_* &= 1 + \frac{\Pi_{WW}(0)}{m_W^2} - \frac{\Pi_{ZZ}(0)}{m_Z^2} 
    - \frac{2s_w}{c_w}\,\frac{\Pi_{\gamma Z}(0)}{m_Z^2} \,.
\end{aligned}
\end{equation}
In the correction terms the lowest-order expressions $s_w^2=g^{\prime\,2}/(g^2+g^{\prime\,2})$ and $c_w^2=g^2/(g^2+g^{\prime\,2})$ can be used. Note that our relation for $s_0^2$ differs from a corresponding relation in \cite{Peskin:1995ev}, where the polarization function $\Pi_{\gamma\gamma}(m_Z^2)$ in the first term has been expanded about $q^2=0$. In a new-physics model containing light new particles, such as ours, such an expansion is not legitimate. We find that, at dimension-6 order, the ALP-induced contributions to the vacuum-polarization functions derived from the effective Lagrangian (\ref{Leff}) involve intermediate $(aV)$ states with $V=\gamma,Z,W$, see the first graph in Figure~\ref{fig:graphs_STU}. These contributions vanish at $q^2=0$, and hence they do not give a contribution to the $\rho_*$ parameter. The individual $\Pi_{AB}(q^2)$ functions are quadratically divergent, however these divergences cancel if we consider the differences between the various definitions of $s_w^2$. In the class of new-physics models in which the non-polynomial operator (\ref{Leffnonpol}) is present, there is an additional contribution to $\Pi_{ZZ}(q^2)$ shown in the second graph in Figure~\ref{fig:graphs_STU}, which does not vanish at $q^2=0$, and hence a contribution to the $\rho_*$ parameter arises in these models. Setting the ALP mass to zero for simplicity, we obtain
\begin{equation}\label{s2Wdifferences}
\begin{aligned}
   s_0^2 - s_*^2\,\big|_{\rm ALP}
   &= - 8\alpha^2\,\frac{m_Z^2}{\Lambda^2}\,\frac{C_{WW}\,C_{BB}}{c_w^2-s_w^2}
    \left( \ln\frac{\mu^2}{m_Z^2} + \delta_2 + 2 + \frac{i\pi}{3} \right) \\
   &\quad\mbox{}- \frac{s_w^2 c_w^2}{c_w^2-s_w^2}\,
    \frac{\big(C_{Zh}^{(5)}\big)^2}{16\pi^2}\,\frac{m_h^2}{\Lambda^2}
    \left[ \left( 1 - \frac{m_Z^2}{3m_h^2} \right) \left( \ln\frac{\mu^2}{m_h^2} + \frac32 \right) 
    - \frac{m_Z^2}{3m_h^2}\,p\bigg( \frac{m_Z^2}{m_h^2} \bigg) \right] , \\
\end{aligned}
\end{equation}
and
\begin{equation}\label{s2Wdifferences2}
\begin{aligned}
   s_W^2 - s_*^2\,\big|_{\rm ALP} 
   &= \frac{16\alpha^2}{3}\,\frac{m_Z^2}{\Lambda^2}\,\frac{c_w^2}{s_w^2}\,C_{WW}^2\!
    \left( \ln\frac{\mu^2}{m_Z^2} + \delta_2 + \frac53 + \frac{c_w^2}{s_w^2} \ln c_w^2 \right) \\
   &\quad\mbox{}- 8\alpha^2\,\frac{m_Z^2}{\Lambda^2}\,\frac{C_{WW}}{s_w^2}
    \left( c_w^2\,C_{WW} - s_w^2\,C_{BB} \right)\!
    \left( \ln\frac{\mu^2}{m_Z^2} + \delta_2 + 2 + \frac{i\pi}{3} \right) \\
   &\quad\mbox{}+ c_w^2\,\frac{\big(C_{Zh}^{(5)}\big)^2}{16\pi^2}\,\frac{m_h^2}{\Lambda^2}
    \left[ \left( 1 - \frac{m_Z^2}{3m_h^2} \right) \left( \ln\frac{\mu^2}{m_h^2} + \frac32 \right) 
    - \frac{m_Z^2}{3m_h^2}\,p\bigg( \frac{m_Z^2}{m_h^2} \bigg) \right] ,
\end{aligned}
\end{equation}
where $\delta_2=-3$, and we have defined
\begin{equation}
   p(x) = \frac{(1-x)^3\ln(1-x)}{x^3} + \frac{1}{x^2} - \frac{5}{2x} + \frac76 \,.
\end{equation} 
The imaginary parts in the above expressions arise from loop graphs containing a photon and an ALP and reflect the existence of the on-shell decay $Z\to\gamma a$ considered in Section~\ref{sec:Zga}. In cross sections these imaginary parts only enter at two-loop order and thus can be omitted here. We can then match the above results with the $S$, $T$, $U$ parameters defined in terms of $\rho_*$ and the quantities given in (\ref{s2Wdifferences}) and (\ref{s2Wdifferences2}) \cite{Peskin:1991sw}. This leads to 
\begin{align}\label{STUres}
   S &= 32\alpha(m_Z)\,\frac{m_Z^2}{\Lambda^2}\,
    C_{WW}\,C_{BB} \left( \ln\frac{\Lambda^2}{m_Z^2} -1 \right) 
    - \frac{\big(C_{Zh}^{(5)}\big)^2}{12\pi}\,\frac{v^2}{\Lambda^2} 
    \left[ \ln\frac{\Lambda^2}{m_h^2} + \frac32 + p\bigg( \frac{m_Z^2}{m_h^2} \bigg) \right] , \nonumber\\
   T &= - \frac{\big(C_{Zh}^{(5)}\big)^2}{16\pi^2\alpha}\,\frac{m_h^2}{\Lambda^2}
    \left( \ln\frac{\Lambda^2}{m_h^2} + \frac32 \right) , \\
   U &= \frac{32\alpha(m_Z)}{3}\,\frac{m_Z^2}{\Lambda^2}\,C_{WW}^2\! 
    \left( \ln\frac{\Lambda^2}{m_Z^2} - \frac13 - \frac{2c_w^2}{s_w^2} \ln c_w^2 \right) 
    + \frac{\big(C_{Zh}^{(5)}\big)^2}{12\pi}\,\frac{v^2}{\Lambda^2} 
    \left[ \ln\frac{\Lambda^2}{m_h^2} + \frac32 + p\bigg( \frac{m_Z^2}{m_h^2} \bigg) \right] , \nonumber
\end{align}
where we have set $\mu=\Lambda$. The coupling $\alpha$ in the $T$ parameter should be evaluated at $q^2=0$. The presence of UV divergences in these expressions signals that additional short-distance contributions from dimension-6 operators not containing the pseudoscalar $a$ are required in order to cancel the scale dependence. Like in Section~\ref{sec:amu}, we will assume that these are small at the new physics scale, since they are not enhanced by the large logarithm $\ln(\Lambda^2/m_Z^2)$.

\begin{figure}[t]
\begin{center}
\includegraphics[width=0.45\textwidth]{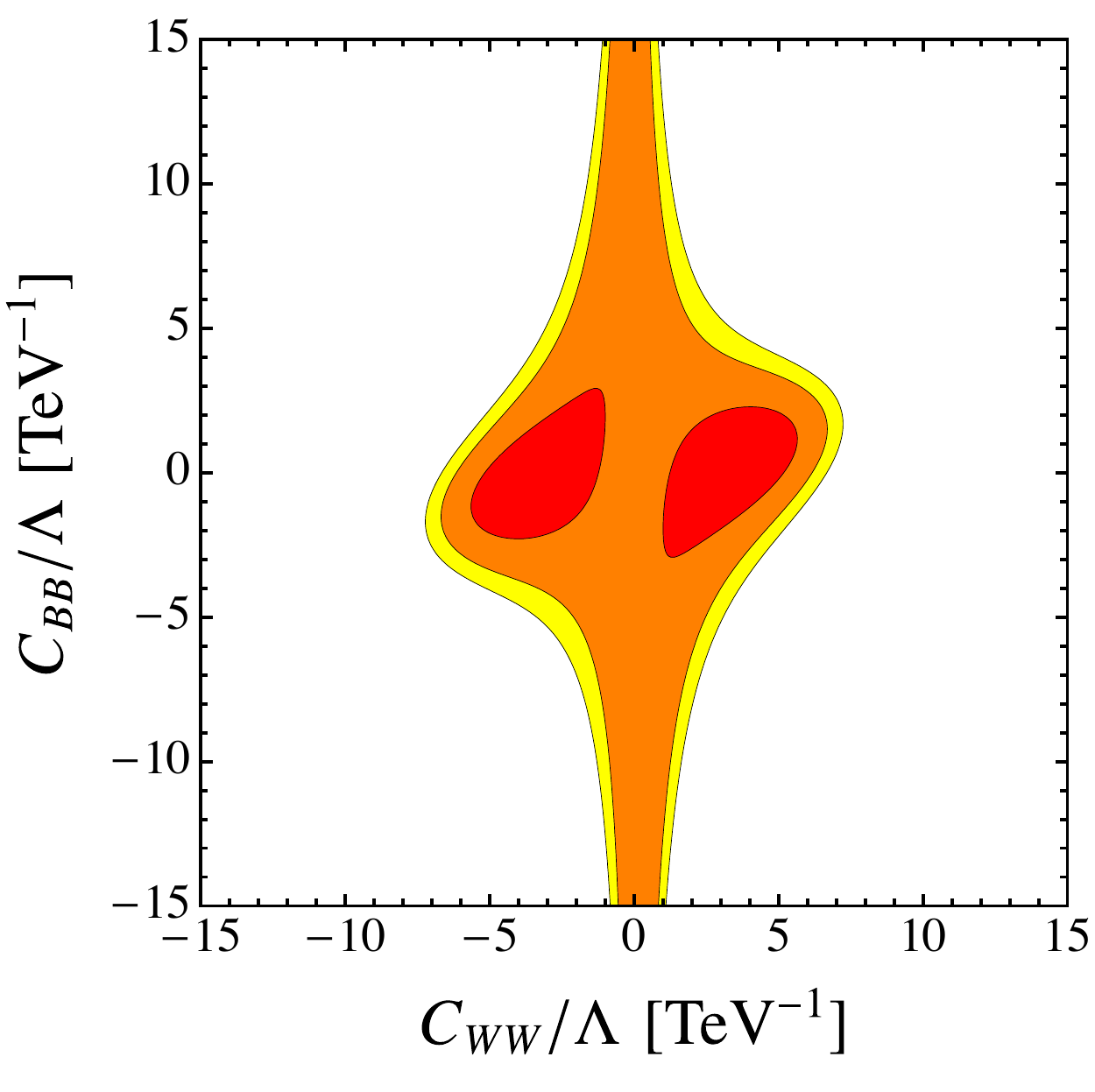}\qquad
\includegraphics[width=0.45\textwidth]{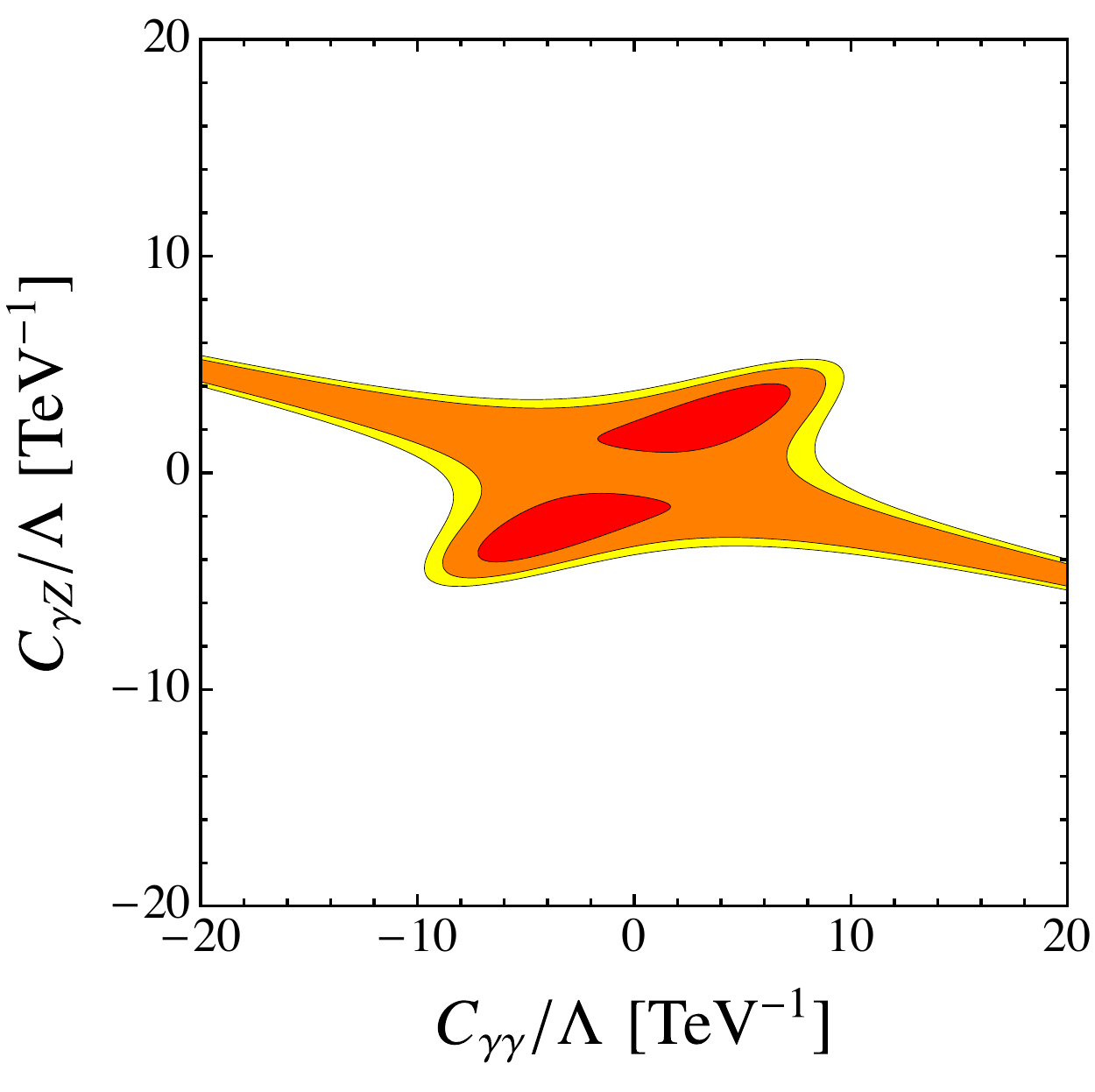}
\end{center}
\vspace{-3mm}
\caption{\label{fig:STU} Allowed regions in the parameters space of the Wilson coefficients $C_{WW}-C_{BB}$ (left) and $C_{\gamma\gamma}-C_{\gamma Z}$ (right) obtained from a global two-parameter electroweak fit \cite{Baak:2014ora} with $C_{Zh}^{(5)}=0$ at 68\% CL (red), 95\% CL (orange) and 99\% CL (yellow). We assume that contributions from dimension-6 operators not containing the ALP field can be neglected at $\Lambda=1$\,TeV.}
\end{figure}

\begin{figure}
\begin{center}
\hspace{-5mm}
\includegraphics[width=0.355\textwidth]{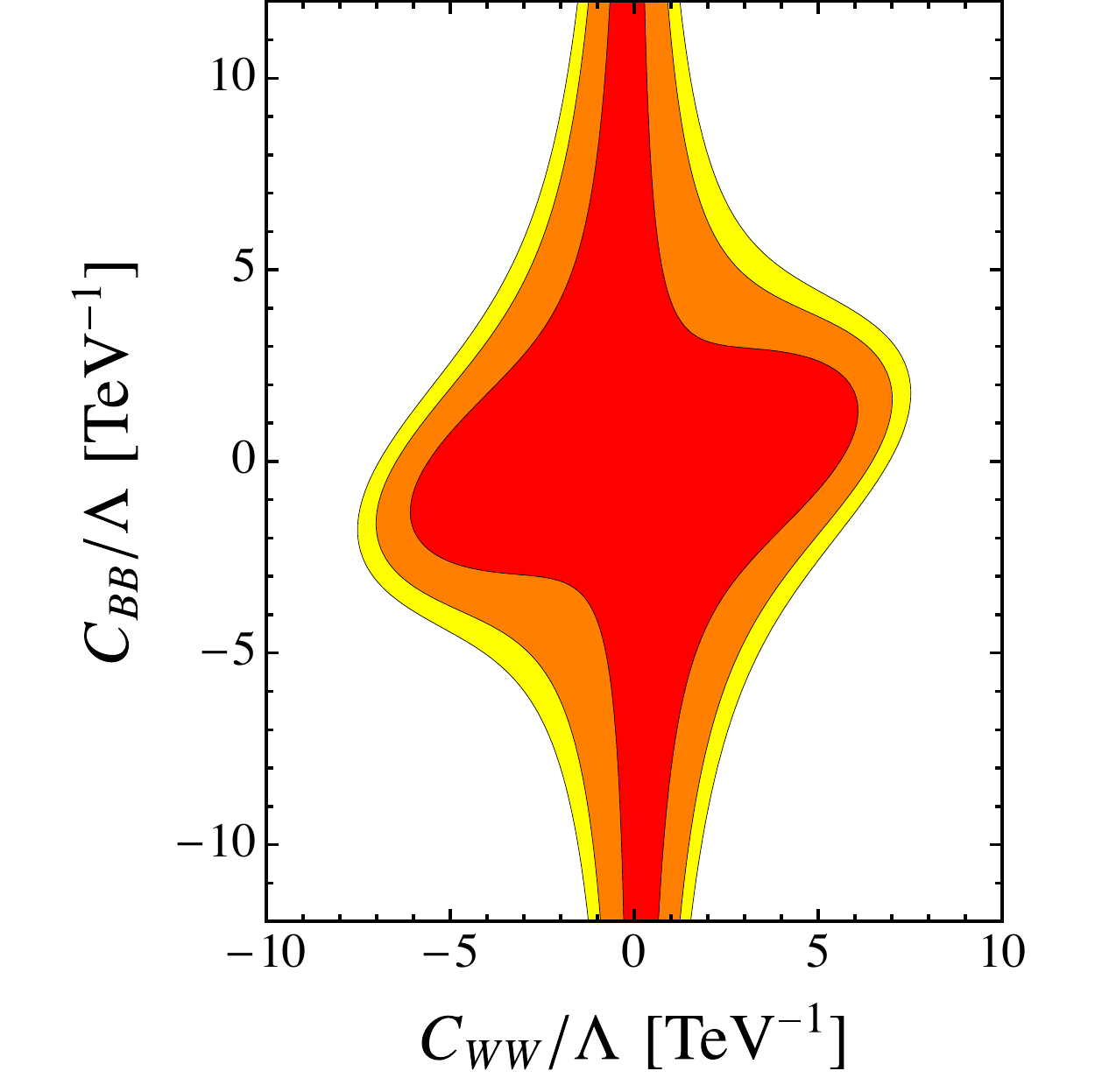}\hspace{-5mm}
\includegraphics[width=0.355\textwidth]{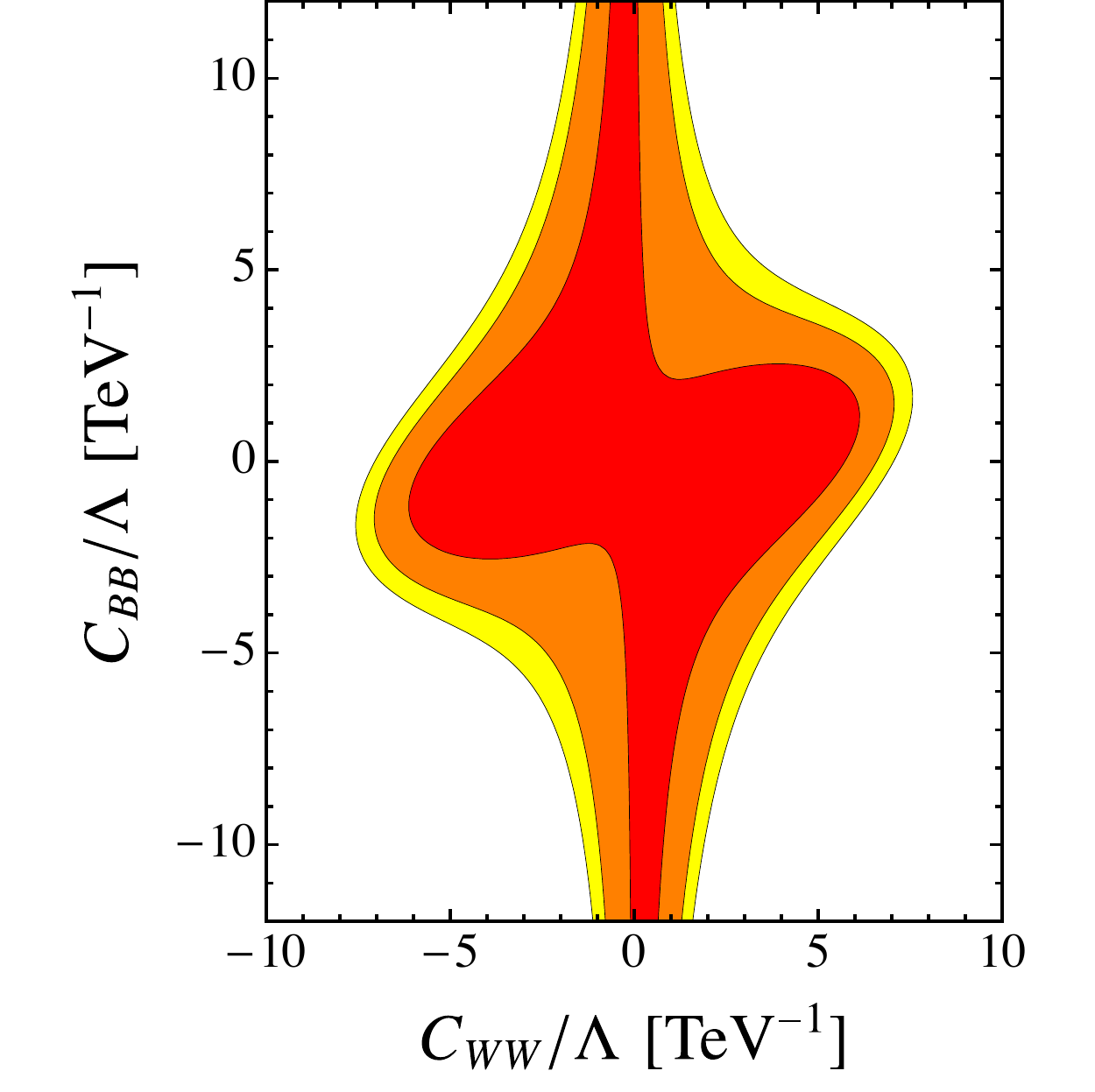}\hspace{-5mm}
\includegraphics[width=0.355\textwidth]{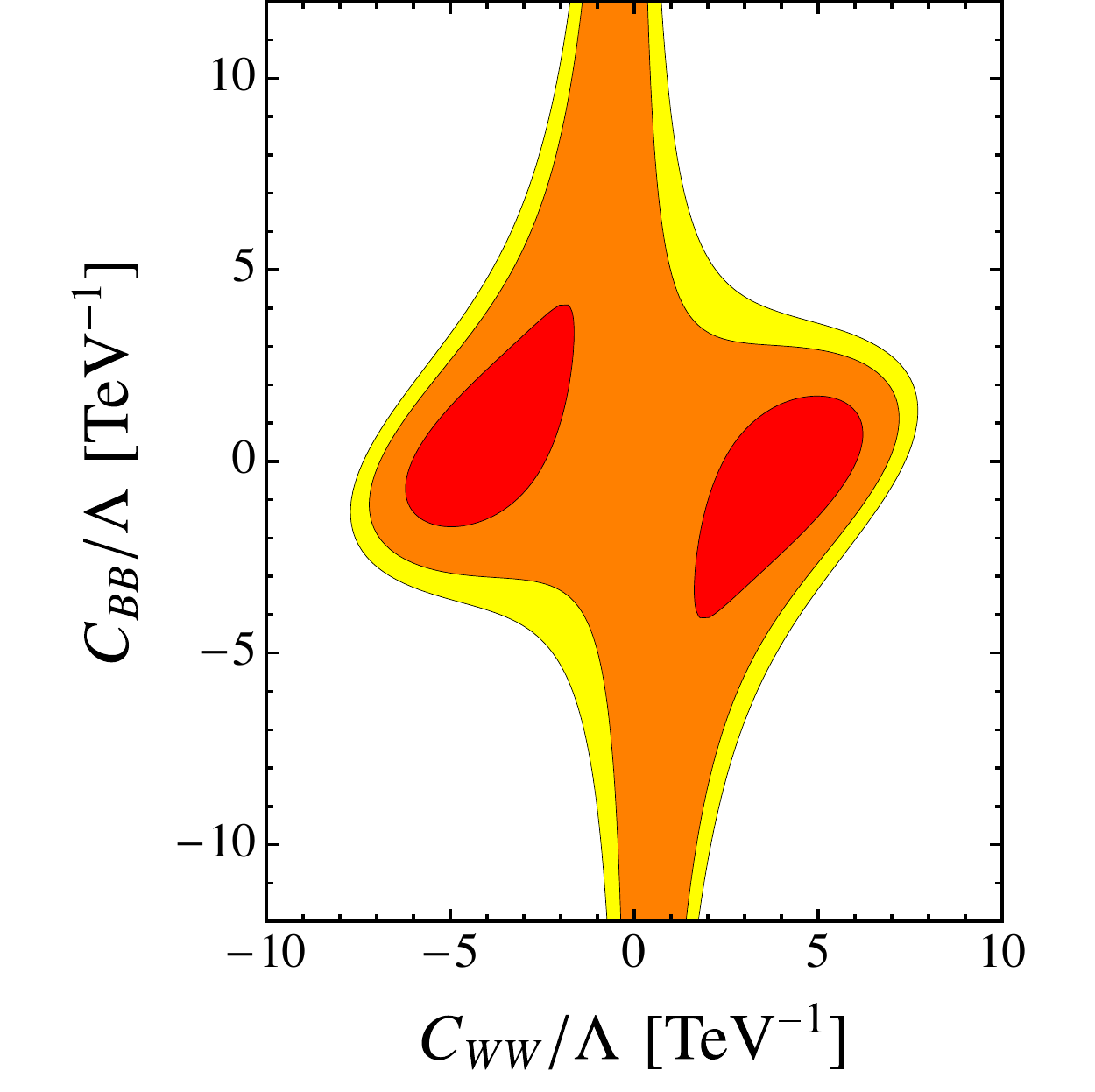}
\hspace{-5mm}
\end{center}
\vspace{-6mm}
\caption{\label{fig:STU2} Allowed regions in the parameters space of the Wilson coefficients $C_{WW}-C_{BB}$ obtained from a global three-parameter electroweak fit \cite{Baak:2014ora} at 68\% CL (red), 95\% CL (orange) and 99\% CL (yellow). The plots show projections onto the planes where $C_{Zh}^{(5)}/\Lambda=0$ (left), $0.36\,\text{TeV}^{-1}$ (center) and $0.72\,\text{TeV}^{-1}$ (right). We assume that contributions from dimension-6 operators not containing the ALP field can be neglected at $\Lambda=1$\,TeV.}
\end{figure}

Figure~\ref{fig:STU} shows the allowed parameter space for $C_{Zh}^{(5)}=0$ in the plane of the Wilson coefficients $C_{\gamma\gamma}-C_{\gamma Z}$ (left) and $C_{WW}-C_{BB}$ (right) obtained from the global electroweak fit \cite{Baak:2014ora}. The various coefficients are related by (\ref{Cgagadef}). We observe that the coefficients $C_{\gamma\gamma}$ and $C_{BB}$ are largely unconstrained, while $C_{\gamma Z}$ and $C_{WW}$ are restricted to relatively narrow ranges. At 99\% CL, we obtain to good approximation $|C_{\gamma Z}|/\Lambda<6\,\text{TeV}^{-1}$ and $|C_{WW}|/\Lambda<8\,\text{TeV}^{-1}$. The flat directions arise because for $C_{WW}=0$ (corresponding to $C_{\gamma Z}=-s_w^2\,C_{\gamma\gamma}$) the contributions to $S$ and $U$ in (\ref{STUres}) become independent of $C_{BB}$ and $C_{\gamma\gamma}$). We have also performed a global fit for three degrees of freedom including the effect of $C_{Zh}^{(5)}$. Its contribution to the $T$-parameter is negative and thus creates a slight tension with the current best fit. To lie within one or two standard deviations of the current best fit point requires $|C_{Zh}^{(5)}|/\Lambda<0.53\,\text{TeV}^{-1}$ and $|C_{Zh}^{(5)}|/\Lambda<1.39\,\text{TeV}^{-1}$ respectively. Given the model-independent bound (\ref{eq:cZhbound}), the tension is therefore very minor. Figure~\ref{fig:STU2} depicts the results of this fit projected onto the planes where $C_{Zh}^{(5)}/\Lambda=0$, $0.36\,\text{TeV}^{-1}$ and $0.72\,\text{TeV}^{-1}$ (i.e.\ maximal). Only for values of $C_{Zh}^{(5)}$ close to the upper bound (\ref{eq:cZhbound}) a slight tension arises for values of $C_{WW}$ and $C_{BB}$ of ${\cal O}(1/\mbox{TeV})$ or less.

\begin{figure}[t]
\begin{center}
\includegraphics[width=0.45\textwidth]{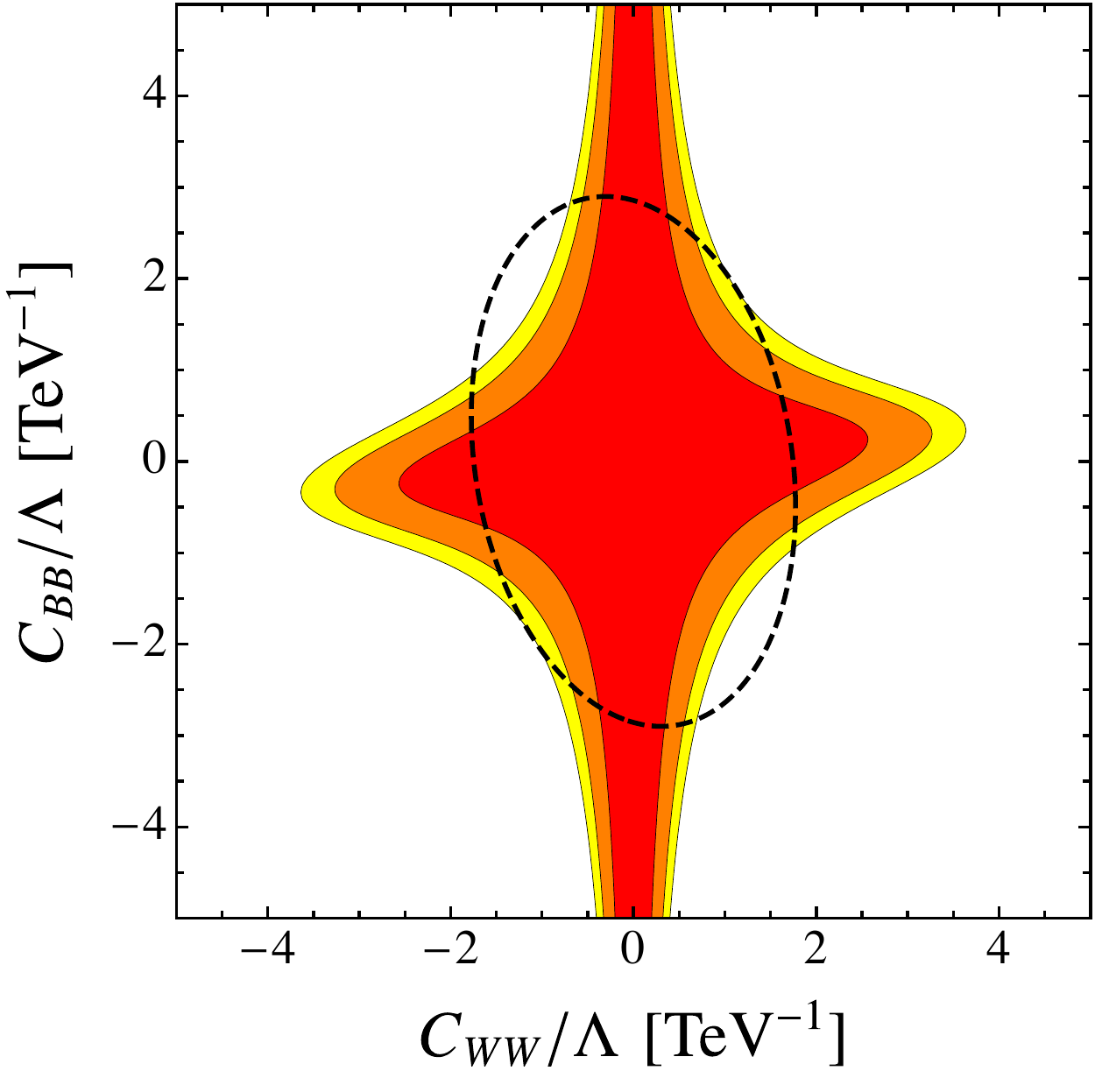}\qquad
\includegraphics[width=0.45\textwidth]{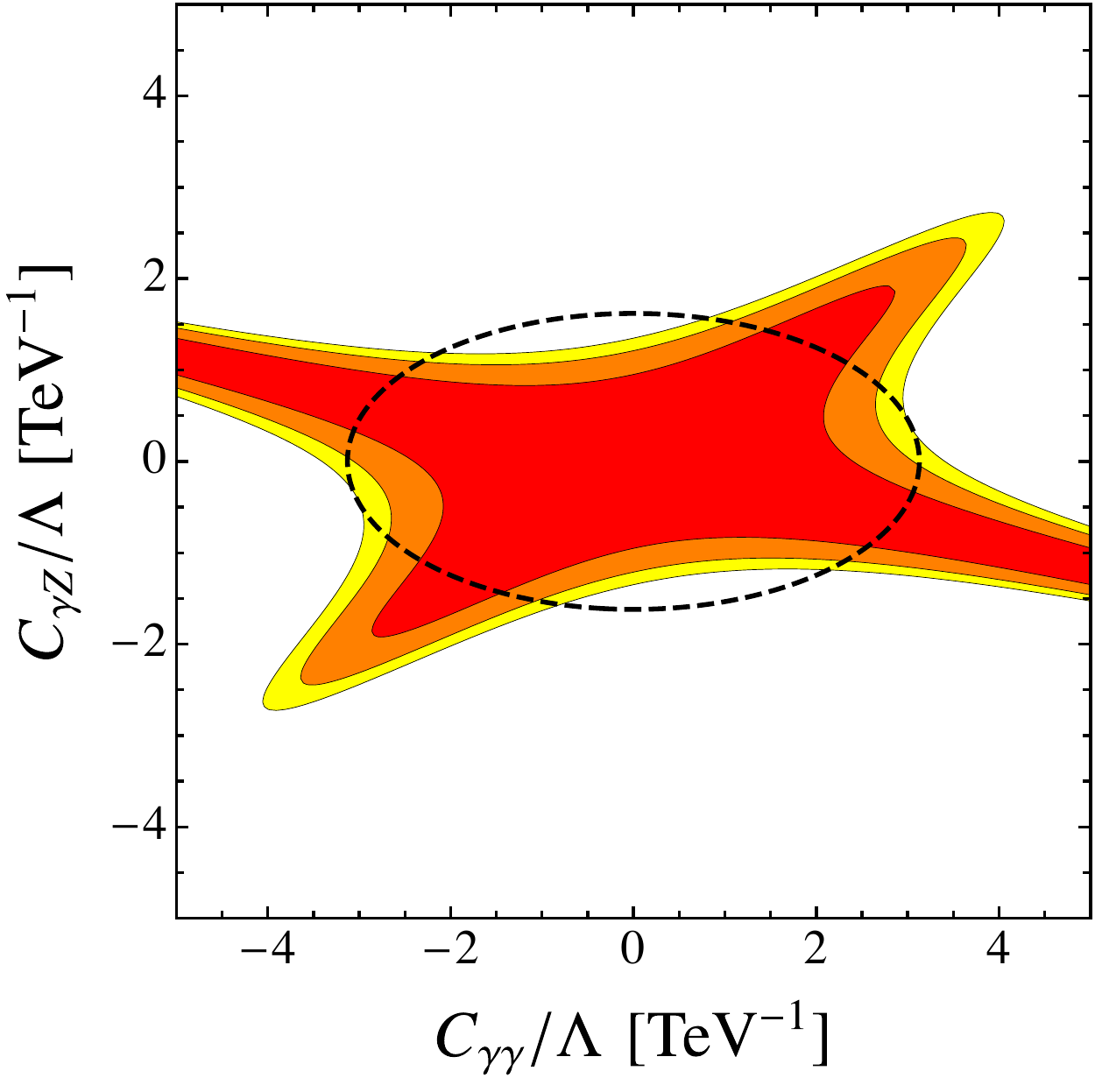}
\end{center}
\vspace{-3mm}
\caption{\label{fig:STUc5} Allowed regions in the parameters space of the Wilson coefficients $C_{WW}-C_{BB}$ (left) and $C_{\gamma\gamma}-C_{\gamma Z}$ (right) obtained from projections for the two-parameter global electroweak fit at a future FCC-ee machine \cite{deBlas:2016ojx} at 68\% CL (red), 95\% CL (orange) and 99\% CL (yellow), setting $C_{Zh}^{(5)}=0$. We assume that contributions from dimension-6 operators not containing the ALP field can be neglected at $\Lambda=1$\,TeV. For the parameter space within the dashed black contour, a FCC-ee measurement of $\alpha(m_Z)$ is within its projected errors at 95\% CL \cite{Janot:2015gjr}.}
\end{figure}

Another precision test can be performed by considering the running of the electromagnetic coupling constant from $q^2=0$ to $q^2=m_Z^2$. In our model we obtain 
\begin{equation}
   \frac{\alpha(0)}{\alpha(m_Z)}
   = \left. \frac{\alpha(0)}{\alpha(m_Z)} \right|_{\rm SM} 
    - \left[ \frac{\Pi_{\gamma\gamma}(m_Z^2)}{m_Z^2} - \Pi_{\gamma\gamma}'(0) \right]_{\rm ALP} ,
\end{equation}
where the vacuum-polarization functions now contain the ALP contribution only. Dropping again a small imaginary part and setting $\mu=\Lambda$, we find
\begin{equation}
   \frac{\alpha(0)}{\alpha(m_Z)}
   = \left. \frac{\alpha(0)}{\alpha(m_Z)} \right|_{\rm SM} 
    + \frac{8\alpha^2}{3}\,\frac{m_Z^2}{\Lambda^2} \left[ 
    C_{\gamma\gamma}^2 \left( \ln\frac{\Lambda^2}{m_Z^2} - \frac13 \right)  
    + \frac{C_{\gamma Z}^2}{s_w^2 c_w^2} \left( \ln\frac{\Lambda^2}{m_Z^2} - \frac{11}{6} \right) \right] .
\end{equation}
A measurement of $\alpha(m_Z)$ has been performed by the OPAL collaboration at a center of mass energy of 193\,GeV \cite{Abbiendi:2003dh}. The precision of this measurement is at the percent level, which is still compatible with values of $C_{WW}$ and $C_{BB}$ of ${\cal O}(30/\mbox{TeV})$. 

A significant improvement on the precision is expected from a future circular $e^+e^-$ collider FCC-ee \cite{Janot:2015gjr}, which will be able to measure $Z$-pole observables with unprecedented precision. In particular, $\alpha(m_Z)$ can be determined with an uncertainty of about $10^{-5}$. In Figure~\ref{fig:STUc5}, we show projections for the two-parameter electroweak fit based on the data obtained at such a machine \cite{deBlas:2016ojx}, assuming that the central values of $C_{WW}$ and $C_{BB}$ vanish. In the same figure, we superimpose the expected 95\% CL bound derived from the measurement of $\alpha(m_Z)$ (dashed contours), assuming that the theoretical error on this quantity will have decreased below the experimental uncertainty by the time the measurement can be performed. Combining these measurements can constrain $|C_{WW}|/\Lambda<2\,\text{TeV}^{-1}$ and $|C_{BB}|/\Lambda<3\,\text{TeV}^{-1}$, or equivalently $|C_{\gamma\gamma}|/\Lambda<2.5\,\text{TeV}^{-1}$ and $|C_{\gamma Z}|/\Lambda<1\,\text{TeV}^{-1}$ (at 95\% CL).

\section{Conclusions}
\label{sec:Conclusions}

Pseudoscalar particles with an approximate shift symmetry, so-called axion-like particles (ALPs), appear as pseudo Nambu--Goldstone bosons in any theory in which a global symmetry is spontaneously broken. If the mass scale of new physics is high, a light pseudo Nambu--Goldstone boson could be a harbinger of a new UV sector, which cannot otherwise be probed directly. The discovery of an ALP would not only confirm the existence of a UV theory beyond the SM, but by measuring its couplings important information on the properties of this theory can be derived. 

Based on the most general effective Lagrangian for a pseudoscalar with an approximate shift symmetry (softly broken only by an explicit mass term), we have computed the partial decay widths for ALPs into pairs of photons, leptons, jets and heavy quarks at one-loop order. Since the decay $a\to\gamma\pi$ is not allowed, relevant hadronic decay modes only open up when the ALP is heavy enough to decay into three pions. We have calculated the $a\to\pi\pi\pi$ partial widths of the ALP in terms of a chiral Lagrangian for the first time. We have emphasized that even loop-suppressed Wilson coefficients can lead to non-negligible branching ratios of ALPs decaying into photons or leptons. The assumption of stable ALPs therefore becomes unrealistic above a certain mass, if sizable couplings $C_{ii}/\Lambda$ of order $(0.01-1)\,\mbox{TeV}^{-1}$ to {\em any\/} SM fields exist. For the same reason, an ALP with such couplings cannot be lighter than about 1\,MeV, since in the presence of loop corrections it is impossible to satisfy the very strong cosmological bounds on the ALP--photon coupling without excessive fine tuning.

Significant insights can be gained by considering the exotic, on-shell decays $h\to Za$, $h\to aa$ and $Z\to\gamma a$. These three decays offer complementary information on a possible UV sector beyond the SM. While $Z$ decays are induced by dimension-5 operators coupling the ALP to electroweak gauge bosons, Higgs decays probe the dimension-6 Higgs portal in the case of $h\to aa$ and dimension-5 or 7 operators in the case of $h\to Z a$. The non-polynomial dimension-5 operator in (\ref{Leffnonpol}) inducing $h\to Za$ decay at Born level only arises if the heavy particles in the UV theory obtain a dominant fraction of their mass from electroweak symmetry breaking. Discovering an ALP in any or a combination of these exotic decays therefore allows us to extract non-trivial details about the underlying UV theory. To see the important role of Higgs decays, consider as a concrete example a scenario in which the only tree-level ALP couplings to SM fields are flavor-universal couplings to the up-type quarks, $c_{uu}=c_{cc}=c_{tt}=\Lambda/\mbox{TeV}$. ALP couplings to other SM particles are induced only by means of quark loops. Assuming $m_a=1$\,GeV, one then finds the Higgs branching ratios $\text{Br}(h\to Za)=2.5\times 10^{-4}$ and $\text{Br}(h\to aa)=8.5\times 10^{-3}$, which are 0.15 and 5.5 times the SM $h\to\gamma Z$ branching ratio, respectively. The loop-induced couplings to electroweak gauge bosons are rather small, and correspondingly $\mbox{Br}(Z\to\gamma a)=4.8\times 10^{-9}$ is absolutely negligible. On the other hand, the loop-induced ALP--photon coupling $C_{\gamma\gamma}^{\rm eff}\approx 0.008$ lies in the range of sensitivity of our approach. In this scenario it would be easy to discover the ALP in $h\to aa$ decay, challenging to probe its couplings in $h\to Za$ decay, and hopeless to see any hints of ALPs in $Z\to\gamma a$ decay.

We have presented a comprehensive discussion of the LHC reach in searches for ALPs in exotic Higgs- and $Z$-boson decays. Taking into account constraints from existing searches, model-independent bounds on non-SM Higgs or $Z$ decays and finite-lifetime effects of light ALPs or ALPs with small couplings, we found LHC searches for the decays $h\to Za$, $h\to aa$ and $Z\to\gamma a$ to be sensitive to new-physics scales as high as 100\,TeV for ALP masses in the GeV range. Depending on the decay mode of the ALP, several striking signatures can be observed. We have especially considered subsequent ALP decays into photons and charged leptons, taking into account the possibility that light boosted ALPs decay into collimated photon jets, which cannot be distinguished from a single photon experimentally. Cosmological bounds, ALP searches with helioscopes, beam-dump experiments and searches for ALPs at lepton and hadron colliders significantly constrain the parameter space for ALPs decaying into di-photons or $e^+e^-$ pairs. Intriguingly, we project the best sensitivity for ALP searches in on-shell Higgs- and $Z$-boson decays at the LHC for ALP masses in the range above approximately 10\,MeV and up to about 90\,GeV, a region of parameter space mostly unconstrained by existing bounds once we assume that the relevant ALP couplings are of order 1/TeV or smaller. For ALPs in the GeV mass range, this reach extends many orders of magnitude beyond current bounds, without the need to assume any large Wilson coefficients. Even with loop-suppressed ALP--Higgs couplings, the bounds on the ALP--photon coupling can be improved by up to five orders of magnitude using searches for the decays $h\to Za\to\ell^+\ell^-+\gamma\gamma$ and $h\to aa\to 4\gamma$. Improvements by several orders of magnitude can also be obtained from a search for the decay $Z\to\gamma a\to 3\gamma$; however, the reach in this case depends on the correlation of the $a\gamma\gamma$ and $a\gamma Z$ couplings, which depends on the underlying UV model. Importantly, these bounds can be derived even if the $a\to\gamma\gamma$ branching ratio is significantly less than~1. In the leptonic decay channels $a\to\ell^+\ell^-$, completely uncharted territory in parameter space can be probed, extending down to ALP--lepton couplings as small as $(10^6\,\mbox{TeV})^{-1}$.

We have further computed the parameter space for which the long-standing $(g-2)_\mu$ anomaly can be explained by ALPs coupling to muons and photons. A possible resolution by a loop contribution from ALPs is largely independent of its mass and requires a sizable coupling to photons and a coupling of similar size (and the correct sign) to muons. For example, a good fit can be found for $m_a=1$\,GeV and $C_{\gamma\gamma}\approx -c_{\mu\mu}\approx 1.5\,(\Lambda/\text{TeV})$. We have translated the bound from a Babar search for a new $Z'$ boson in the $e^+e^-\to\mu^+\mu^-+\mu^+\mu^-$ channel into a constraint on the $c_{\mu\mu}-C_{\gamma\gamma}$ plane, thereby directly constraining a possible explanation of $(g-2)_\mu$ by ALP exchange. We find that future searches for $e^+e^-\to\mu^+\mu^-+\mu^+\mu^-$ as well as $e^+e^-\to\mu^+\mu^-+\gamma\gamma$ at Belle~II have the potential to discover or exclude an ALP explanation of the anomaly for $2m_\mu<m_a\lesssim 2$\,GeV. Remarkably, the complete unconstrained parameter space for which an ALP can explain the muon anomaly can be probed by the exotic Higgs- and $Z$-boson decays studied in this paper. Barring for scenarios in which the $a\gamma\gamma$ coupling is very large, whereas the $aZh$, $aah$ and $a\gamma Z$ couplings are all more than one-loop suppressed, searches for ALPs in exotic decays of on-shell Higgs and $Z$ bosons at the LHC can therefore exclude or confirm an ALP explanation of $(g-2)_\mu$.  Electroweak precision tests constrain the ALP couplings to electroweak gauge bosons and to $Zh$. These coefficients control the $Z\to\gamma a$ and  $h\to Za$ decay rates, respectively. We have computed the one-loop corrections to the oblique parameters and to $\alpha(m_Z)$ and derived the corresponding bounds on the Wilson coefficients from the global electroweak fit, finding that they are rather weak. We have also presented projections for a future FCC-ee machine, where it will be possible to probe ALP couplings to electroweak gauge bosons of order 1/TeV.

The LHC has an unprecedented reach in searching for ALPs in exotic, on-shell decays of Higgs and $Z$ bosons. We strongly encourage experimental searches in the full mass range and in all three channels discussed in this paper. A UFO file for the ALP model discussed in the present work is available from the authors upon request.

\subsubsection*{Acknowledgments}

We are grateful to Thomas Becher, Konstantin Chetyrkin, David Curtin, Aneesh Manohar, Maxim Pospelov, J\"org J\"ackel, Joachim Kopp and Pedro Schwaller for useful discussions. M.N.\ thanks the Aspen Center for Physics, where this work has been completed,  for hospitality and support. The research reported here has been supported by the Advanced Grant EFT4LHC of the European Research Council (ERC), the Cluster of Excellence {\em Precision Physics, Fundamental Interactions and Structure of Matter\/} (PRISMA -- EXC 1098), and grant 05H12UME of the German Federal Ministry for Education and Research (BMBF). 

\subsubsection*{Note added in proof}

We would like thank the referee for valuable comments on the Edelweiss and BaBar bounds and for encouraging us to include Figures 18 and 20. After the submission of this paper, two new analyses discussing first experimental results in heavy-ion collisions \cite{Knapen:2017ebd} and bounds from beam-dump searches \cite{Dolan:2017osp} have been submitted to the arXiv. These results supersede some of the constraints on the ALP--photon coupling shown in our work, but they do not change any of the conclusions derived in this paper.

\newpage
\begin{appendix}

\section{Naive dimensional analysis estimates}
\label{app:NDA}
\renewcommand{\theequation}{A.\arabic{equation}}
\setcounter{equation}{0}

Here we collect order-of-magnitude estimates for the Wilson coefficients in the effective Lagrangians (\ref{Leff}) and (\ref{LeffD>5}) based on naive dimensional analysis. Using the counting rules derived in \cite{Manohar:1983md,Luty:1997fk,Cohen:1997rt}, one obtains
\begin{equation}
   \bm{C}_F = 4\pi\,\bar{\bm{C}}_F \,, \qquad
   C_{VV} = \frac{\bar C_{VV}}{4\pi} \,, \qquad
   C_{ah}^{(\prime)} = (4\pi)^2\,\bar C_{ah}^{(\prime)} \,, \qquad
   C_{Zh}^{(7)} = (4\pi)^3\,\bar C_{Zh}^{(7)} \,,
\end{equation}
where the subscript $V$ is the second relation can be $G$, $W$ or $B$. The barred coefficients on the right-hand sides of these relations can naturally be of ${\cal O}(1)$ in strongly coupled theories. When the effective Lagrangians are rewritten in terms of a parameter $f$ defined such that $4\pi f\equiv\Lambda$ (this parameter is related to the ALP decay constant $f_a$ by $f=-2\bar C_{GG} f_a$), one obtains expressions analogous to (\ref{Leff}) and (\ref{LeffD>5}), in which the Wilson coefficients are replaced by the barred Wilson coefficients and $\Lambda$ is replaced by $f$. The only exception are the ALP--gauge-boson couplings, which are given by $\bar C_{VV}/(4\pi)^2$. It would therefore have been more natural to introduce a loop factor $1/(4\pi)^2$ in the three terms shown in the second line of (\ref{Leff}).\footnote{A similar argument applies for the coefficient of the non-polynomial operator in (\ref{Leffnonpol}), for which one should assign an extra factor $1/(4\pi)^2$, since a loop is needed to generate a logarithmic dependence on the Higgs field. This leads to the counting rule $C_{Zh}^{(5)}=\bar C_{Zh}^{(5)}/(4\pi)$, in analogy with the ALP--boson couplings.}  
Following a standard practice in the ALP literature, we have refrained from doing so.

In light of these remarks, it becomes evident that an explanation of the $(g-2)_\mu$ anomaly requires a somewhat unnaturally large value of the ALP--photon coupling. From Figure~\ref{fig:g-2} we see that we typically need $|C_{\gamma\gamma}|/\Lambda\gtrsim 0.5/\mbox{TeV}$, corresponding to $|\bar C_{\gamma\gamma}|/\Lambda\gtrsim 6/\mbox{TeV}$. Generating such a large coefficient may require to have a large multiplicity of new TeV-scale particles in a loop or lowering $\Lambda$ below the TeV scale, but it does not appear to be impossible.

Extensions of the SM in which the electroweak symmetry is realized non-linearly provide an explicit example of strongly coupled models, in which operators of higher dimension in the effective Lagrangian are suppressed by powers of $1/f$ rather than $1/\Lambda$ \cite{Feruglio:1992wf}. In realistic composite Higgs scenarios the ratio $\xi=v^2/f^2$ is tightly constrained by electroweak precisions tests, implying $\xi<0.05$ \cite{Grojean:2013qca}, and Higgs phenomenology, yielding $\xi<0.1$ \cite{Aad:2015pla}, both at 95\% CL. As a result, it is unlikely that $f$ can be significantly below the TeV scale in these models~\cite{Panico:2015jxa}.

\section{Couplings of a light ALP to hadrons}
\label{app:ChPT}
\renewcommand{\theequation}{B.\arabic{equation}}
\setcounter{equation}{0}

At energies below a few GeV, the effective Lagrangian (\ref{Leff}) supplemented by the QCD Lagrangian gives rise to the terms
\begin{equation}\label{eq:a1}
\begin{aligned}
   {\cal L}_{\rm eff}
   &\ni \frac12 \left( \partial_\mu a\right)\!\left( \partial^\mu a\right) - \frac{m_{a,0}^2}{2}\,a^2
    + \bar q\,(i\rlap{\,/}D - m_q)\,q 
    + \frac{\partial^\mu a}{2\Lambda}\,\bar q\,c_{qq}\,\gamma_\mu\gamma_5\,q \\
   &\quad\mbox{}+ g_s^2\,C_{GG}\,\frac{a}{\Lambda}\,G_{\mu\nu}^A\,\tilde G^{\mu\nu,A}
    + e^2\,C_{\gamma\gamma}\,\frac{a}{\Lambda}\,F_{\mu\nu}\,\tilde F^{\mu\nu} \,,
\end{aligned}
\end{equation}
where $m_{a,0}$ denotes a possible ALP mass term resulting from an explicit breaking of the shift symmetry. We will for simplicity only consider the two light $u$ and $d$ quarks. We use a compact matrix notation, where in the mass basis $m_q=\mbox{diag}(m_u,m_d)$ and $c_{qq}=\mbox{diag}(c_{uu},c_{dd})$ are diagonal hermitian matrices. Before mapping this expression onto an effective chiral Lagrangian, it is convenient to remove the ALP--gluon coupling by means of the chiral rotation
\begin{equation}
   q\to \exp\bigg( i\kappa_q\,\frac{a}{2f_a}\,\gamma_5 \bigg)\,q \,, 
\end{equation}
where $\kappa_q$ is a diagonal matrix satisfying $\mbox{tr}\,\kappa_q=1$, and $f_a$ is referred to as the ALP decay constant. Under the chiral rotation the measure of the path integral is not invariant \cite{Fujikawa:1979ay,Fujikawa:1980eg}, and this generates extra terms adding to the anomalous couplings in (\ref{eq:a1}). In order to remove the ALP--gluon coupling we need to require that
\begin{equation}
   \frac{1}{f_a} = - 32\pi^2\,\frac{C_{GG}}{\Lambda} \,.
\end{equation}
This leads to
\begin{equation}
\begin{aligned}
   {\cal L}_{\rm eff}
   &\ni \frac12 \left( \partial_\mu a\right)\!\left( \partial^\mu a\right) - \frac{m_{a,0}^2}{2}\,a^2
    + \bar q\,\big[i\rlap{\,/}D - \hat m_q(a)\big]\,q 
    + \frac{\partial^\mu a}{2\Lambda}\,\bar q\,\hat c_{qq}\,\gamma_\mu\gamma_5\,q \\
   &\quad\mbox{}+ e^2\,\Big( C_{\gamma\gamma} - 2N_c\,C_{GG}\,\mbox{tr}[\kappa_q\,Q_q^2] \Big)\,
    \frac{a}{\Lambda}\,F_{\mu\nu}\,\tilde F^{\mu\nu} \,,
\end{aligned}
\end{equation}
where  
\begin{equation}\label{hatmq}
   \hat m_q(a) = \exp\bigg( i\kappa_q\,\frac{a}{2f_a}\,\gamma_5 \bigg)\,m_q\,
    \exp\bigg( i\kappa_q\,\frac{a}{2f_a}\,\gamma_5 \bigg) \,, \qquad
   \hat c_{qq} = c_{qq} + 32\pi^2\kappa_q\,C_{GG} \,.
\end{equation}
Matching the above effective Lagrangian onto a chiral Lagrangian, one obtains \cite{Georgi:1986df,Bardeen:1986yb,Krauss:1986bq}
\begin{equation}
\begin{aligned}
   {\cal L}_{\chi PT} 
   &= \frac12\,\partial^\mu a\,\partial_\mu a - \frac{m_{a,0}^2}{2}\,a^2
    + \frac{f_\pi^2}{8}\,\mbox{tr}\big[ D^\mu\Sigma\,D_\mu\Sigma^\dagger \big] 
    + \frac{f_\pi^2}{4}\,B_0\,\mbox{tr}\big[ \Sigma\,\hat m_q^\dagger(a) + \hat m_q(a)\,\Sigma^\dagger \big] \\
   &\quad\mbox{}+ \frac{if_\pi^2}{4}\,\frac{\partial^\mu a}{2\Lambda}\,
    \mbox{tr}\big[ \hat c_{qq} (\Sigma^\dagger D_\mu\Sigma - \Sigma\,D_\mu\Sigma^\dagger)\big]
    + e^2\,\Big( C_{\gamma\gamma} - 2N_c\,C_{GG}\,\mbox{tr}[\kappa_q\,Q_q^2] \Big)\,
    \frac{a}{\Lambda}\,F_{\mu\nu}\,\tilde F^{\mu\nu} \,,
\end{aligned}
\end{equation} 
where $\Sigma$ containing the pion fields has been defined after (\ref{LeffChpT}). Note that now $\hat m_q(a)$ is evaluated by replacing $\gamma_5$ in (\ref{hatmq}) by its eigenvalue $+1$. The covariant derivative is defined as $D^\mu\Sigma=\partial^\mu\Sigma-ieA^\mu\,[Q,\Sigma]$, where $Q$ contains the quark electric charges in units of $e$.

Even if the explicit mass term $m_{a,0}$ is absent, QCD dynamics generates a mass for the ALP \cite{Bardeen:1978nq,Shifman:1979if,DiVecchia:1980yfw}, thereby breaking the continuous shift symmetry. Expanding the terms in the first line to quadratic order in the pion and ALP fields, one finds the mass eigenvalues 
\begin{equation}
\begin{aligned}
   m_\pi^2 &= B_0\,(m_u+m_d) + {\cal O}\bigg( \frac{m_\pi^2\,f_\pi^2}{f_a^2} \bigg) \,, \\
   m_a^2 &= m_{a,0}^2 + \frac{m_\pi^2\,f_\pi^2}{2f_a^2}\,\frac{m_u m_d}{(m_u+m_d)^2}
    + {\cal O}\bigg( \frac{m_\pi^2\,f_\pi^4}{f_a^4} \bigg) \,,
\end{aligned}
\end{equation} 
where we have adopted the choice (\ref{kappaqvals}) for the $\kappa_q$ parameters, which eliminates the mass mixing of the ALP with the neutral pion. This choice leads to the effective chiral Lagrangian given in (\ref{LeffChpT}). The coefficient in front of the ALP--photon coupling now takes the form
\begin{equation}\label{E/N}
   \left[ C_{\gamma\gamma} - \frac23\,\frac{4m_d+m_u}{m_u+m_d} C_{GG} \right] 
   = \left[ \frac{E}{N} - \frac53 - \frac{m_d-m_u}{m_u+m_d} \right] C_{GG} 
   \approx \left[ \frac{E}{N} - 2.02 \right] C_{GG} \,,
\end{equation}
where $E/N=C_{\gamma\gamma}/C_{GG}$ and we have used that $m_u/m_d\approx 0.48$ (see e.g.\ \cite{Horsley:2015eaa,Basak:2015lla} for two recent lattice determination of this ratio). The term proportional to the explicit isospin breaking caused by the mass difference between up and down quarks results from the coupling of the neutral pion to $G_{\mu\nu}^A\,\tilde G^{\mu\nu,A}$. The corresponding matrix element has been evaluated in \cite{Beneke:2000ry} and is found to be
\begin{equation}
   \big\langle\pi^0\big|\,\frac{\alpha_s}{4\pi}\,G_{\mu\nu}^A\,\tilde G^{\mu\nu,A}\,\big|0\big\rangle
   = - \frac{m_d-m_u}{m_d+m_u}\,\frac{f_\pi\,m_\pi^2}{\sqrt2} \,.
\end{equation}
The pion then decays into two photons via the axial anomaly. The contribution 5/3 arises from an analogous coupling the flavor-singlet meson $\varphi^0$ (the analogue of $\eta_1$ in flavor $SU(3)$) \cite{Kaiser:2000gs}. Next-to-leading order corrections to the result (\ref{E/N}) have been worked out in \cite{diCortona:2015ldu}. They lead to a coefficient $[E/N-(1.92\pm 0.04)]$, which we use in our numerical analysis.

\section{Technical details of the loop calculations}
\label{app:A}
\renewcommand{\theequation}{C.\arabic{equation}}
\setcounter{equation}{0}

The loop function $g(\tau)$ entering the expression for the effective ALP--lepton coupling in (\ref{clleff}) is given by the parameter integral
\begin{equation}
   g(\tau) = 5 + \frac43 \int_0^1\!dx\,\frac{1-4\tau(1-x)^2-2x+4x^2}{\sqrt{\tau(1-x)^2-x^2}}\,
   \arctan\bigg(\frac{x}{\sqrt{\tau(1-x)^2-x^2}}\bigg) \,,
\end{equation}
where $\tau=4m_\ell^2/m_a^2-i0$. The asymptotic expansions for small and large values of $\tau$ have been shown in (\ref{gfunasy}).

The scheme-dependent constant $\delta_1$ in (\ref{clleff}) arises from the treatment of the Levi--Civita symbol in $d$ dimensions. We follow the standard procedure of expressing the product $\epsilon^{\alpha\beta\gamma\delta}\,\epsilon^{\mu\nu\rho\sigma}$ in terms of the determinant of a $4\times 4$ matrix consisting of elements of the metric tensor \cite{Larin:1993tq}. In this way, we obtain the relations (with $d=4-2\varepsilon$)
\begin{equation}\label{Levi--Civita}
\begin{aligned}
   \epsilon^{\alpha\beta\gamma\delta} \gamma_\beta\gamma_\gamma\gamma_\delta 
   &= i(d-3)(d-2)(d-1)\,\gamma^\alpha\gamma_5 
    = 6i \left( 1 + \varepsilon\delta_1 + \dots \right) \gamma^\alpha\gamma_5 \,, \\
   \epsilon^{\alpha\beta\gamma\delta}\,\epsilon^{\mu}{}_{\beta\gamma\delta}
   &= -(d-3)(d-2)(d-1)\,g^{\alpha\mu} \,, \\
   \epsilon^{\alpha\beta\gamma\delta} \gamma_\gamma\gamma_\delta 
   &= - \frac{i}{2}\,(d-3)(d-2)\,[\gamma^\alpha,\gamma^\beta]\,\gamma_5
    = -i \left( 1 + \varepsilon\delta_2 + \dots \right) [\gamma^\alpha,\gamma^\beta]\,\gamma_5 \,, \\
   \epsilon^{\alpha\beta\gamma\delta}\,\epsilon^{\mu\nu}{}_{\gamma\delta}
   &= (d-3)(d-2) \left( g^{\alpha\nu} g^{\beta\mu} - g^{\alpha\mu} g^{\beta\nu} \right) , \\
\end{aligned}
\end{equation}
where $\delta_1=-\frac{11}{3}$ and $\delta_2=-3$. In a scheme where instead the Levi--Civita symbol is treated as a 4-dimensional object, one would have $\delta_1=\delta_2=0$.

\section{Effect of a finite ALP lifetime}
\label{app:B}
\renewcommand{\theequation}{D.\arabic{equation}}
\setcounter{equation}{0}

The two event fractions defined in (\ref{eq:29}) obey the exact relations
\begin{equation}
   f_{\rm dec}^{Za} = F\bigg(\frac{L_{\rm det}}{L_a}\bigg) \,, \qquad
   f_{\rm dec}^{aa} = 2 F\bigg(\frac{L_{\rm det}}{L_a}\bigg) - F\bigg(\frac{2L_{\rm det}}{L_a}\bigg) \,,
\end{equation}
where the function $F(x)$ is given by
\begin{equation}
   F(x) = 1 - \int_x^\infty\!dy\,\sqrt{1-\frac{x^2}{y^2}}\,e^{-y} \,.
\end{equation}
It obeys the asymptotic expansions
\begin{equation}
   F(x) = \left\{
    \begin{array}{cl} 
    \displaystyle \frac{\pi}{2}\,x - \frac{x^2}{2} \left( \frac32 + \ln 2-\gamma_E-\ln x\right) 
     + \dots \,; &~ x\ll 1 \,, \\[4mm]
    \displaystyle 1 - \sqrt{\frac{\pi}{2x}}\,e^{-x} + \dots \,; &~ x\gg 1 \,. \\
    \end{array} \right.
\end{equation}
Using the first result, we have obtained the asymptotic relations given in (\ref{eq:fdecasy}).

\end{appendix}


\newpage
\begin{thebibliography}{199}

\bibitem{Peccei:1977hh} 
  R.~D.~Peccei and H.~R.~Quinn,
  Phys.\ Rev.\ Lett.\ {\bf 38}, 1440 (1977).
  
\bibitem{Peccei:1977ur} 
  R.~D.~Peccei and H.~R.~Quinn,
  Phys.\ Rev.\ D {\bf 16}, 1791 (1977).
    
\bibitem{Weinberg:1977ma} 
  S.~Weinberg,
  Phys.\ Rev.\ Lett.\ {\bf 40}, 223 (1978).

\bibitem{Wilczek:1977pj} 
  F.~Wilczek,
  Phys.\ Rev.\ Lett.\ {\bf 40}, 279 (1978).

\bibitem{Kim:1979if} 
  J.~E.~Kim,
  Phys.\ Rev.\ Lett.\  {\bf 43}, 103 (1979).

\bibitem{Shifman:1979if} 
  M.~A.~Shifman, A.~I.~Vainshtein and V.~I.~Zakharov,
  Nucl.\ Phys.\ B {\bf 166}, 493 (1980).
    
\bibitem{Zhitnitsky:1980tq} 
  A.~R.~Zhitnitsky,
  Sov.\ J.\ Nucl.\ Phys.\  {\bf 31}, 260 (1980)
  [Yad.\ Fiz.\  {\bf 31}, 497 (1980)].

\bibitem{Dine:1981rt} 
  M.~Dine, W.~Fischler and M.~Srednicki,
  Phys.\ Lett.\  {\bf 104B}, 199 (1981).

\bibitem{Dolan:2014ska} 
  M.~J.~Dolan, F.~Kahlhoefer, C.~McCabe and K.~Schmidt-Hoberg,
  JHEP {\bf 1503}, 171 (2015);
  Erratum: [JHEP {\bf 1507}, 103 (2015)]
  [\hhref{1412.5174} [hep-ph]].
    
\bibitem{Chang:2000ii} 
  D.~Chang, W.~F.~Chang, C.~H.~Chou and W.~Y.~Keung,
  Phys.\ Rev.\ D {\bf 63}, 091301 (2001)
  [\hhref{hep-ph/0009292}].
    
\bibitem{Marciano:2016yhf} 
  W.~J.~Marciano, A.~Masiero, P.~Paradisi and M.~Passera,
  Phys.\ Rev.\ D {\bf 94}, no. 11, 115033 (2016)
  [\hhref{1607.01022} [hep-ph]].

\bibitem{Krasznahorkay:2015iga} 
  A.~J.~Krasznahorkay {\it et al.},
  Phys.\ Rev.\ Lett.\ {\bf 116}, no. 4, 042501 (2016)
  [\hhref{1504.01527} [nucl-ex]].  
    
\bibitem{Feng:2016ysn} 
  J.~L.~Feng, B.~Fornal, I.~Galon, S.~Gardner, J.~Smolinsky, T.~M.~P.~Tait and P.~Tanedo,
  Phys.\ Rev.\ D {\bf 95}, no. 3, 035017 (2017)
  [\hhref{1608.03591} [hep-ph]].

\bibitem{Ellwanger:2016wfe} 
  U.~Ellwanger and S.~Moretti,
  JHEP {\bf 1611}, 039 (2016)
  [\hhref{1609.01669} [hep-ph]].

\bibitem{Boehm:2014hva} 
  C.~Boehm, M.~J.~Dolan, C.~McCabe, M.~Spannowsky and C.~J.~Wallace,
  JCAP {\bf 1405}, 009 (2014)
  [\hhref{1401.6458} [hep-ph]].

\bibitem{Berlin:2014tja} 
  A.~Berlin, D.~Hooper and S.~D.~McDermott,
  Phys.\ Rev.\ D {\bf 89}, no. 11, 115022 (2014)
  [\hhref{1404.0022} [hep-ph]].

\bibitem{Jaeckel:2010ni} 
  J.~Jaeckel and A.~Ringwald,
  Ann.\ Rev.\ Nucl.\ Part.\ Sci.\ {\bf 60}, 405 (2010)
  [\hhref{1002.0329} [hep-ph]].
  
\bibitem{Arias:2012az} 
  P.~Arias, D.~Cadamuro, M.~Goodsell, J.~Jaeckel, J.~Redondo and A.~Ringwald,
  JCAP {\bf 1206}, 013 (2012)
  [\hhref{1201.5902} [hep-ph]].
  
\bibitem{Irastorza:2013dav} 
  I.~Irastorza {\it et al.} [IAXO Collaboration],
  CERN-SPSC-2013-022.
  
\bibitem{Alekhin:2015byh} 
  S.~Alekhin {\it et al.},
  Rept.\ Prog.\ Phys.\ {\bf 79}, no. 12, 124201 (2016)
  [\hhref{1504.04855} [hep-ph]].

\bibitem{Dobrich:2015jyk} 
  B.~D\"obrich, J.~Jaeckel, F.~Kahlhoefer, A.~Ringwald and K.~Schmidt-Hoberg,
  JHEP {\bf 1602}, 018 (2016)
  [JHEP {\bf 1602}, 018 (2016)]
  [\hhref{1512.03069} [hep-ph]].
  
\bibitem{Kleban:2005rj} 
  M.~Kleban and R.~Rabadan,
  [\hhref{hep-ph/0510183}].
  
\bibitem{Mimasu:2014nea} 
  K.~Mimasu and V.~Sanz,
  JHEP {\bf 1506}, 173 (2015)
  [\hhref{1409.4792} [hep-ph]].

\bibitem{Jaeckel:2015jla} 
  J.~Jaeckel and M.~Spannowsky,
  Phys.\ Lett.\ B {\bf 753}, 482 (2016)
  [\hhref{1509.00476} [hep-ph]].
  
\bibitem{Knapen:2016moh} 
  S.~Knapen, T.~Lin, H.~K.~Lou and T.~Melia,
  Phys.\ Rev.\ Lett.\ {\bf 118}, no. 17, 171801 (2017)
  [\hhref{1607.06083} [hep-ph]].
      
\bibitem{Brivio:2017ije} 
  I.~Brivio, M.~B.~Gavela, L.~Merlo, K.~Mimasu, J.~M.~No, R.~del Rey and V.~Sanz,
  [\hhref{1701.05379} [hep-ph]].

\bibitem{Kim:1989xj} 
  J.~E.~Kim and U.~W.~Lee,
  Phys.\ Lett.\ B {\bf 233}, 496 (1989).
  
\bibitem{Djouadi:1990ms} 
  A.~Djouadi, P.~M.~Zerwas and J.~Zunft,
  Phys.\ Lett.\ B {\bf 259}, 175 (1991).

\bibitem{Rupak:1995kg} 
  G.~Rupak and E.~H.~Simmons,
  Phys.\ Lett.\ B {\bf 362}, 155 (1995)
  [\hhref{hep-ph/9507438}].
  
\bibitem{Izaguirre:2016dfi} 
  E.~Izaguirre, T.~Lin and B.~Shuve,
  [\hhref{1611.09355} [hep-ph]].

\bibitem{Dobrescu:2000jt} 
  B.~A.~Dobrescu, G.~L.~Landsberg and K.~T.~Matchev,
  Phys.\ Rev.\ D {\bf 63}, 075003 (2001)
  [\hhref{hep-ph/0005308}].

\bibitem{Dobrescu:2000yn} 
  B.~A.~Dobrescu and K.~T.~Matchev,
  JHEP {\bf 0009}, 031 (2000)
  [\hhref{hep-ph/0008192}].

\bibitem{Chang:2006bw} 
  S.~Chang, P.~J.~Fox and N.~Weiner,
  Phys.\ Rev.\ Lett.\ {\bf 98}, 111802 (2007)
  [\hhref{hep-ph/0608310}].
 
\bibitem{Draper:2012xt} 
  P.~Draper and D.~McKeen,
  Phys.\ Rev.\ D {\bf 85}, 115023 (2012)
  [\hhref{1204.1061} [hep-ph]].
  
\bibitem{Curtin:2013fra} 
  D.~Curtin {\it et al.},
  Phys.\ Rev.\ D {\bf 90}, no. 7, 075004 (2014)
  [\hhref{1312.4992} [hep-ph]].

\bibitem{Chatrchyan:2012cg} 
  S.~Chatrchyan {\it et al.} [CMS Collaboration],
  Phys.\ Lett.\ B {\bf 726}, 564 (2013)
  [\hhref{1210.7619} [hep-ex]].

\bibitem{CMS:2015iga} 
  CMS Collaboration,
  CMS-PAS-HIG-14-022.

\bibitem{CMS:2016cel} 
  CMS Collaboration,
  CMS-PAS-HIG-14-041.
 
\bibitem{Aad:2015bua} 
  G.~Aad {\it et al.} [ATLAS Collaboration],
  Eur.\ Phys.\ J.\ C {\bf 76}, no. 4, 210 (2016)
  [\hhref{1509.05051} [hep-ex]].
  
\bibitem{Khachatryan:2015nba} 
  V.~Khachatryan {\it et al.} [CMS Collaboration],
  JHEP {\bf 1601}, 079 (2016)
  [\hhref{1510.06534} [hep-ex]].

\bibitem{CMS:2016tgd} 
  CMS Collaboration,
  CMS-PAS-HIG-16-035.
  
\bibitem{Khachatryan:2017mnf} 
  V.~Khachatryan {\it et al.} [CMS Collaboration],
  [\hhref{1701.02032} [hep-ex]].

\bibitem{Khachatryan:2016are} 
  V.~Khachatryan {\it et al.} [CMS Collaboration],
  Phys.\ Lett.\ B {\bf 759}, 369 (2016)
  [\hhref{1603.02991} [hep-ex]].

\bibitem{Aad:2015sva} 
  G.~Aad {\it et al.} [ATLAS Collaboration],
  Phys.\ Rev.\ D {\bf 92}, no. 9, 092001 (2015)
  [\hhref{1505.07645} [hep-ex]].

\bibitem{Branco:2011iw} 
  G.~C.~Branco, P.~M.~Ferreira, L.~Lavora, M.~N.~Rebelo, M.~Sher and J.~P.~Silva,
  Phys.\ Rept.\  {\bf 516}, 1 (2012)
  [\hhref{1106.0034} [hep-ph]].

\bibitem{Bauer:2017nlg} 
  M.~Bauer, M.~Neubert and A.~Thamm,
  Phys.\ Rev.\ Lett.\  {\bf 119}, no. 3, 031802 (2017)
  [\hhref{1704.08207} [hep-ph]].
  
\bibitem{Bauer:2016ydr} 
  M.~Bauer, M.~Neubert and A.~Thamm,
  [\hhref{1607.01016} [hep-ph]];
  M.~Bauer, M.~Neubert and A.~Thamm,
  Phys.\ Rev.\ Lett.\ {\bf 117}, 181801 (2016)
  [\hhref{1610.00009} [hep-ph]].

\bibitem{Toro:2012sv} 
  N.~Toro and I.~Yavin,
  Phys.\ Rev.\ D {\bf 86}, 055005 (2012)
  [\hhref{1202.6377} [hep-ph]].

\bibitem{Chou:2016lxi} 
  J.~P.~Chou, D.~Curtin and H.~J.~Lubatti,
  Phys.\ Lett.\ B {\bf 767}, 29 (2017)
  [\hhref{1606.06298} [hep-ph]].

\bibitem{Curtin:2017izq} 
  D.~Curtin and M.~E.~Peskin,
  \hhref{1705.06327} [hep-ph].

\bibitem{Georgi:1986df} 
  H.~Georgi, D.~B.~Kaplan and L.~Randall,
  Phys.\ Lett.\ {\bf 169B}, 73 (1986).

\bibitem{Bauer:2016lbe} 
  M.~Bauer, C.~H\"orner and M.~Neubert,
  JHEP {\bf 1607}, 094 (2016)
  [\hhref{1603.05978} [hep-ph]].

\bibitem{Chala:2017sjk} 
  M.~Chala, G.~Durieux, C.~Grojean, L.~de Lima and O.~Matsedonskyi,
  [\hhref{1703.10624} [hep-ph]].

\bibitem{Bardeen:1978nq} 
  W.~A.~Bardeen, S.-H.~H.~Tye and J.~A.~M.~Vermaseren,
  Phys.\ Lett.\  {\bf 76B}, 580 (1978).
  
\bibitem{DiVecchia:1980yfw} 
  P.~Di Vecchia and G.~Veneziano,
  Nucl.\ Phys.\ B {\bf 171}, 253 (1980).

\bibitem{inprep}
  M.~Bauer, M.~Neubert and A.~Thamm, in preparation.
  
\bibitem{Spira:1995rr} 
  M.~Spira, A.~Djouadi, D.~Graudenz and P.~M.~Zerwas,
  Nucl.\ Phys.\ B {\bf 453}, 17 (1995)
  [\hhref{hep-ph/9504378}].

\bibitem{diCortona:2015ldu} 
  G.~Grilli di Cortona, E.~Hardy, J.~Pardo Vega and G.~Villadoro,
  JHEP {\bf 1601}, 034 (2016)
  [\hhref{1511.02867} [hep-ph]].

\bibitem{Bardeen:1986yb} 
  W.~A.~Bardeen, R.~D.~Peccei and T.~Yanagida,
  Nucl.\ Phys.\ B {\bf 279}, 401 (1987).
  
\bibitem{Larin:1993tq} 
  S.~A.~Larin,
  Phys.\ Lett.\ B {\bf 303}, 113 (1993)
  [\hhref{hep-ph/9302240}].
  
\bibitem{Poggio:1975af} 
  E.~C.~Poggio, H.~R.~Quinn and S.~Weinberg,
  Phys.\ Rev.\ D {\bf 13}, 1958 (1976).
  
\bibitem{Shifman:2000jv} 
  M.~A.~Shifman,
  [\hhref{hep-ph/0009131}].

\bibitem{Calibbi:2017uvl} 
  L.~Calibbi and G.~Signorelli,
  \hhref{1709.00294} [hep-ph].

\bibitem{Cadamuro:2011fd} 
  D.~Cadamuro and J.~Redondo,
  JCAP {\bf 1202}, 032 (2012)
  [\hhref{1110.2895} [hep-ph]].
  
\bibitem{Millea:2015qra} 
  M.~Millea, L.~Knox and B.~Fields,
  Phys.\ Rev.\ D {\bf 92}, no. 2, 023010 (2015)
  [\hhref{1501.04097} [astro-ph]].

\bibitem{Raffelt:1985nk} 
  G.~G.~Raffelt,
  Phys.\ Rev.\ D {\bf 33}, 897 (1986).
  
\bibitem{Raffelt:1987yu} 
  G.~G.~Raffelt and D.~S.~P.~Dearborn,
  Phys.\ Rev.\ D {\bf 36}, 2211 (1987).

\bibitem{Raffelt:2006cw} 
  G.~G.~Raffelt,
  Lect.\ Notes Phys.\ {\bf 741}, 51 (2008)
  [\hhref{hep-ph/0611350}].
  
\bibitem{Payez:2014xsa} 
  A.~Payez, C.~Evoli, T.~Fischer, M.~Giannotti, A.~Mirizzi and A.~Ringwald,
  JCAP {\bf 1502}, no. 02, 006 (2015)
  [\hhref{1410.3747} [astro-ph]].

\bibitem{Jaeckel:2017tud} 
  J.~Jaeckel, P.~C.~Malta and J.~Redondo,
  arXiv:1702.02964 [hep-ph].

\bibitem{Inoue:2008zp} 
  Y.~Inoue, Y.~Akimoto, R.~Ohta, T.~Mizumoto, A.~Yamamoto and M.~Minowa,
  Phys.\ Lett.\ B {\bf 668}, 93 (2008)
  [\hhref{0806.2230} [astro-ph]].
  
\bibitem{Arik:2008mq} 
  E.~Arik {\it et al.} [CAST Collaboration],
  JCAP {\bf 0902}, 008 (2009)
  [\hhref{0810.4482} [hep-ex]].

\bibitem{Graham:2015ouw} 
  P.~W.~Graham, I.~G.~Irastorza, S.~K.~Lamoreaux, A.~Lindner and K.~A.~van Bibber,
  Ann.\ Rev.\ Nucl.\ Part.\ Sci.\ {\bf 65}, 485 (2015)
  [\hhref{1602.00039} [hep-ex]].

\bibitem{Riordan:1987aw} 
  E.~M.~Riordan {\it et al.},
  Phys.\ Rev.\ Lett.\ {\bf 59}, 755 (1987).
  
\bibitem{Bjorken:1988as} 
  J.~D.~Bjorken {\it et al.},
  Phys.\ Rev.\ D {\bf 38}, 3375 (1988).
 
\bibitem{Balest:1994ch} 
  R.~Balest {\it et al.} [CLEO Collaboration],
  Phys.\ Rev.\ D {\bf 51}, 2053 (1995).

\bibitem{delAmoSanchez:2010ac} 
  P.~del Amo Sanchez {\it et al.} [BaBar Collaboration],
  Phys.\ Rev.\ Lett.\ {\bf 107}, 021804 (2011)
  [\hhref{1007.4646} [hep-ex]].

\bibitem{Aaboud:2017bwk} 
  M.~Aaboud {\it et al.} [ATLAS Collaboration],
  \hhref{1702.01625} [hep-ex].

\bibitem{Armengaud:2013rta} 
  E.~Armengaud {\it et al.},
  JCAP {\bf 1311}, 067 (2013)
  [\hhref{1307.1488} [astro-ph]].

\bibitem{Essig:2010gu} 
  R.~Essig, R.~Harnik, J.~Kaplan and N.~Toro,
  Phys.\ Rev.\ D {\bf 82}, 113008 (2010)
  [\hhref{1008.0636} [hep-ph]].

\bibitem{Merkel:2014avp}
 H.~Merkel {\it et al.},
 Phys.\ Rev.\ Lett.\  {\bf 112}, no. 22, 221802 (2014)
 [\hhref{1404.5502} [hep-ex]].

\bibitem{Lees:2014xha}
 J.~P.~Lees {\it et al.} [BaBar Collaboration],
 Phys.\ Rev.\ Lett.\  {\bf 113}, no. 20, 201801 (2014)
 [\hhref{1406.2980} [hep-ex]].

\bibitem{Liu:2017htz}
 Y.~S.~Liu and G.~A.~Miller,
 Phys.\ Rev.\ D {\bf 96}, no. 1, 016004 (2017)
 [\hhref{1705.01633} [hep-ph]].

\bibitem{TheBABAR:2016rlg} 
  J.~P.~Lees {\it et al.} [BaBar Collaboration],
  Phys.\ Rev.\ D {\bf 94}, no. 1, 011102 (2016)
  [arXiv:1606.03501 [hep-ex]].

\bibitem{Bennett:2006fi} 
  G.~W.~Bennett {\it et al.} [Muon g-2 Collaboration],
  Phys.\ Rev.\ D {\bf 73}, 072003 (2006)
  [\hhref{hep-ex/0602035}].
  
\bibitem{Davier:2016iru} 
  M.~Davier,
  [\hhref{1612.02743} [hep-ph]].

\bibitem{Jegerlehner:2017lbd} 
  F.~Jegerlehner,
  [\hhref{1705.00263} [hep-ph]].
  
\bibitem{Leveille:1977rc} 
  J.~P.~Leveille,
  Nucl.\ Phys.\ B {\bf 137}, 63 (1978).

\bibitem{Haber:1978jt} 
  H.~E.~Haber, G.~L.~Kane and T.~Sterling,
  Nucl.\ Phys.\ B {\bf 161}, 493 (1979).

\bibitem{Chatrchyan:2013vaa} 
  S.~Chatrchyan {\it et al.} [CMS Collaboration],
  Phys.\ Lett.\ B {\bf 726}, 587 (2013)
  [\hhref{1307.5515} [hep-ex]].

\bibitem{Aad:2014fia} 
  G.~Aad {\it et al.} [ATLAS Collaboration],
  Phys.\ Lett.\ B {\bf 732}, 8 (2014)
  [\hhref{1402.3051} [hep-ex]].
  
\bibitem{Dittmaier:2011ti} 
  S.~Dittmaier {\it et al.} [LHC Higgs Cross Section Working Group],
  [\hhref{1101.0593} [hep-ph]].

\bibitem{Han:2003wu} 
  T.~Han, H.~E.~Logan, B.~McElrath and L.~T.~Wang,
  Phys.\ Rev.\ D {\bf 67}, 095004 (2003)
  [\hhref{hep-ph/0301040}].

\bibitem{Perelstein:2003wd} 
  M.~Perelstein, M.~E.~Peskin and A.~Pierce,
  Phys.\ Rev.\ D {\bf 69}, 075002 (2004)
  [\hhref{hep-ph/0310039}].

\bibitem{Dedes:2014hga} 
  A.~Dedes and D.~Karamitros,
  Phys.\ Rev.\ D {\bf 89}, no. 11, 115002 (2014)
  [\hhref{1403.7744} [hep-ph]].

\bibitem{Freitas:2015hsa} 
  A.~Freitas, S.~Westhoff and J.~Zupan,
  JHEP {\bf 1509}, 015 (2015)
  [\hhref{1506.04149} [hep-ph]].

\bibitem{Pierce:2006dh} 
  A.~Pierce, J.~Thaler and L.~T.~Wang,
  JHEP {\bf 0705}, 070 (2007)
  [\hhref{hep-ph/0609049}].

\bibitem{Khachatryan:2016vau} 
  G.~Aad {\it et al.} [ATLAS and CMS Collaborations],
  JHEP {\bf 1608}, 045 (2016)
  [\hhref{1606.02266} [hep-ex]].
  
\bibitem{ATL-PHYS-PUB-2014-016}
  ATLAS Collaboration,
  public note ATL-PHYS-PUB-2014-016 
  [https://cds.cern.ch/record/1956710].

\bibitem{Aad:2015pla} 
  G.~Aad {\it et al.} [ATLAS Collaboration],
  JHEP {\bf 1511}, 206 (2015)
  [\hhref{1509.00672} [hep-ex]].
  
\bibitem{Khachatryan:2016whc} 
  V.~Khachatryan {\it et al.} [CMS Collaboration],
  JHEP {\bf 1702}, 135 (2017)
  [\hhref{1610.09218} [hep-ex]].

\bibitem{ATLAS:2012soa} 
  ATLAS Collaboration,
  ATLAS-CONF-2012-079.
  
\bibitem{Gonzalez-Alonso:2014rla} 
  M.~Gonzalez-Alonso and G.~Isidori,
  Phys.\ Lett.\ B {\bf 733}, 359 (2014)
  [\hhref{1403.2648} [hep-ph]].

\bibitem{Chala:2015cev} 
  M.~Chala, M.~Duerr, F.~Kahlhoefer and K.~Schmidt-Hoberg,
  Phys.\ Lett.\ B {\bf 755}, 145 (2016)
  [\hhref{1512.06833} [hep-ph]].

\bibitem{Aad:2015oqa} 
  G.~Aad {\it et al.} [ATLAS Collaboration],
  Phys.\ Rev.\ D {\bf 92}, no. 5, 052002 (2015)
  [\hhref{1505.01609} [hep-ex]].

\bibitem{Aaboud:2016oyb} 
  M.~Aaboud {\it et al.} [ATLAS Collaboration],
  Eur.\ Phys.\ J.\ C {\bf 76}, no. 11, 605 (2016)
  [\hhref{1606.08391} [hep-ex]].

\bibitem{Anastasiou:2016cez} 
  C.~Anastasiou, C.~Duhr, F.~Dulat, E.~Furlan, T.~Gehrmann, F.~Herzog, A.~Lazopoulos and B.~Mistlberger,
  JHEP {\bf 1605}, 058 (2016)
  [\hhref{1602.00695} [hep-ph]].

\bibitem{Gomez-Ceballos:2013zzn} 
  M.~Bicer {\it et al.} [TLEP Design Study Working Group],
  JHEP {\bf 1401}, 164 (2014)
  [\hhref{1308.6176} [hep-ex]].
  
\bibitem{ALEPH:2005ab} 
  S.~Schael {\it et al.} [ALEPH and DELPHI and L3 and OPAL and SLD Collaborations and LEP Electroweak Working Group and SLD Electroweak Group and SLD Heavy Flavour Group],
  Phys.\ Rept.\  {\bf 427}, 257 (2006)
  [\hhref{hep-ex/0509008}].

\bibitem{Acciarri:1994gb} 
  M.~Acciarri {\it et al.} [L3 Collaboration],
  Phys.\ Lett.\ B {\bf 345}, 609 (1995).
  
\bibitem{Abreu:1994du} 
  P.~Abreu {\it et al.} [DELPHI Collaboration],
  Phys.\ Lett.\ B {\bf 327}, 386 (1994).

\bibitem{Aaltonen:2013mfa} 
  T.~A.~Aaltonen {\it et al.} [CDF Collaboration],
  Phys.\ Rev.\ Lett.\  {\bf 112}, 111803 (2014)
  [\hhref{1311.3282} [hep-ex]].
  
\bibitem{Acton:1991dq} 
  P.~D.~Acton {\it et al.} [OPAL Collaboration],
  Phys.\ Lett.\ B {\bf 273}, 338 (1991).

\bibitem{Aad:2016naf} 
  G.~Aad {\it et al.} [ATLAS Collaboration],
  Phys.\ Lett.\ B {\bf 759}, 601 (2016)
  [\hhref{1603.09222} [hep-ex]].
  
\bibitem{Peskin:1991sw} 
  M.~E.~Peskin and T.~Takeuchi,
  Phys.\ Rev.\ D {\bf 46}, 381 (1992).

\bibitem{Peskin:1995ev} 
  M.~E.~Peskin and D.~V.~Schroeder,
  {\em An Introduction to Quantum Field Theory} (Addison Wesley, 1995).

\bibitem{Baak:2014ora} 
  M.~Baak {\it et al.} [Gfitter Group Collaboration],
  Eur.\ Phys.\ J.\ C {\bf 74}, 3046 (2014)
  [\hhref{1407.3792} [hep-ph]].

\bibitem{Abbiendi:2003dh} 
  G.~Abbiendi {\it et al.} [OPAL Collaboration],
  Eur.\ Phys.\ J.\ C {\bf 33}, 173 (2004)
  [\hhref{hep-ex/0309053}].

\bibitem{Janot:2015gjr} 
  P.~Janot,
  JHEP {\bf 1602}, 053 (2016)
  [\hhref{1512.05544} [hep-ph]].
  
\bibitem{deBlas:2016ojx} 
  J.~de Blas, M.~Ciuchini, E.~Franco, S.~Mishima, M.~Pierini, L.~Reina and L.~Silvestrini,
  JHEP {\bf 1612}, 135 (2016)
  [\hhref{1608.01509} [hep-ph]].

\bibitem{Manohar:1983md} 
  A.~Manohar and H.~Georgi,
  Nucl.\ Phys.\ B {\bf 234}, 189 (1984).

\bibitem{Luty:1997fk} 
  M.~A.~Luty,
  Phys.\ Rev.\ D {\bf 57}, 1531 (1998)
  [\hhref{hep-ph/9706235}].

\bibitem{Cohen:1997rt} 
  A.~G.~Cohen, D.~B.~Kaplan and A.~E.~Nelson,
  Phys.\ Lett.\ B {\bf 412}, 301 (1997)
  [\hhref{hep-ph/9706275}].

\bibitem{Feruglio:1992wf} 
  F.~Feruglio,
  Int.\ J.\ Mod.\ Phys.\ A {\bf 8}, 4937 (1993)
  [\hhref{hep-ph/9301281}].

\bibitem{Grojean:2013qca} 
  C.~Grojean, O.~Matsedonskyi and G.~Panico,
  JHEP {\bf 1310}, 160 (2013)
  [\hhref{1306.4655} [hep-ph]].

\bibitem{Panico:2015jxa} 
  G.~Panico and A.~Wulzer,
  Lect.\ Notes Phys.\  {\bf 913}, pp.1 (2016)
  [\hhref{1506.01961} [hep-ph]].

\bibitem{Fujikawa:1979ay} 
  K.~Fujikawa,
  Phys.\ Rev.\ Lett.\  {\bf 42}, 1195 (1979).

\bibitem{Fujikawa:1980eg} 
  K.~Fujikawa,
  Phys.\ Rev.\ D {\bf 21}, 2848 (1980)
  Erratum: [Phys.\ Rev.\ D {\bf 22}, 1499 (1980)].

\bibitem{Krauss:1986bq} 
  L.~M.~Krauss and M.~B.~Wise,
  Phys.\ Lett.\ B {\bf 176}, 483 (1986).

\bibitem{Horsley:2015eaa} 
  R.~Horsley {\it et al.},
  J.\ Phys.\ G {\bf 43}, no. 10, 10LT02 (2016)
  [\hhref{1508.06401} [hep-lat]].

\bibitem{Basak:2015lla} 
  S.~Basak {\it et al.} [MILC Collaboration],
  J.\ Phys.\ Conf.\ Ser.\  {\bf 640}, no. 1, 012052 (2015)
  [\hhref{1510.04997} [hep-lat]].

\bibitem{Beneke:2000ry} 
  M.~Beneke, G.~Buchalla, M.~Neubert and C.~T.~Sachrajda,
  Nucl.\ Phys.\ B {\bf 591}, 313 (2000)
  [\hhref{hep-ph/0006124}].

\bibitem{Kaiser:2000gs} 
  R.~Kaiser and H.~Leutwyler,
  Eur.\ Phys.\ J.\ C {\bf 17}, 623 (2000)
  [\hhref{hep-ph/0007101}].

\bibitem{Knapen:2017ebd} 
  S.~Knapen, T.~Lin, H.~K.~Lou and T.~Melia,
  arXiv:1709.07110 [hep-ph].

\bibitem{Dolan:2017osp} 
  M.~J.~Dolan, T.~Ferber, C.~Hearty, F.~Kahlhoefer and K.~Schmidt-Hoberg,
  arXiv:1709.00009 [hep-ph].
  
\end{thebibliography}
\end{document}